\RequirePackage{fix-cm}
\documentclass[smallextended]{svjour3}       
\smartqed  
\usepackage[utf8]{inputenc}
\usepackage{graphicx, float}
\usepackage{amsmath, amssymb, amsfonts}
\usepackage{mathtools,hyperref}
\usepackage{xcolor,verbatim}
\definecolor{darkgreen}{rgb}{0.0, 0.5, 0.0}

\DeclarePairedDelimiter{\floor}{\lfloor}{\rfloor}
\begin{document}

\title{On the probability of finding a nonphysical solution through shadowing\footnote{Preprint available at \cite{preprint}}}
\date{June 2020}


\author{Nisha Chandramoorthy \and Qiqi Wang}

\institute{N. Chandramoorthy, Corresponding Author \at
              Mechanical Engineering and Center for Computational Science and Engineering \\
              77 Massachusetts Avenue, Cambridge MA
               \email{nishac@mit.edu}           
           \and
           Q. Wang \at
              Department of Aeronautics and Astronautics and Center for Computational Science and Engineering \\
              77 Massachusetts Avenue, Cambridge MA \\
              \email{qiqi@mit.edu}
}

\maketitle

\begin{abstract}
This paper proves that shadowing solutions can be almost surely nonphysical.
This finding invalidates the argument that small perturbations in a chaotic system can only have a small impact on its statistical behavior.
This theoretical finding has implications for many applications
in which chaotic mechanics plays an important role.
It suggests, for example, that we can control the climate through subtle perturbations. It also suggests that numerical simulations of chaotic dynamics, such as turbulent flows, may fail to predict the true long-term or statistical behavior.
\end{abstract}

\maketitle
\section{Introduction}
In this paper, we show that a large mismatch is possible in the statistical or long-time averages measured in a pair of chaotic solutions to slightly different models. We construct numerical and analytical examples that illustrate this feature of chaotic dynamics. The analysis of the long-term behavior of {\em shadowing} solutions in the majority of the chaotic models constructed in this paper reveals that shadowing solutions are almost surely {\em nonphysical}: they do not represent the behavior of a physically observed solution starting almost everywhere. This also leads us to the conclusion that shadowing-based sensitivities \cite{qiqi-lss} of long-time averages to parameter changes can be incorrect. Thus, the main conjecture of this paper urges us to reconsider the use of chaotic numerical simulations, and statistical sensitivity analysis on these simulations. These implications are particularly important for climate studies, where chaotic models are widely used to make long-term predictions. In particular, three implications of the evidence in this paper may be stated, the third of which harnesses the statistical inconsistency between slightly different models:
\begin{itemize}
    \item we cannot trust the long-time averages generated by numerical simulations of chaotic processes, even when we eliminate model uncertainties stemming from imperfect modeling of the underlying physics, and statistical noise due to finite-time averaging;
    \item we cannot trust the sensitivities of long-time averages or statistical averages to small parameter changes computed by using shadowing-based methods for sensitivity computation;
    \item it is possible for two models separated by tiny changes in parameter values to have drastically different statistical behavior. This means that the Lorenzian butterfly can significantly alter the climate (statistical/long-term average) of Texas, and not just its weather (short-term events).
\end{itemize}
The paper is organized as follows. In section \ref{sec:shadowingIntro}, we introduce the dynamical systems concept of shadowing and its applications in the context of chaotic numerical simulations. In section \ref{sec:nonphysical}, we define the notion of physicality and give examples to illustrate the different ways in which to obtain nonphysical solutions. Section \ref{sec:shadowingnonphysical} is dedicated to several examples of perturbed tent maps that have nonphysical shadowing solutions. In the discussion in section \ref{sec:discussion}, we illustrate with an example that a small parameter perturbation can change the statistics significantly. The supplementary material, section \ref{sec:supp}, contains the numerical procedure to generate the probability distributions of the shadowing solutions.

\section{Shadowing and applications}
\label{sec:shadowingIntro}
Our main analysis tool in advancing our conjecture is {\em shadowing}. Shadowing and other dynamical systems concepts used in this paper are introduced in detail, where they appear, so as to make our presentation accessible to the large audience of computational scientists across different disciplines. Shadowing refers to the relationship between a pair of solutions to slightly different governing equations. The difference between the governing equations can be due to parameter perturbation or numerical error. The solution to one governing equation is said to shadow a solution to a second, slightly different equation if
the first solution stays close to the second solution for some amount of time.

When the slightly different governing equations are uniformly hyperbolic, it has been long known that infinitely 
long shadowing solutions exist \cite{anosov}\cite{bowen}.
Numerical experiments in non-hyperbolic systems have shown that shadowing
solutions may still exist \cite{grebogi}\cite{hammel} for a 
finite amount of time.

The existence of shadowing trajectories underlies many applications.
Numerical simulations of turbulent flows have been widely used to study its statistical behavior.
It is argued, based on shadowing, that such numerical simulations of chaotic dynamical systems can be
useful, despite the \emph{butterfly effect} that causes infinitesimal errors in the state to grow exponentially in time. Because of numerical and modeling errors, there is invariably a small difference between the true governing physics and the equations solved on a computer. By nature of chaos, this difference increases exponentially, as the system is evolved forward in time. Thus, the fidelity of numerical solutions of chaotic systems, such
as turbulent flows, to the true physics is called into question.  The existence of shadowing solutions
is used to argue for the usefulness of such numerical solutions.

When certain conditions for shadowing theorems are met, the numerical solution would be an
approximation to a ``true'' solution that satisfies the real governing physics.

Shadowing is also used in sensitivity analysis of chaotic dynamical systems.  In particular,
it is used in computing how long-time-averages in a chaotic system
respond to small perturbations in the governing equation \cite{qiqi-lss}\cite{angxiu-lss}\cite{lasagna}. In this application, the derivative
of the long-time-averages is computed using a solution to the perturbed governing equation
that shadows a solution to the unperturbed equation.  The Least Squares Shadowing \cite{qiqi-lss}\cite{angxiu-lss} method
uses this concept.

There is an implicit assumption in both of these applications.  The assumption is that
the shadowing trajectory is a \emph{physical} trajectory, a trajectory on which
the long-time-average of a quantity is equal to the ensemble average.
Not all solutions satisfying the physical governing equation are considered physical in this sense.  In high-Reynolds number fluid flows, for example, a steady-state, laminar flow solution may satisfy the Navier-Stokes equation. But such a solution would never be observed in reality because it is unstable, and any small perturbation
would trip it into turbulence. Unstable steady-state solutions are not the only
nonphysical solution. Many chaotic dynamical systems have infinitely many periodic
solutions that are, similar to their steady counterpart, unstable. These trajectories can have a probability distribution that is remarkably different from that of a typically observed solution of the governing equation. It is also possible to effect significant change in the statistics of the true governing dynamics by introducing a minor parameter perturbation. Thus, although trajectories of governing equations at slightly different parameters may shadow each other, the long-time-averages that are typically observed at one parameter may not be close to that of the other.    
As we will see
in the next section, even solutions that look ``chaotic'', i.e., unsteady and aperiodic, may not be physical. 

\section{What are physical and nonphysical solutions?}
\label{sec:nonphysical}
Intuitively, we call a solution to a governing equation \emph{physical} if
it represents what one would observe in a physical experiment.  In particular, the statistics measured from a physical solution match the statistics observed
in an experiment.   Not all solutions are physical.  A laminar flow
solution, despite satisfying the governing equation, does not produce the
turbulent statistics one would observe in a high-Reynolds number experiment. Such solutions are thus called \emph{nonphysical}.

In this section, we first describe what distinguishes, mathematically, a physical
and a nonphysical solution.  We also explain why it is theoretically unlikely
to observe a nonphysical solution in an experiment.  We then give a few
examples of these almost-never-observed nonphysical solutions.

\subsection{Physical solutions}
What we have intuitively defined as physical solutions to a governing equation satisfy the following
two criteria: 
\begin{enumerate}
    \item Time-averaged quantities converge in the limit of infinite averaging time.
    Consider $u(t)$ as the solution to a chaotic governing equation, then for a regular observable
    of interest $J(u(t))$,
    \begin{equation} \label{longtimeavg}
        \lim_{T\to\infty}\frac1T \int_0^T J(u(t))\,dt
    \end{equation}
    exists.
    \item For almost any small perturbation to the initial condition $u(0)\to u'(0)$, the perturbed
    solution $u'(t)$ should have the same statistics, i.e.,
    \begin{equation}
        \lim_{T\to\infty}\frac1T \int_0^T J(u(t))\,dt
      = \lim_{T\to\infty}\frac1T \int_0^T J(u'(t))\,dt
    \end{equation}
\end{enumerate}

\begin{figure} \centering
\includegraphics[width=0.48\textwidth]{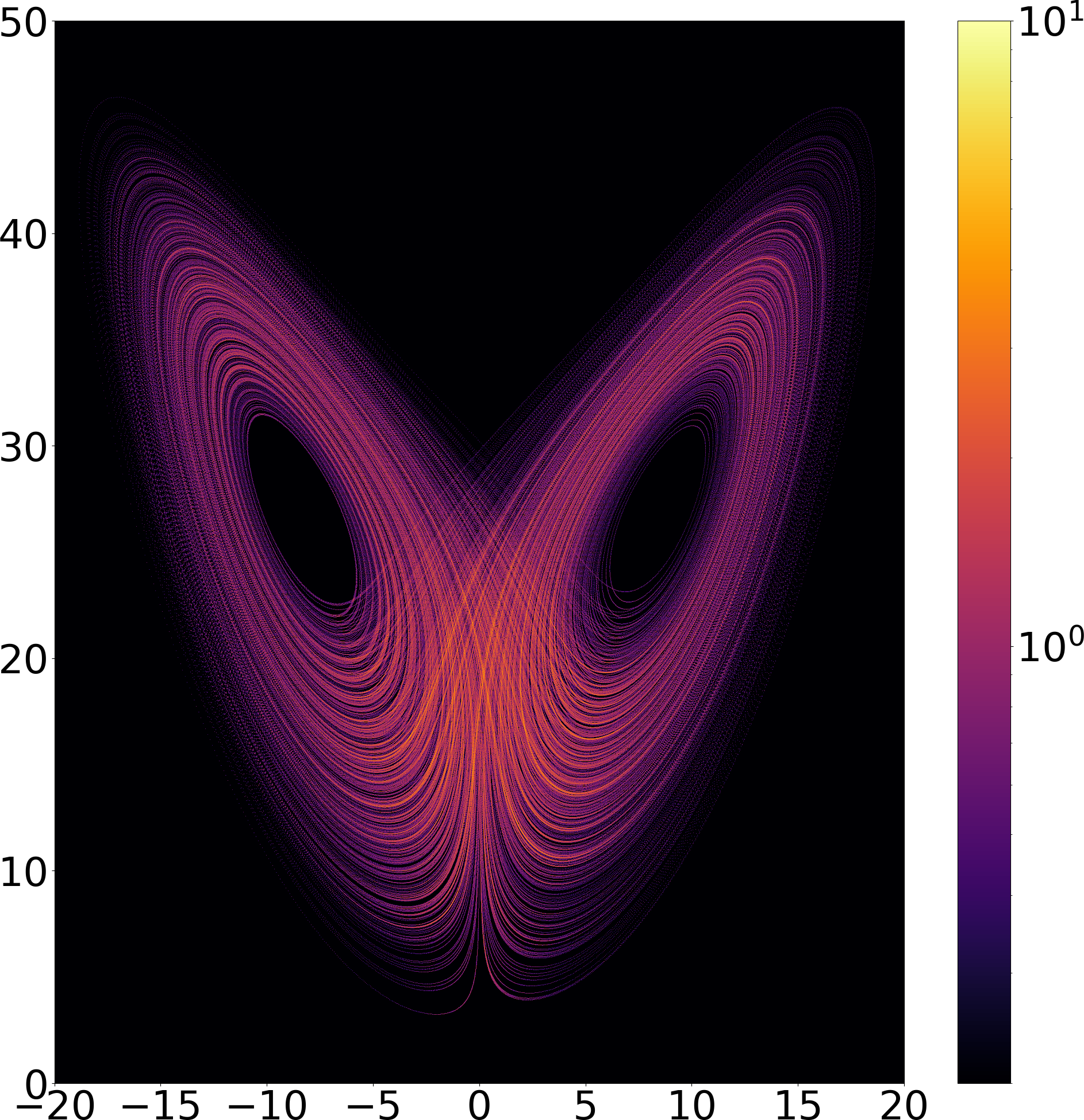}
\hspace{0.02\textwidth}
\includegraphics[width=0.48\textwidth]{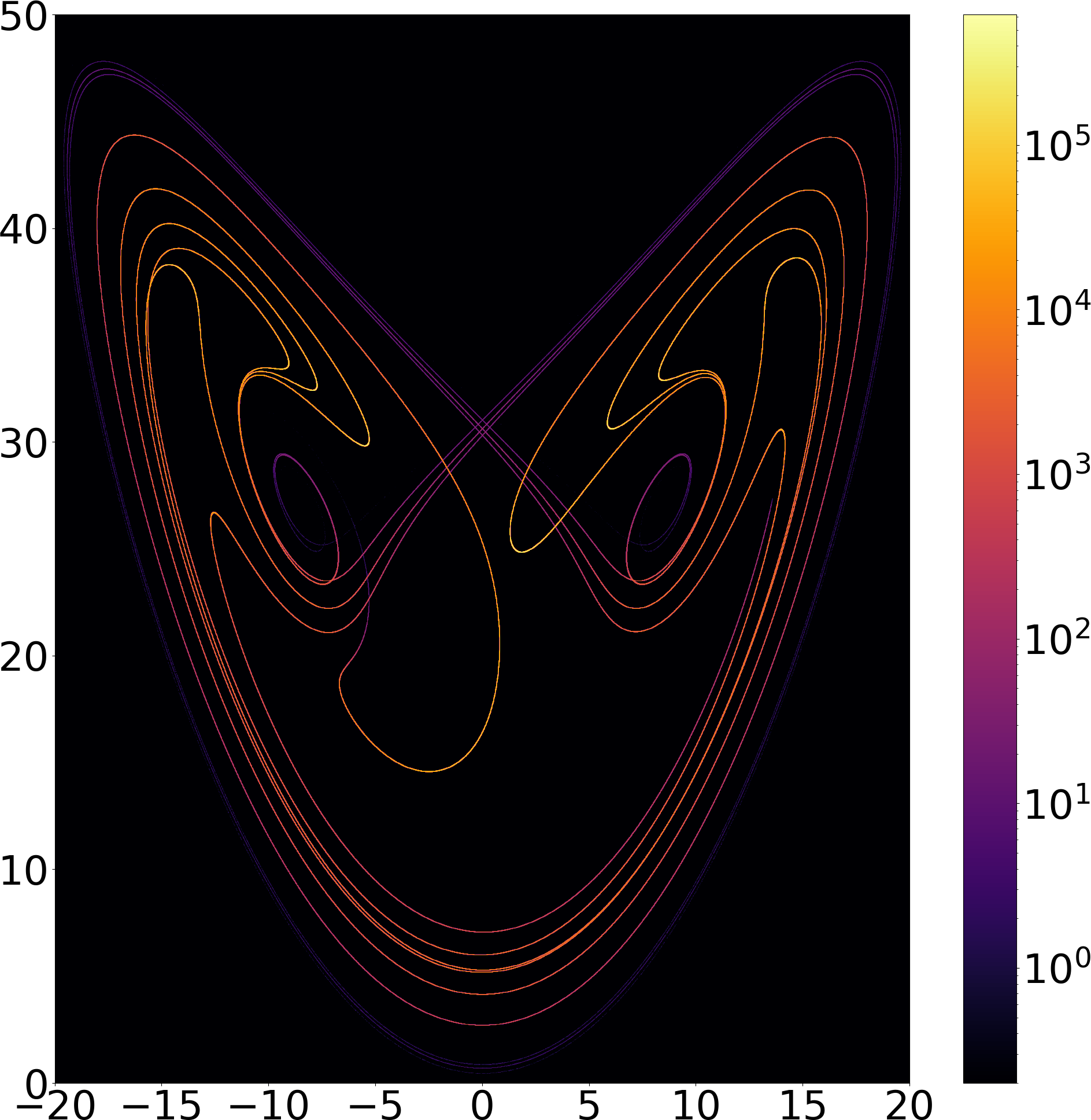}
\caption{
L: An ensemble of initial conditions distributed uniformly in the box $u_1\in[0,1], u_2\in[0,1], u_3\in[28,39]$ 
after 10 time units of evolution, shown on the $u_1-u_3$ plane.
R: distribution of a trajectory of $1000$ time units in length.
In both plots, the color represents the number of samples in a
$2048\times2048$ uniform grid.  The trajectory is sampled every 0.001 time units.
}
\label{fig:lorenz_ergodicity1}
\end{figure}
\begin{figure} \centering
\includegraphics[width=0.48\textwidth]{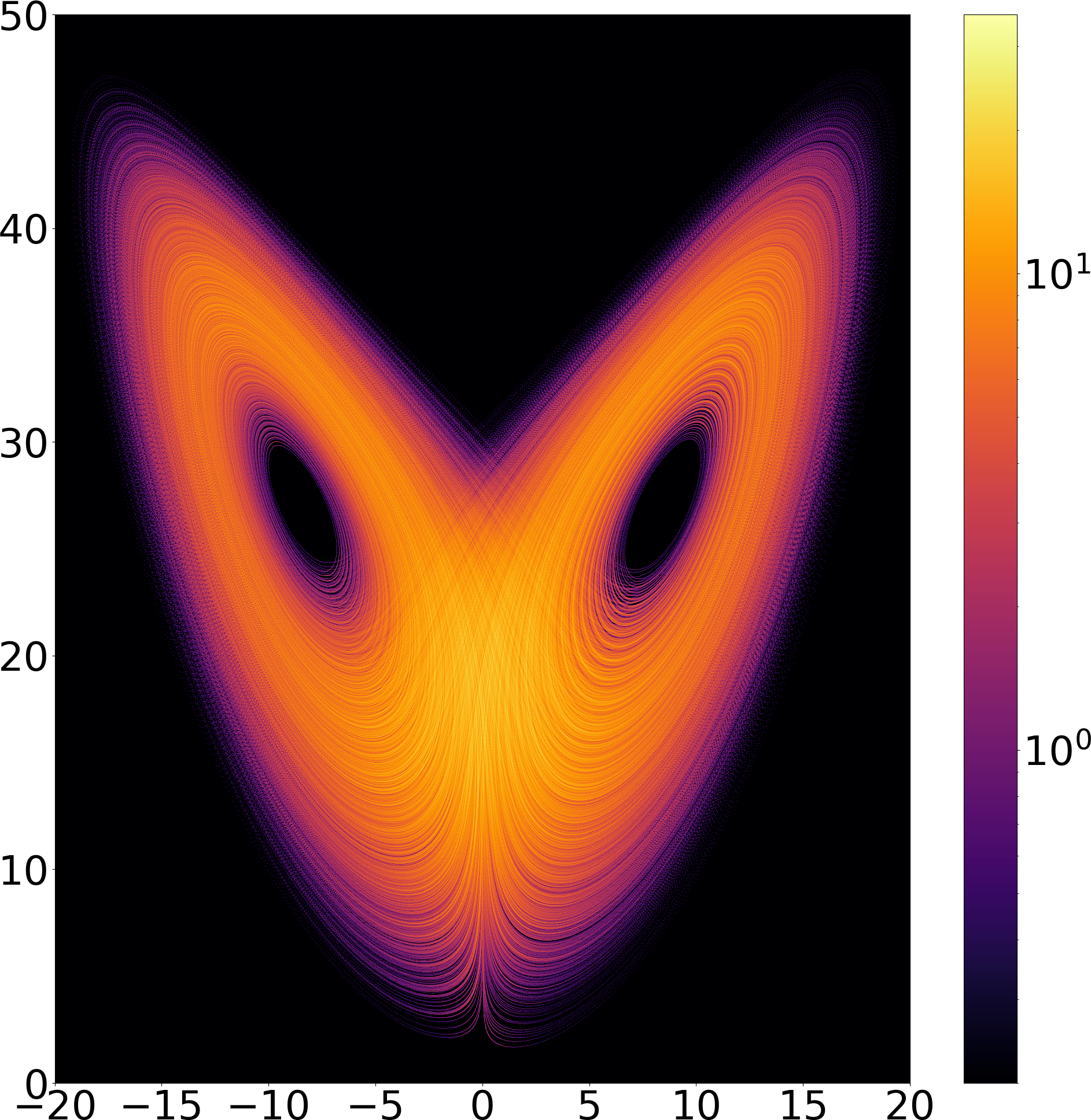}
\hspace{0.02\textwidth}
\includegraphics[width=0.48\textwidth]{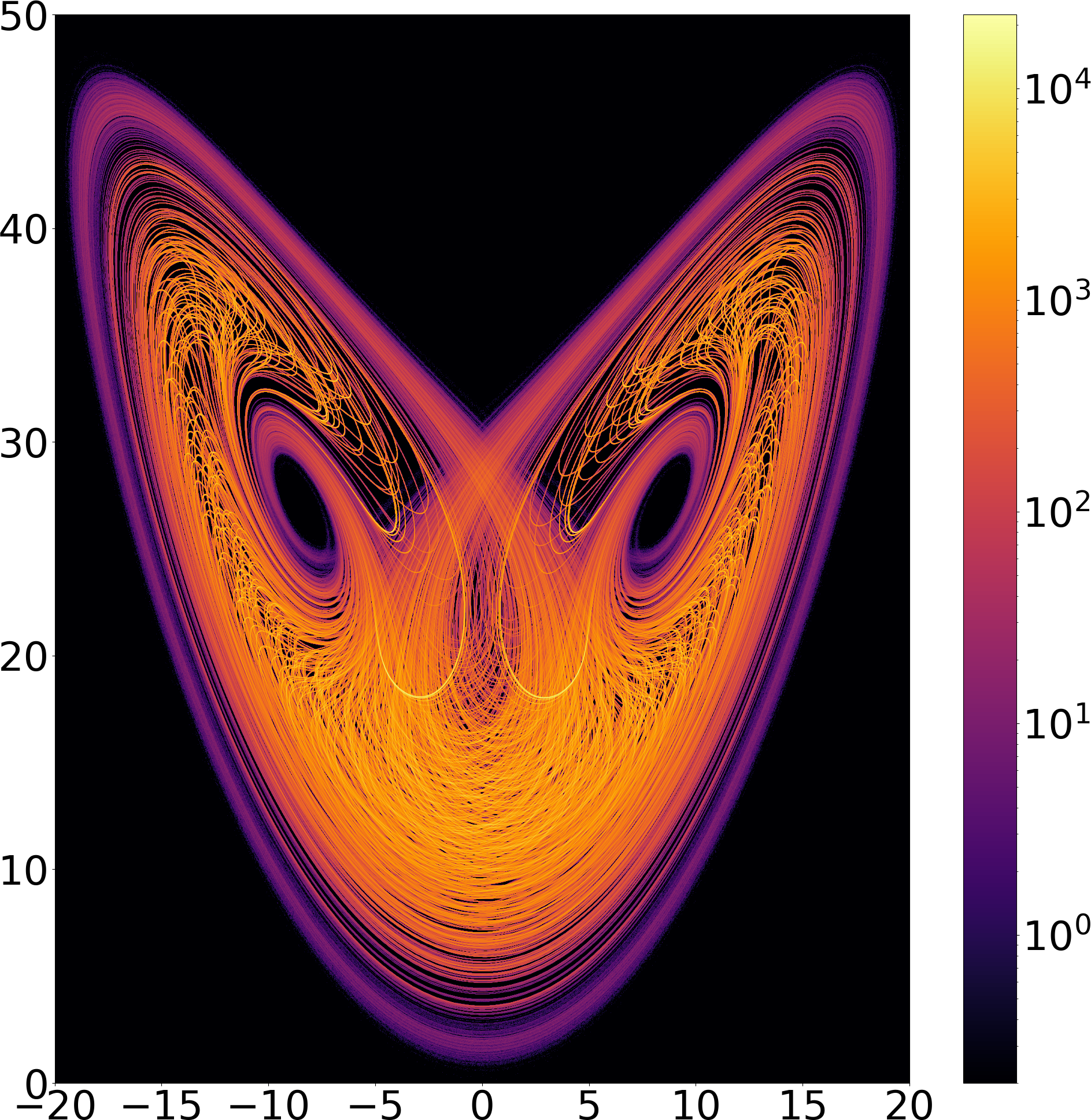}
\caption{
L: distribution of the same ensemble as in Figure \ref{fig:lorenz_ergodicity1}
after another 5 time units of evolution.
R: distribution of the same trajectory as in Figure \ref{fig:lorenz_ergodicity1}
evolved for $10,000$ time units.
}
\label{fig:lorenz_ergodicity2}
\end{figure}
\begin{figure} \centering
\includegraphics[width=0.48\textwidth]{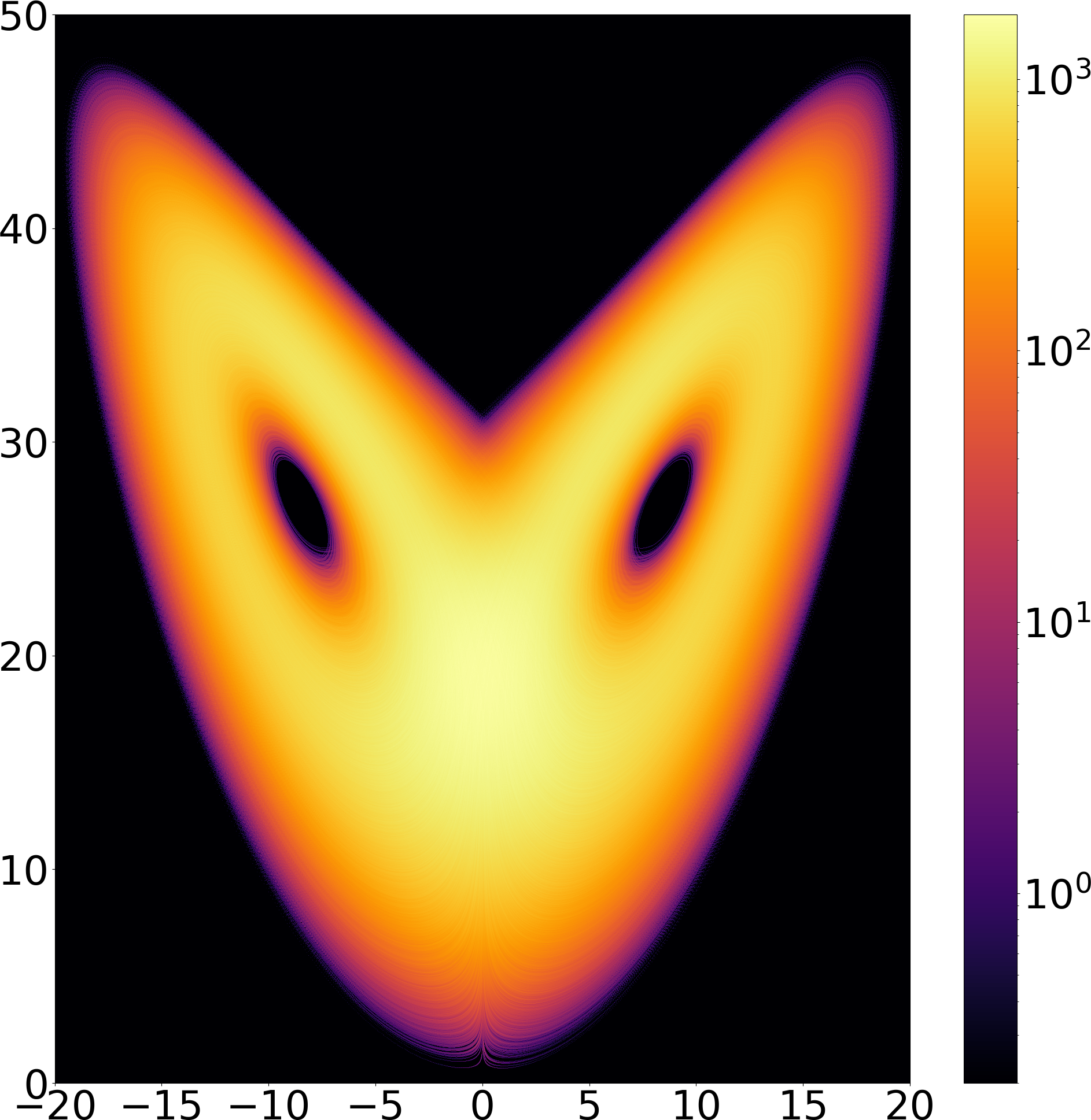}
\hspace{0.02\textwidth}
\includegraphics[width=0.48\textwidth]{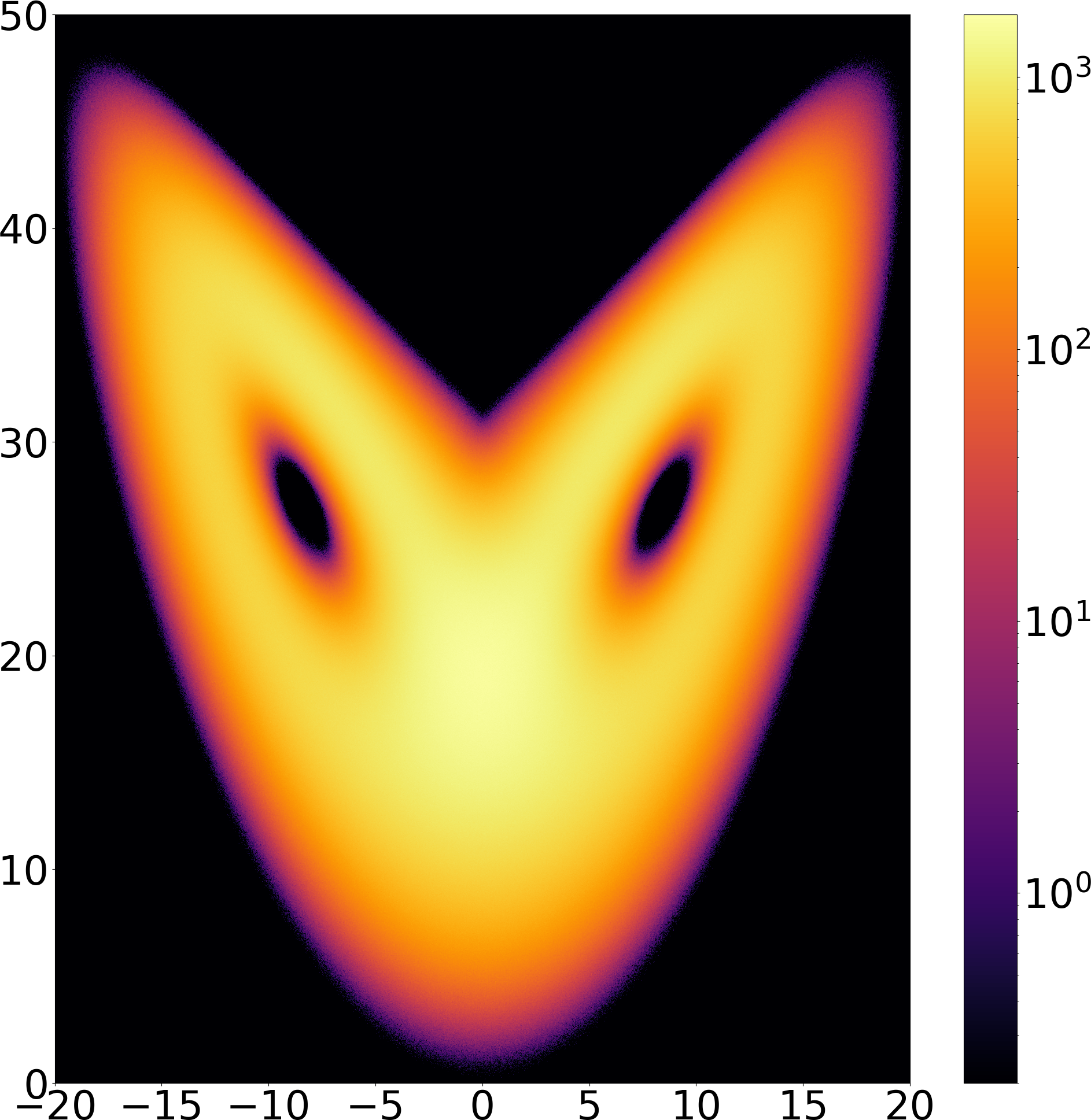}
\caption{
L: distribution of the same ensemble as in Figures \ref{fig:lorenz_ergodicity1}
and \ref{fig:lorenz_ergodicity2} after a total of 50 time units of evolution.
R: distribution of the same trajectory as in Figures \ref{fig:lorenz_ergodicity1}
and \ref{fig:lorenz_ergodicity2} evolved for $100,000$ time units.
}
\label{fig:lorenz_ergodicity3}
\end{figure}

The left column of Figures \ref{fig:lorenz_ergodicity1}-\ref{fig:lorenz_ergodicity3}
illustrates a physical solution of the Lorenz system,
\begin{equation}
\label{eqn:lorenz}
\frac{du}{dt} =
    \frac{d}{dt}\begin{pmatrix}u_1\\u_2\\u_3\end{pmatrix} =
    \begin{pmatrix}
        \sigma (u_2-u_1) \\
        u_1(\rho -u_3) - u_2 \\
        u_1\,u_2 - \beta\,u_3
\end{pmatrix}
\end{equation}
where $\sigma=10,\rho=28$, and $\beta=8/3$.  The solution starts at the initial
condition $u(0)=(0.01, 0.01, 28)$.  After a long time evolution, we observe that
the solution has visited a large portion of the $u_1-u_3$ plane with a varying, but well-defined, frequency.  We can use a probability distribution, $\mu$, to quantify
how frequently a physical solution visits portions of the phase space.
Specifically, for a subset $A$ of the phase space, $\mu(A)$ measures the
fraction of time a very long physical solution spends inside the subset $A$.
With this distribution defined, the infinite-time average of any quantity $J(u)$ can be represented as an average of $J$ over the entire phase space $U$, weighted
by this statistical distribution $\mu$.  Mathematically,
\begin{equation}
    \frac1T \int_0^T J(u(t))\,dt \xrightarrow{T\to\infty}
    \int_U J(u)\: d\mu(u)
\end{equation}

Remarkably, the distribution $\mu$ not only characterizes the history of
a long physical solution, but also describes the settled state of
an ensemble of solutions.
This is illustrated on the right column of Figures \ref{fig:lorenz_ergodicity1}--\ref{fig:lorenz_ergodicity3}.
We generate these plots by starting from an ensemble of about one billion initial
conditions, randomly and uniformly spaced in a small three-dimensional box.  All these billion solutions are evolved by solving Eq. \ref{eqn:lorenz} for 10, 15, and 50 time units, to obtain the plots. 
We observe that, as time evolves, the ensemble spreads over an increasingly
larger portion of the $u_1-u_3$ plane. After a long time, the ensemble settles into a time-invariant \emph{attractor} that contains the \emph{unstable manifold}, which appears as filaments forming the attractor. The probability distribution of the ensemble on the attractor becomes identical to the distribution of a single, very long physical
solution, which is also contained in the attractor (Figure \ref{fig:lorenz_ergodicity3}).

This remarkable agreement has been thoroughly studied under the subject
of ergodic theory. Under surprisingly weak conditions, a solution starting from
{\bf almost} any initial condition, chosen randomly from a set enclosing the attractor, is a physical solution \cite{young}.   
Meanwhile, an ensemble of trajectories starting from any distribution with a 
finite density also evolves towards the same final distribution, $\mu$. Due to expansion of a volume of solutions tangent to the attractor filaments, a finite density under long-time evolution becomes absolutely continuous on the unstable manifold, i.e., the likelihood of a trajectory visiting any set not intersecting the attractor filaments is zero. The 
stationary distribution achieved on long-time evolution of the ensemble that has the absolute continuity property is called the Sinai-Ruelle-Bowen
(SRB) measure. The absolute continuity property is sufficient to ensure that the SRB measure is the same $\mu.$ That is, the SRB measure is physically observed in the sense that physical solutions produce long-time-averages which are expectations with respect to the SRB measure. 

Note that almost any, not any, initial condition leads to a physical solution. A set of special initial conditions
contained in a neighborhood of the attractor may exist starting from which physical solutions are not generated. These initial conditions do not produce the same statistics as the physical solutions. This special set of initial conditions is Lebesgue measure zero -- one
has zero chance of finding such an initial condition by randomly sampling.
Nonphysical solutions thus take an effort to find.  Nevertheless, they
turn out to be important when discussing shadowing, the topic of this paper.
We first introduce a type of obviously nonphysical solutions in the next
subsection, before discussing a less obvious type in section \ref{sec:quasiphysical}.

\subsection{Nonphysical solutions Type I: Periodic Solutions}

\begin{figure}\centering
\includegraphics[width=0.48\textwidth]{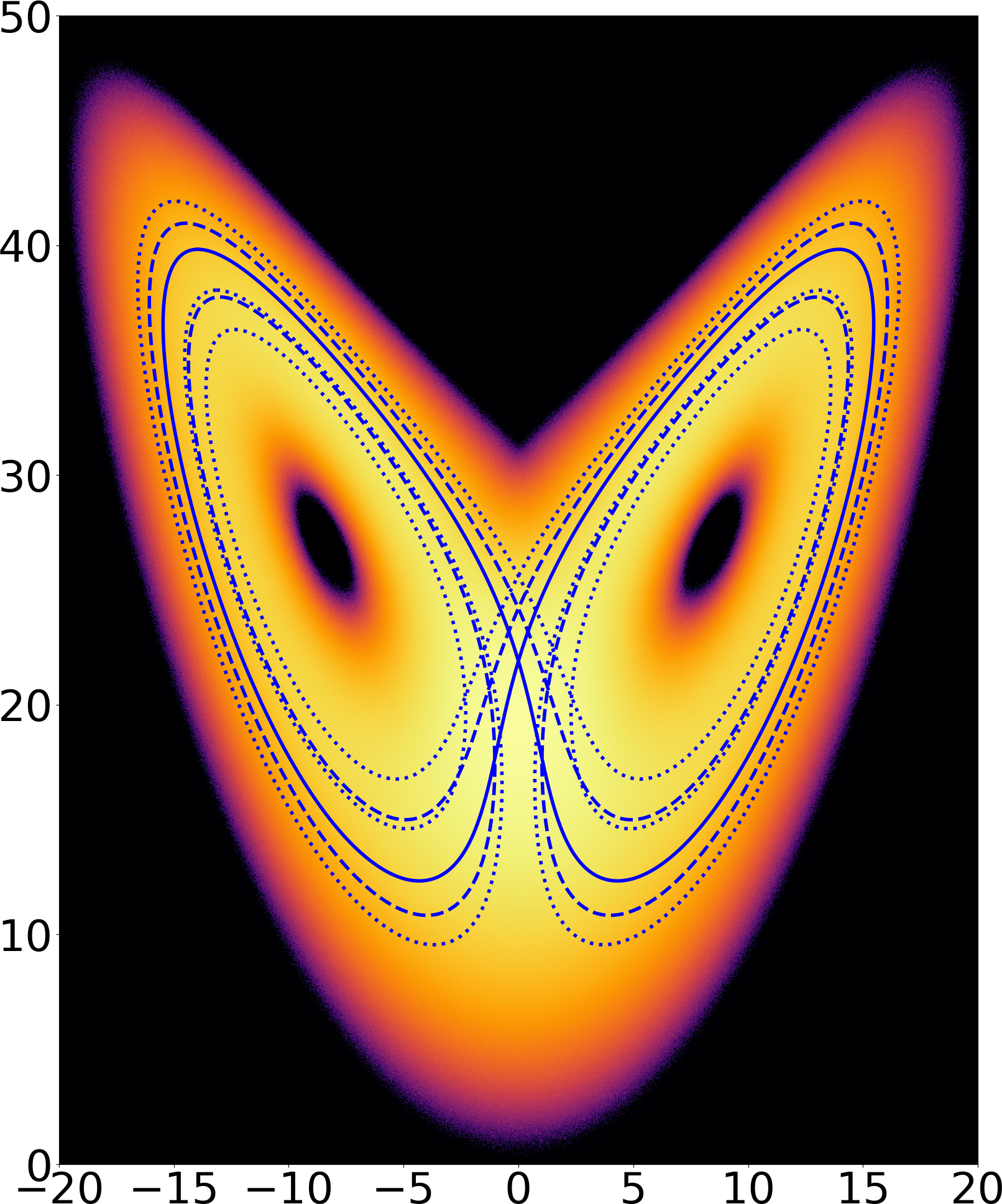}
\caption{Periodic solutions of the Lorenz equation, overlaid on top of
its SRB distribution.  The solid line, dashed line, and dotted line
represent three distinct periodic solutions.}
\label{fig:lorenz_periodic}
\end{figure}

A periodic solution is nonphysical because it 
does not visit as much
of the phase space as a physical solution does.
Figure \ref{fig:lorenz_periodic} shows
a few periodic solutions of the Lorenz equation. Comparing these solutions
to the physical solution visualized in Figures \ref{fig:lorenz_ergodicity1}--\ref{fig:lorenz_ergodicity3},
we see that the periodic solutions are significantly more limited in their
extent of exploration.

Because periodic solutions have a more limited extent in the phase space, their
statistics are different from physical solutions.
Here we illustrate the difference using the mean of two
quantities of interest
\begin{align}
		J_1(u) = u_3\;{\rm and}\;\quad J_2(u) = e^{-\dfrac{u_3^2}2}.
\end{align}
Table \ref{tab:lorenz_periodic_stats} shows these quantities
of interest averaged over the three periodic solutions shown in Figure
\ref{fig:lorenz_periodic}, compared against those averaged over a
physical solution.  Here, Periodic \#1, \#2, and \#3 correspond to
the solid, dashed, and dotted lines, respectively.
We observe from the table that the mean of $J_1$ over the periodic
solutions is different but comparable to the mean over physical solutions.  The mean of $J_2$ over the periodic solutions, on the other hand,
is orders of magnitude different from that of the physical solutions.
These differences disqualify the periodic solutions from being physical.
\begin{table}[H]
    \centering
    \begin{tabular}{c|c|c|c|c}
             & Periodic \#1 & Periodic \#2
             & Periodic \#3 & Physical solutions \\
    \hline
        $J_1$& 23.05 &
               23.19 &
               23.37 &
               23.67 \\
        $J_2$&$4\times 10^{-35}$&
              $7\times 10^{-28}$&
              $3\times 10^{-22}$&
              $1.58\times 10^{-05}$
    \end{tabular}
    \caption{Comparison of statistics computed from periodic solutions of the Lorenz equation with the statistics computed from physical solutions}
    \label{tab:lorenz_periodic_stats}
\end{table}

\begin{figure}\centering
\includegraphics[width=0.48\textwidth]{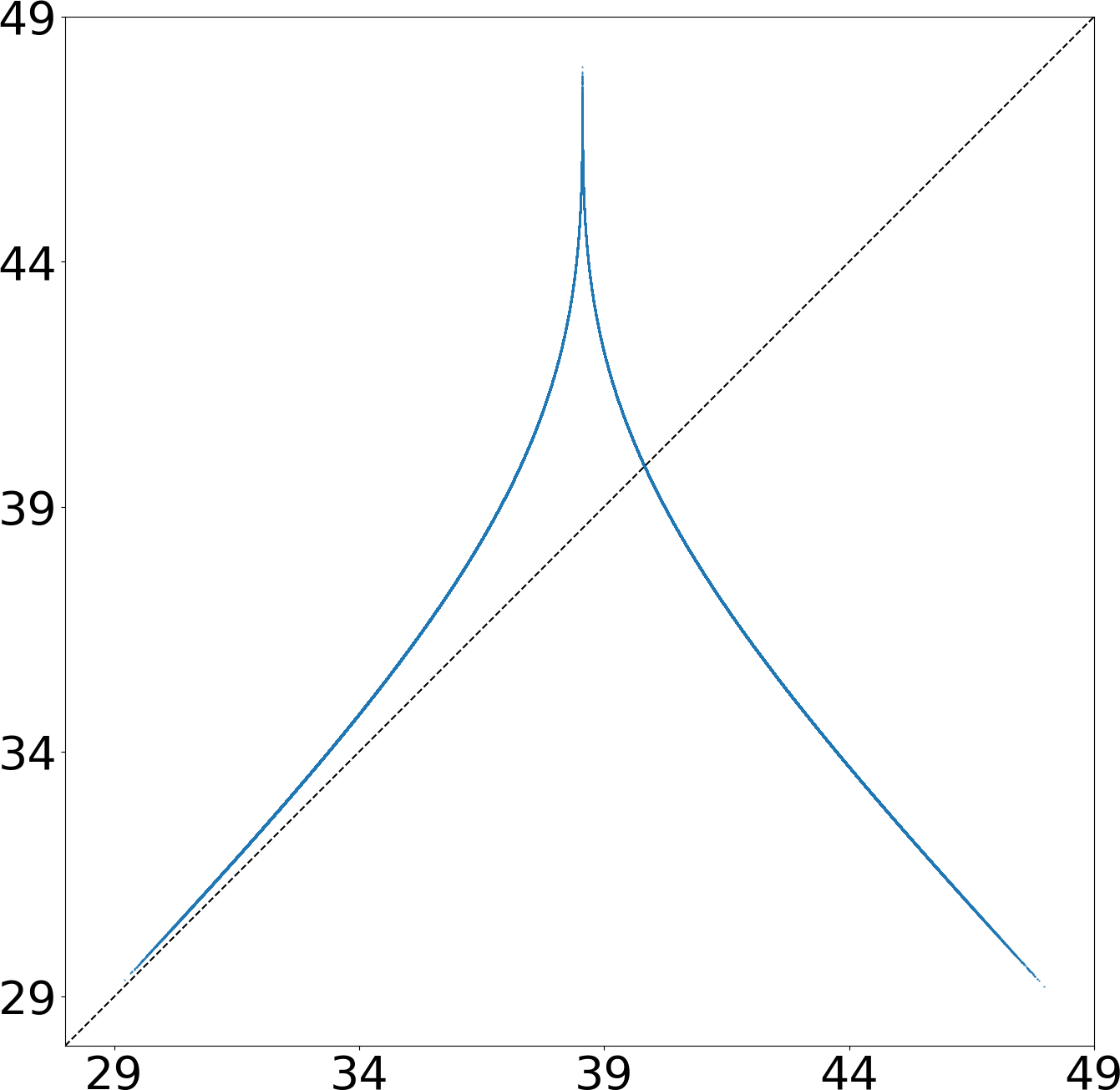}
\hspace{0.02\textwidth}
\includegraphics[width=0.48\textwidth]{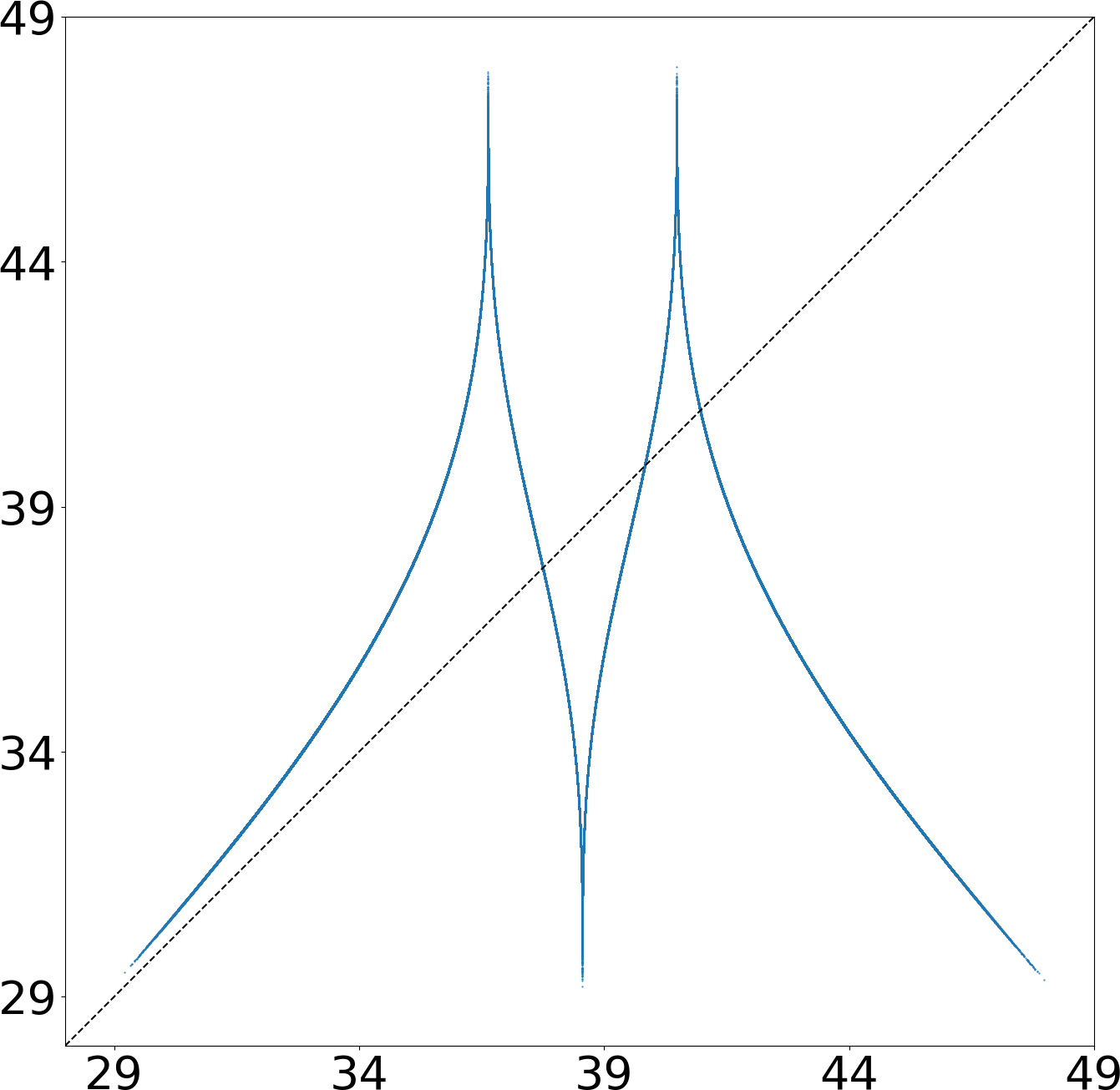}
\caption{L: the Lorenz map.  The x-axis is the $n$-th local maximum of $u_3(t)$ over a
long solution; the y-axis is the $(n+1)$-th local maximum of $u_3(t)$.  The intersection of this curve
		with the dashed line ($y=x$) indicates the initial condition for the solid line
in Figure \ref{fig:lorenz_periodic}.
R: the Lorenz map iterated twice.  The x-axis is the $n$-th local maximum
of $u_3(t)$; the y-axis is the $(n+2)$-th maximum.  The intersections with the diagonal dashed
line indicate the initial conditions for both the dotted and dashed lines in
Figure \ref{fig:lorenz_periodic}.}
\label{fig:lorenz_map}
\end{figure}

While periodic solutions are generally difficult to find,
the Lorenz equation has a special feature that makes the task significantly
easier.  In a typical solution to the Lorenz equation, the 
$u_3(t)$ component oscillates in a pattern that appears neither regular nor random.  Lorenz
observed that the height of one peak in the oscillation can predict the height of the next peak.  He quantified his observation by the Lorenz map,
as shown in Figure \ref{fig:lorenz_map}.

The Lorenz map provides us a tool to find as many periodic solutions as we want.
By intersecting the map with a diagonal line, we can find a local maximum of $u_3(t)$
for which the next maximum is almost the same value.  We then look up the values of
$u_1(t)$ and $u_2(t)$ when $u_3(t)$ achieves this maximum.  This gives us an initial
condition starting from which the solution is nearly periodic.   We can similarly
intersect the second iterate of the Lorenz map (right plot of Figure \ref{fig:lorenz_map}),
and the third iterate, etc, with a diagonal line, to find increasingly
complex periodic solutions. 

This Lorenz map is more than a tool to study the Lorenz equation.  It is a chaotic
dynamical system all by itself. Unlike the Lorenz equation, which is a continuous-time
dynamical system in three dimensions, the Lorenz map is a discrete-time dynamical
system in one dimension.  It exhibits the same sensitivity to initial condition
that characterizes the Lorenz equation.  One can readily observe in the right
plot of Figure \ref{fig:lorenz_map} that a small perturbation in the x-axis
can lead to a large change in the y-axis. This sensitivity grows exponentially
for iterates of this map. A solution to the Lorenz map can be obtained
by extracting the consecutive local maxima of a solution to the Lorenz equation.
If we extract from a physical solution to the Lorenz equation, we obtain a physical solution of the Lorenz map.  It will visit the
interval between 29 and 49 with varying, well-defined frequencies.
By contrast, if one extracts a solution to the Lorenz map from a periodic solution
to the Lorenz equation, the solution will visit only a discrete set of points.
It is thus a periodic, nonphysical solution.  What we learned about
the Lorenz equation could have all been learned from the Lorenz map.

Having discussed periodic solutions in this section,
we move to a second type of nonphysical solutions.
This type is more difficult to find and study than the periodic solutions.
To make it easier, we switch our example to a one-dimensional, 
discrete-time dynamical
systems like the Lorenz map.  Because the Lorenz map lacks a closed form,
we construct, in the next subsection, a one-dimensional discrete-time dynamical system
with a closed form, one that qualitatively resembles the Lorenz map.  This map
will help us, in section \ref{sec:quasiphysical}, to study a more insidious type
of nonphysical solutions that hides in shadowing solutions.

\subsection{Tent map: periodic and physical solutions}
\label{sec:tent}
\begin{figure}\centering
\includegraphics[width=0.48\textwidth]{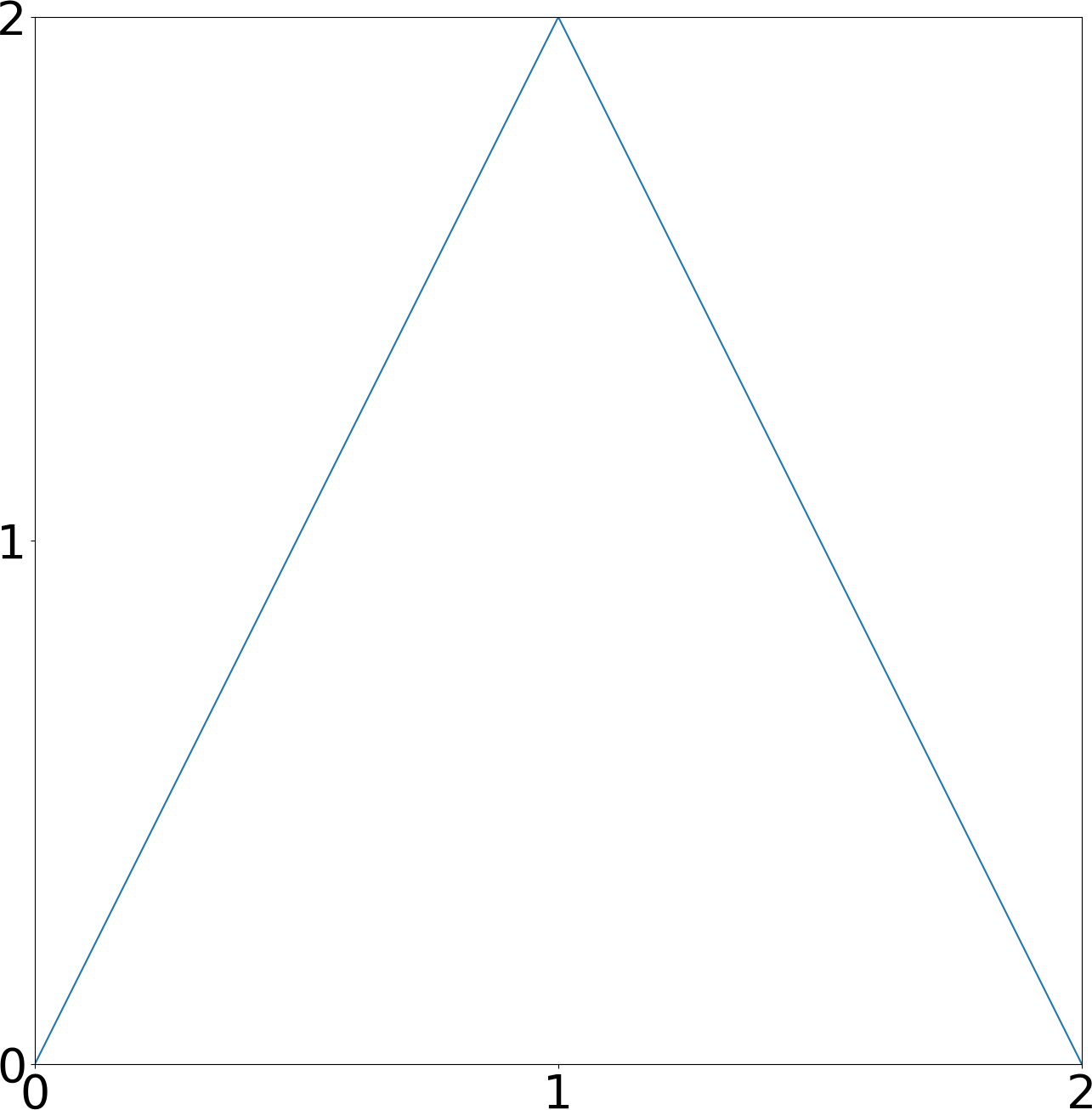}
\caption{The tent map $\varphi$.}
\label{fig:tent_map}
\end{figure}

The tent map is qualitatively similar
to the Lorenz map,
\begin{equation} \label{tentmap}
    \varphi(x) := \begin{cases}
    2x \quad & x < 1 \\
    4 - 2x \quad & 2 \ge x \ge 1
    \end{cases}.
\end{equation}
Due to its simple analytical form, the properties of the tent map have been studied extensively \cite{tent1}\cite{tent2}. Figure \ref{fig:tent_map} shows the tent map.

The tent map $\varphi$ is chaotic because a trajectory $x_0,x_1,\ldots$ satisfying
$x_{i+1} = \varphi(x_i)$ exhibits exponential sensitivity to initial condition. An infinitesimal perturbation of absolute value $\delta x$, applied to an initial condition $x_0$, generates a trajectory that is $2 \delta x$ away from $x_1,$ $4 \delta x$ from $x_2$, $8 \delta x$ from $x_3$ and so on. This exponential divergence of two trajectories that start infinitesimally apart is the butterfly effect that characterizes
chaotic dynamics.

It is easy to find periodic solutions for the tent map.
$\frac43$ maps to itself; $\frac45$ and $\frac85$ map to each other;
$\frac47, \frac87$, and $\frac{12}7$ map circularly.  In fact, any rational number evolves into periodic solutions that visit
only a finite set of rational numbers with the same denominator.
This is compatible
with ergodic theory because all rational numbers in $[0,2]$ comprise a subset of Lebesgue measure zero. That is, we would have zero likelihood of getting a
rational number if we sampled the Lebesgue measure (the uniform distribution) on $[0,2]$.  Instead, with a 100\% probability, one would get an irrational number
that, when iterated under the tent map, leads to a physical trajectory that distributes uniformly on the interval $[0,2]$.

To understand why a physical solution visits the interval $[0,2]$ at a uniform
frequency, it is helpful to view the tent map from a different perspective.
For a randomly chosen $x_0\in[0,2]$, we can represent it in binary form:
\begin{equation}
    x_0 = \sum_{j=0}^{\infty} \frac{x_{0,j}}{2^j}
\end{equation}
where $x_{0,0}\in\{0,1\}$ is the integer component of $x_0$ and $x_{0,j}\in\{0,1\}$ is the $j$th bit after the binary point.
Let $x_{i+1} = \varphi(x_i)$, then it is straightforward to verify from the definition
of the map that its binary representation
\begin{equation}
    x_i = \sum_{j=0}^{\infty} \frac{x_{i,j}}{2^j}
\end{equation}
satisfies
\begin{equation} \label{tent_xor}
    x_{i+1,j} = x_{i,0} \veebar x_{i,j+1},
\end{equation}
where $\veebar$ is the xor operator. To see why, note that 
multiplication by 2 is a left-shift operator in binary, and 
subtraction from 4 flips every bit after the binary point. If $x_0$ is chosen uniformly in $[0,2]$, then each bit $x_{0,j},j=0,1,\ldots$ of $x_0$
has equal probability of being 0 or 1, and each bit is independent of other bits.
It follows from Eq. \ref{tent_xor} that each bit $x_{i,j},j=0,1,\ldots$ of $x_i,
i=1,2,\ldots$ has equal probability of being 0 or 1,
and each bit is still independent of the other bits.
As the map iterates starting from almost any $x_0$, a physical solution explores the entire interval $[0,2]$ uniformly.

\subsection{Nonphysical solutions Type II: Quasi-physical Solutions}
\label{sec:quasiphysical}
The simplicity of the tent map, as well as its binary form (Eq. \ref{tent_xor}),
enables us to study nonphysical solutions that are not periodic.
As in the last section, consider an
\begin{equation}
    x_0 = \sum_{j=0}^{\infty} \frac{x_{0,j}}{2^j}
\end{equation}
in which the bits $x_{0,j}$ are not independent of each other.  Instead,
suppose each bit is more likely to be identical to the previous bit than to be different.
That is, $x_{0,j+1}=x_{0,j}$ with probability $p>\frac12$ for $j=0,1,2,\ldots$.
Then, by Eq. \ref{tent_xor}, the bits of every $x_i, i=1,2,\ldots$ follow the same pattern, namely, each bit repeats the previous bit
with probability $p$. Moreover, consider Eq. \ref{tent_xor} 
for $j=0$ and any $i$:
\begin{equation}
    x_{i+1,0} = x_{i,0} \veebar x_{i,1}.
\end{equation}
Because $x_{i,0}=x_{i,1}$ with probability $p>\frac12$, $x_{i+1,0}=0$ with probability 
$p>\frac12$.  Starting from $x_1$, this solution visits $[0,1]$ with
probability $p>\frac12$.  Instead of visiting $[0,1]$ and $[1,2]$ with
equal probability, as a physical solution does, this solution favors $[0,1]$. Since it provably visits the interval $[0,2]$ with a different frequency from that
of a physical solution, this is a nonphysical solution.

If $p = 1$, the bits of $x_0$ are either all zeros or all ones, which is the same as all zeros in mod 2 arithmetic. So, all further iterates when $x_0$ is 0, are 0, and this is a trivial nonphysical solution of type 1.  Now when $p > \frac12$ but strictly less than 1, 
the solutions we just constructed are both nonphysical and aperiodic.
The bits of $x_0$, though correlated with each other,
still can exhibit an infinite variety of patterns. 
This implies that the solution may not eventually converge to 
any fixed point, nor follow any strict period.  
 As the map iterates, the nonrepetitive bit patterns
shift towards more significant digits, and the solution visits an infinite set of
points. 
 Nevertheless, the solutions observed along a trajectory also 
 do not conform to a uniform distribution on $[0,2]$ since they 
 preferentially visit the first half of this interval.
We call such aperiodic nonphysical solutions ``quasi-physical'' solutions.

\begin{figure}\centering
\includegraphics[width=0.48\textwidth]{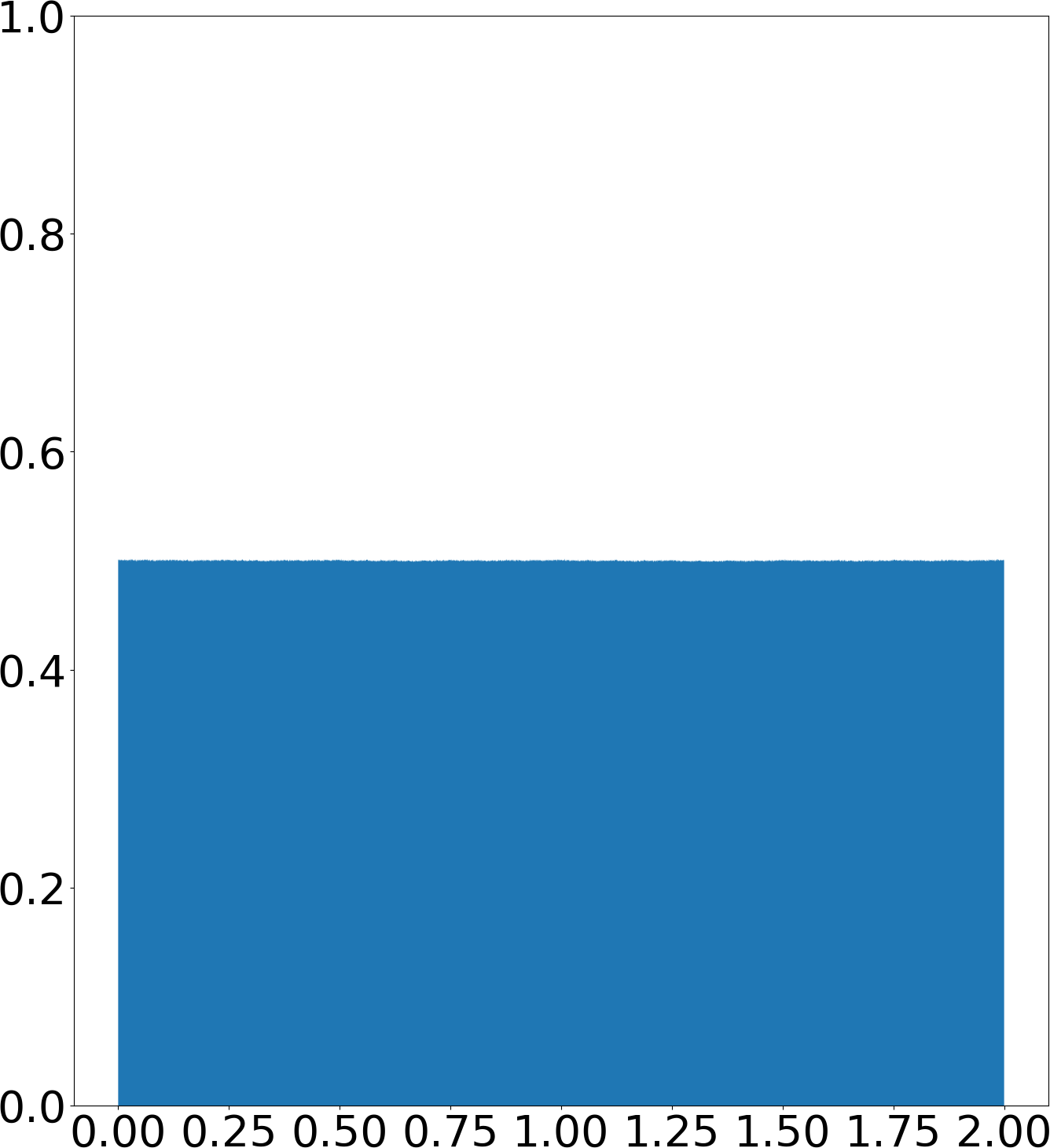}
\hspace{0.02\textwidth}
\includegraphics[width=0.48\textwidth]{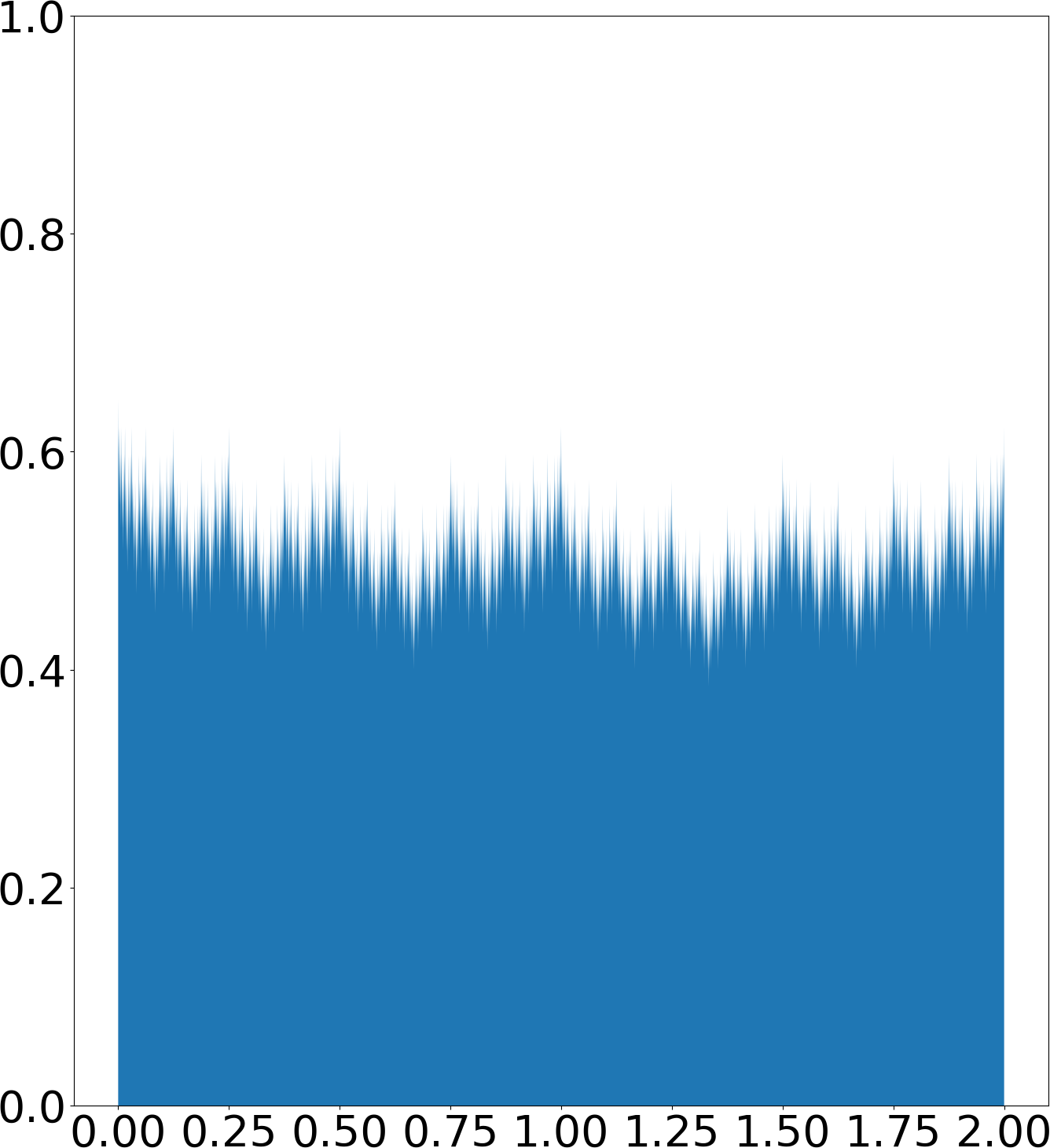}
\\ \vspace{0.05\textwidth}
\includegraphics[width=0.48\textwidth]{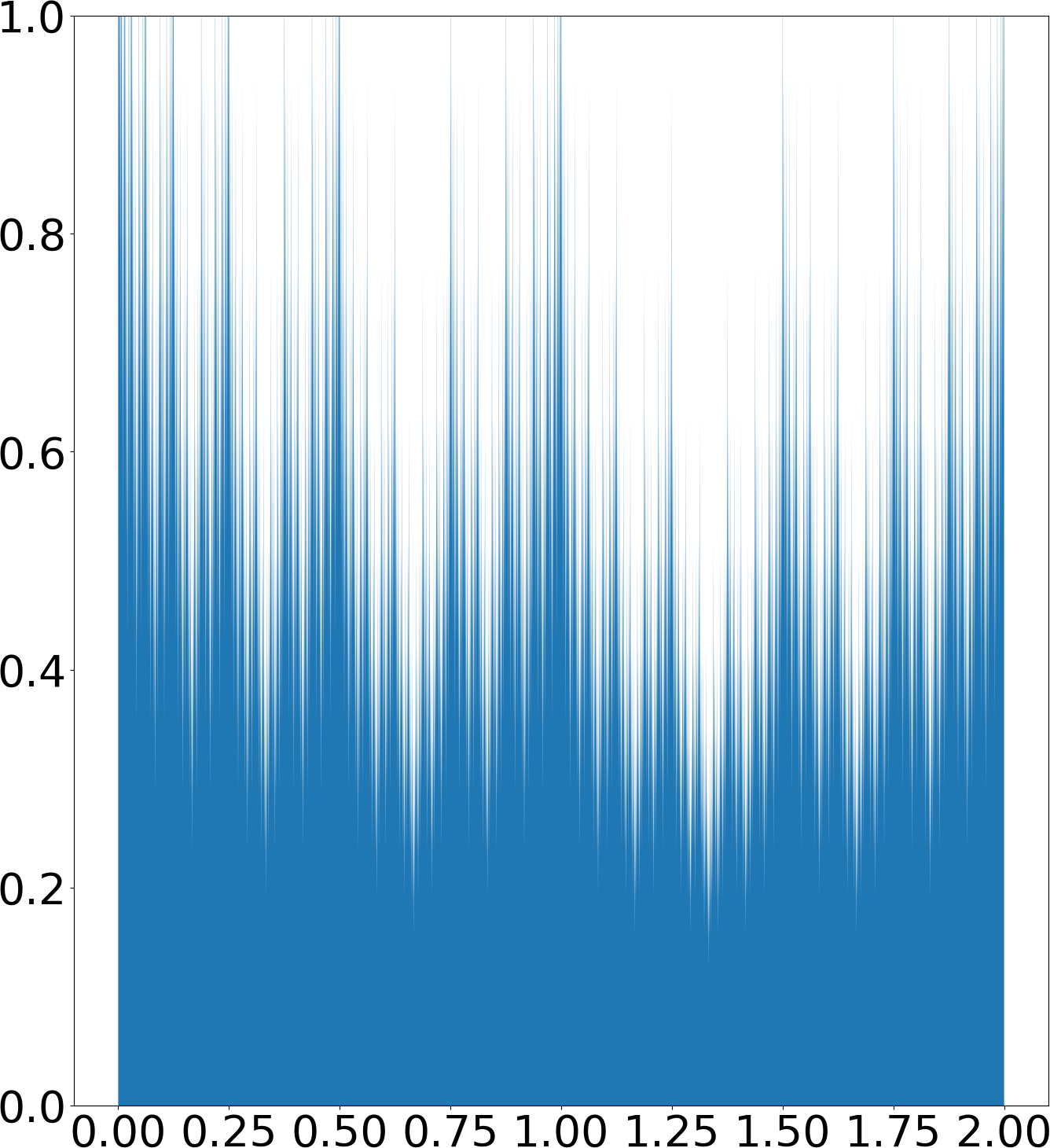}
\hspace{0.02\textwidth}
\includegraphics[width=0.48\textwidth]{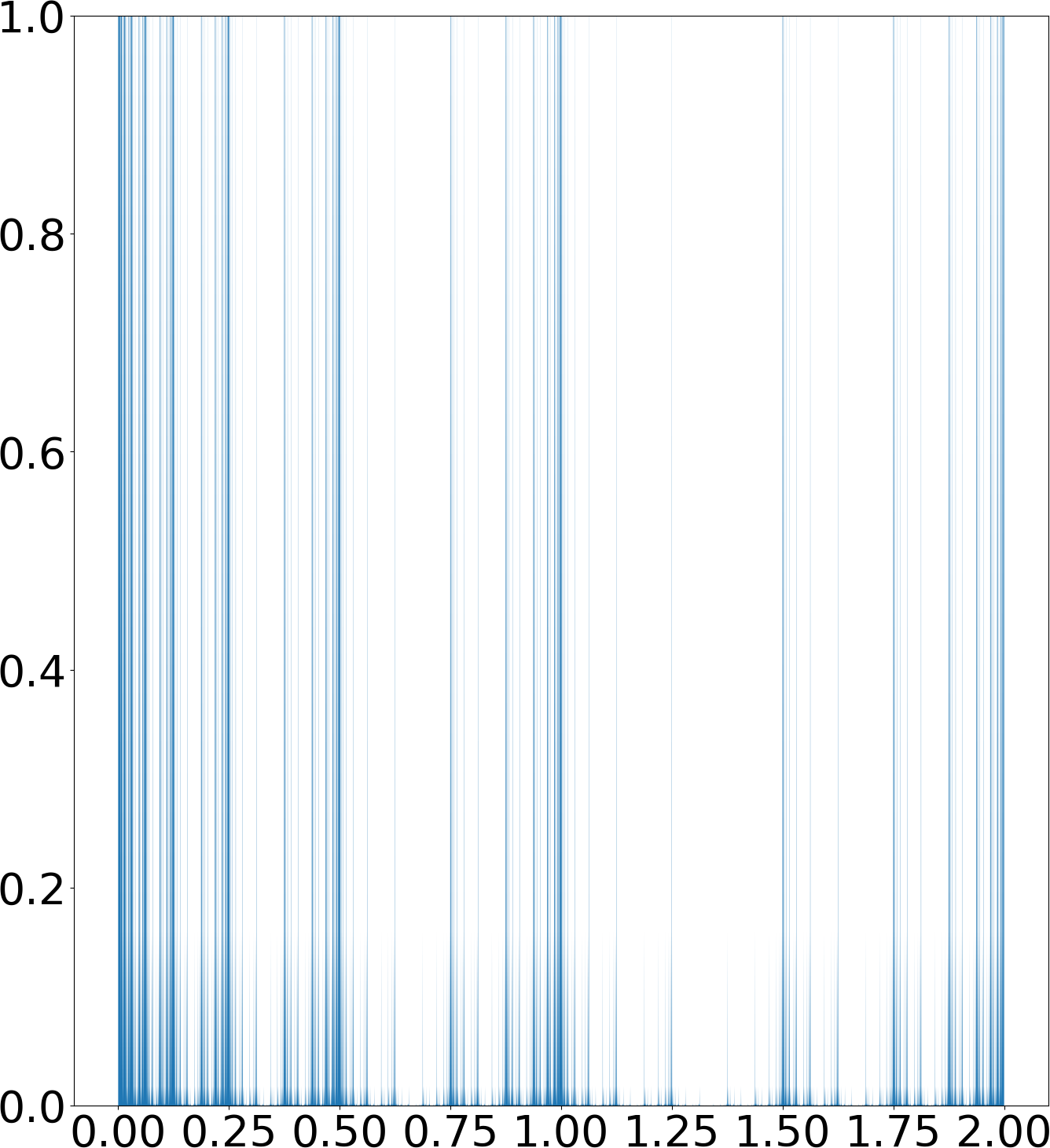}
\caption{Empirical probability distribution of long solutions (a billion steps)
starting from four points whose binary digits have probability $p$ of
repeating the previous digit.
The solution shown in the top-left plot starts from a point with $p=0.5001$;
top-right: $p=0.51$; bottom-left: $p=0.55$; bottom-right: $p=0.9$.
}
\label{fig:tent_quasiphysical}
\end{figure}

Figure \ref{fig:tent_quasiphysical} shows the empirical density functions of
four such quasi-physical solutions.  When $p=0.5001$,
the statistical distribution of the quasi-physical solution
is approximately uniform.  Recall that a physical solution explores $[0,2]$
uniformly.  In this case, the quasi-physical solution has very similar
statistical behavior as a physical solution.  When $p=0.51$, the empirical
distribution becomes ``hairy''.  An apparently fractal pattern emerges.
This fractal pattern further amplifies when $p=0.55$.  Meanwhile,
the density on the left sub-interval, $[0,1]$, becomes obviously higher than
the density on the right sub-interval, $[1,2]$.  This is consistent with our
theoretical analysis at the beginning of this subsection.
When $p=0.9$, the fractal pattern is so intensified that most
of the solution seems to concentrate in a collection of tiny intervals.
These plots expose the diversity of quasi-physical solutions.
Their statistical distribution can be either
very similar or completely different from that of physical solutions.

\begin{figure}\centering
    \centering
    \includegraphics[width=0.48\textwidth]{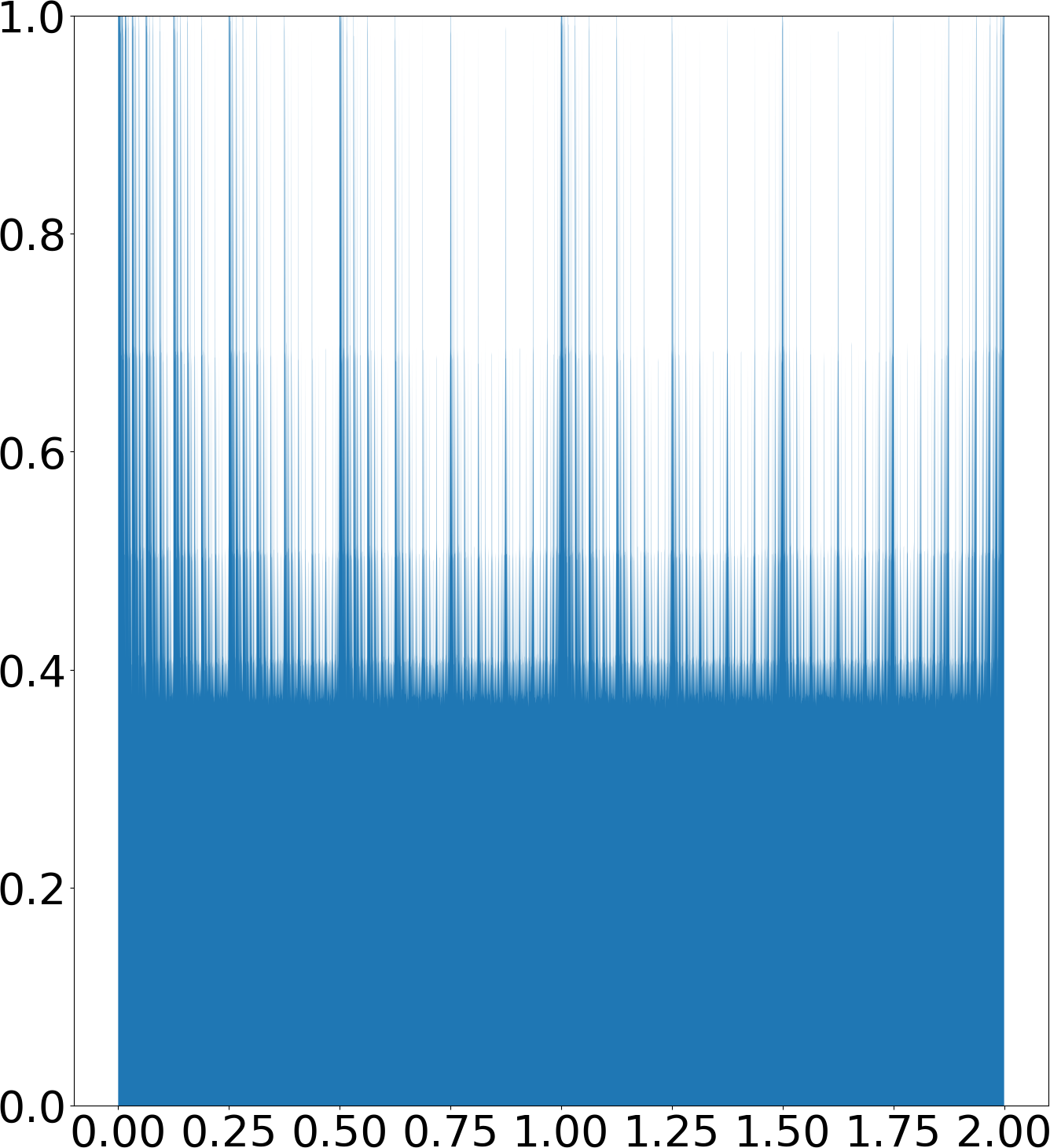}
    \hspace{0.02\textwidth}
    \includegraphics[width=0.48\textwidth]{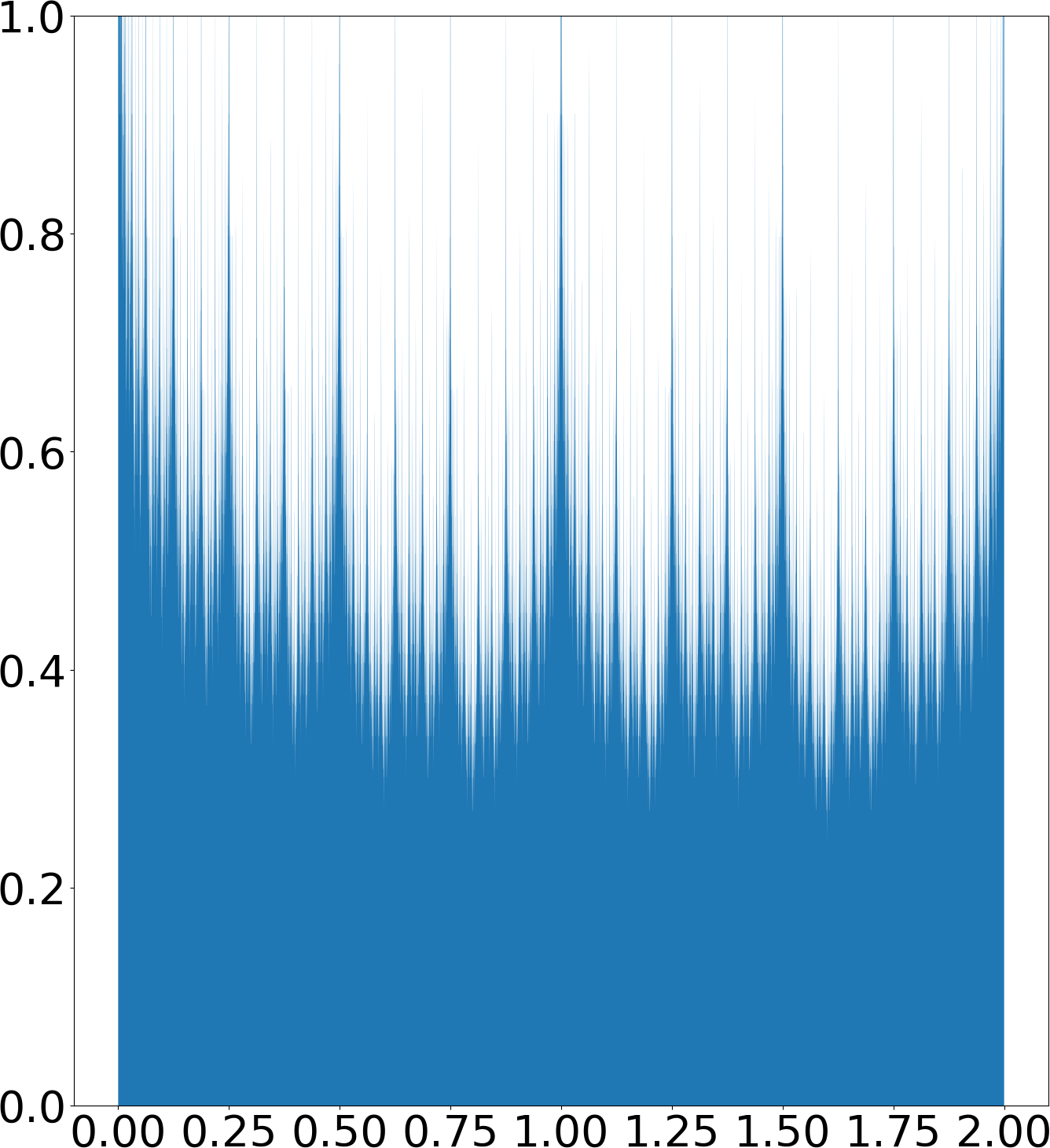}
    \caption{Empirical distribution function of other quasi-physical solutions.
    The solution shown in the left plot starts from an initial condition
    whose bits are independent and have probability 0.6 of being 0.
    The solution shown in the right plot starts from an initial condition
    whose bits have a probability of 0.6 of being 1 only following two
    consecutive 1's; otherwise a bit is 0 or 1 with equal probability.}
    \label{fig:tent_alternative}
\end{figure}

So far we have constructed one class of quasi-physical solutions.
It is noteworthy that there are infinite ways to 
construct quasi-physical solutions.  In the binary representation 
of the initial condition $x_0$, any statistical
deviation from equal probability of 0 or 1, or any statistical
dependence among digits would lead to quasi-physical solutions.
One could, for example, construct an $x_0$ in which a bit is
more likely to be 1 than 0 only if it follows two consecutive 0's.
Such an $x_0$ would lead to a nonphysical solution.  Its statistical
distribution would differ from any of the plots in Figure
\ref{fig:tent_quasiphysical}.  Nevertheless, as can be
observed in Figure \ref{fig:tent_alternative}, it shows
a remarkable resemblance in its ``hairiness''; some kind of fractal
pattern emerges from the distribution.  A fractal distribution
appears to be a signature of quasi-physical solutions.

We have only seen quasi-physical solutions for the tent map.
It is difficult to analytically construct quasi-physical solutions
to the Lorenz equation and other, more complex, governing
equations which typically produce chaotic solutions.  Nevertheless, the similarity 
between the tent map and the Lorenz map \cite{tent2}, shown in Figure \ref{fig:lorenz_map}, suggests that
quasi-physical solutions may exist for the Lorenz map, and by
extension, the Lorenz equation.  It is then natural to conjecture
that such aperiodic, nonphysical, quasi-physical solutions
exist in general for chaotic dynamical systems.

These quasi-physical solutions raise doubts over the usefulness of shadowing
in some applications.  For example, even if a numerical
solution is shadowed by a solution to the true governing physics,
is the shadowing solution physical or quasi-physical? At first glance, this may 
seem to be a nonissue because almost all solutions are physical. 
It seems reasonable to argue that because the set of
all nonphysical solutions is Lebesgue measure-zero, the probability of finding
a nonphysical solution through shadowing is zero percent.  Such an argument,
as we show in the next section, is wrong.  The probability of
finding a nonphysical solution through shadowing can be, instead of
zero percent, one hundred percent.

\section{Are shadowing solutions physical?}
\label{sec:shadowingnonphysical}
\subsection{Example of a shadowing solution for the tent map}
\label{sec:sub:shadow-example}
\begin{figure}   \centering
    \includegraphics[width=0.48\textwidth]{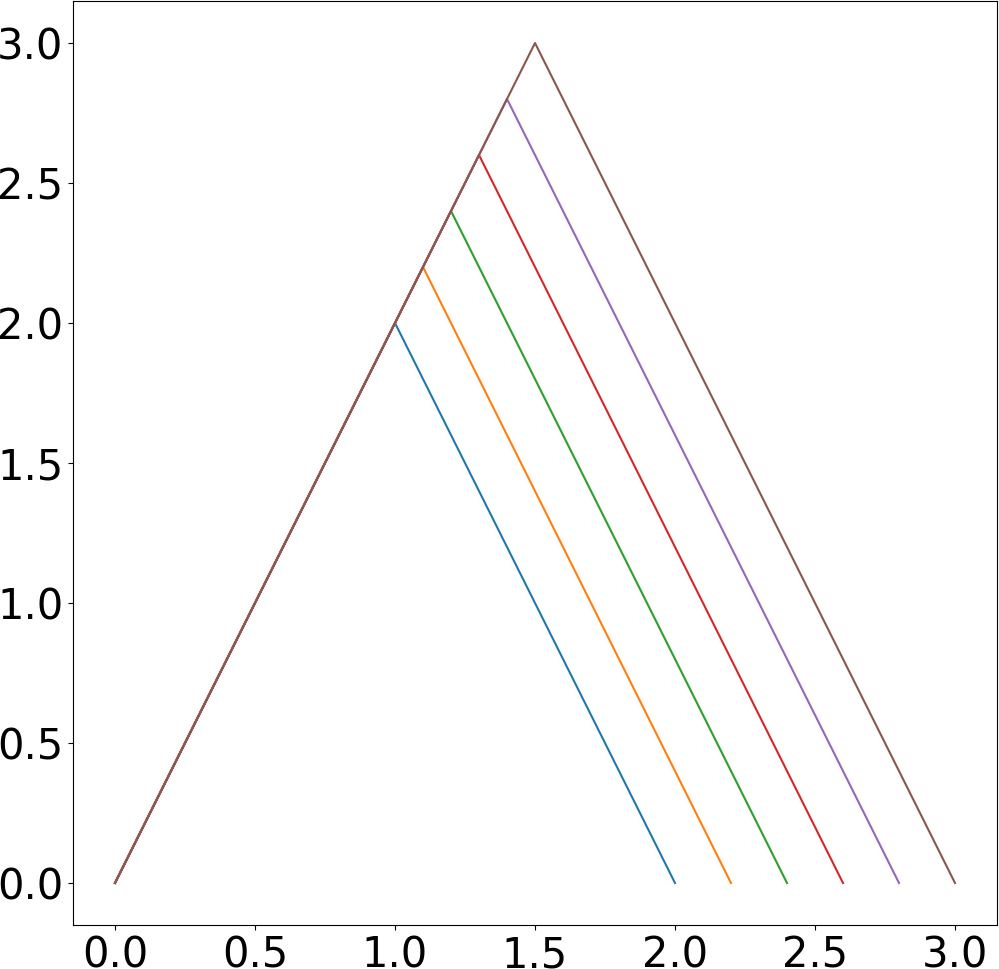}
    \hspace{0.02\textwidth}
    \includegraphics[width=0.48\textwidth]{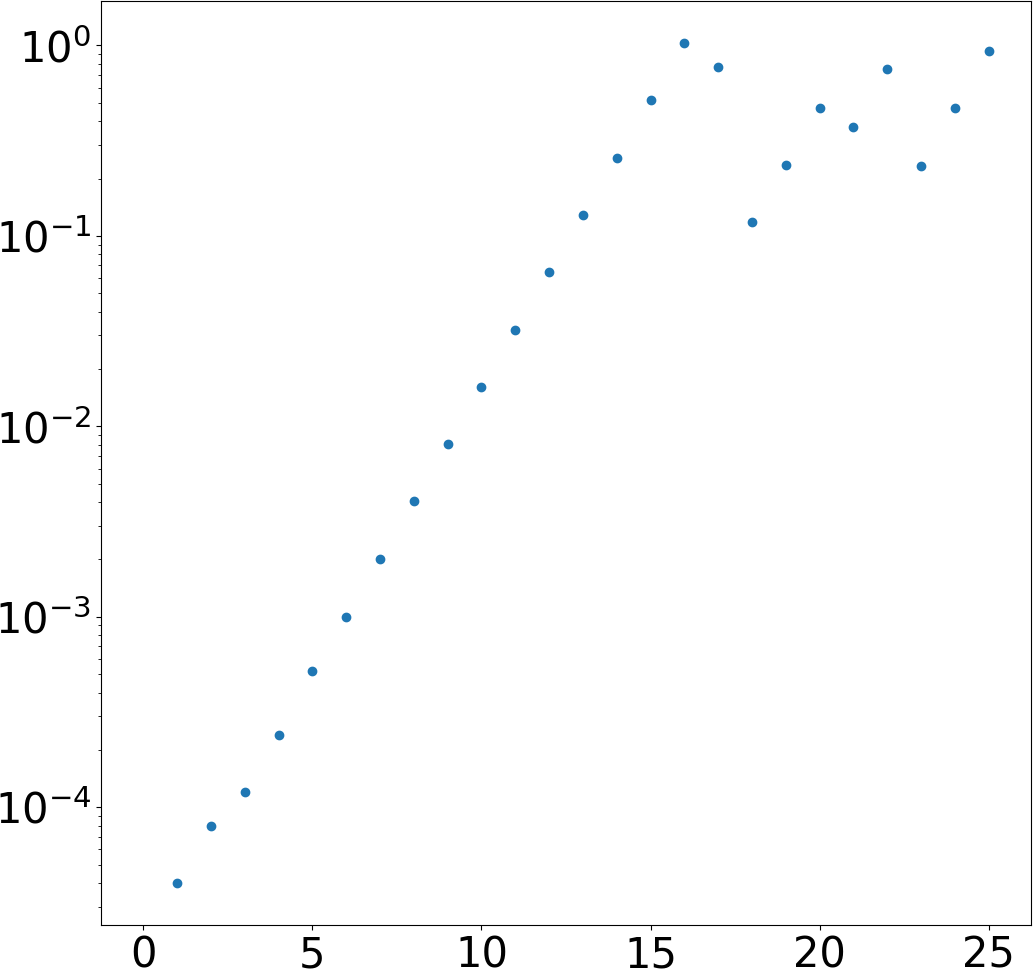}
    \caption{L: The scaled tent map $\hat{\varphi}_s$ at different values of $s$ between 0 and 1. R: sensitivity to small perturbation in the governing
    equation.
    The y-axis shows the absolute value of the difference between
    two solutions, one satisfying Eq. \ref{tent_scaled},
    one for $s=0$ and the other for $s=10^{-5}$.
    The x-axis shows the iteration number.  The initial condition
    is at $x_0=\pi/2$.
    }
    \label{fig:tent_ivp}
\end{figure}

To illustrate the concept of shadowing, consider the tent map, defined
in Eq. \ref{tentmap}, and a scaled version of the map, defined by
\begin{equation} \label{tent_scaled}
    \hat\varphi_s(x) := \begin{cases}
    2x \quad & x < 1+s \\
    4(1+s) - 2x \quad & x \ge 1+s
    \end{cases}
\end{equation}
where $s<<1$.  Note that a small change in $s$ can lead to drastic differences 
between solutions starting from the same initial condition.
As demonstrated in Figure \ref{fig:tent_ivp}, this sensitivity to small
perturbations reflects the chaotic nature of the governing equation.

We can avoid this sensitivity to the governing equation using a coordinated perturbation to the initial condition.
Consider two solutions satisfying Eq. \ref{tent_scaled} at different values of $s$.  Instead of starting from the same
initial condition, these two solutions start from $x_0 (1 + s)$ with
the same $x_0$ but their respective values of $s$.  It can be shown that
the solution of these two equations would be
$x_i (1+s), i=0,1,\ldots$ with their respective values of $s$,
where $x_{i+1}=\varphi(x_i)$ satisfies the original
tent map (Eq. \ref{tentmap}).  When the values of $s$ are similar between
the two solutions, this pair of solutions stays uniformly close to each
other forever.  Such a true solution of the governing equation
that stays close to a given perturbed solution over a long time, is known as a \emph{shadowing} solution. In this example, the perturbed solution satisfies a 
slightly different governing equation.

For every solution $\hat{x}_i,i=0,\ldots$ satisfying Eq. \ref{tent_scaled}, there is a shadowing solution satisfying Eq. \ref{tentmap}: $x_i:= \hat{x}_i/(1+s)$.  The map,
\begin{equation} \label{conjugate_scaled}
    \hat{h}_s(x):=x/(1+s)
\end{equation}
is known as a \emph{conjugacy} between the two maps $\varphi$ and $\hat{\varphi_s}$ 
because it satisfies
\begin{equation} \label{conjugate_def}
    \varphi(\hat{h}_s(x)) = \hat{h}_s(\hat\varphi_s(x))\;,\quad \forall\: x
\end{equation}
or equivalently, $\varphi\circ \hat{h}_s \equiv \hat{h}_s\circ\hat\varphi_s$.
Such conjugacy maps can generally help us construct shadowing solutions.
For every solution satisfying
$\hat{\varphi}_s$, $h_s$ maps it to a shadowing solution satisfying $\varphi$ because iterating Eq. \ref{conjugate_def}, we get $\varphi^n \circ \hat{h}_s (x) = \hat{h}_s \circ \hat{\varphi}_s^n(x)$, for $n = 1,2,3\cdots, $ where $f^n$ stands for the function composition of $f$ $n$-times.

Is the shadowing solution a physical solution?  In this example,
the answer is almost surely positive.  We can demonstrate that
almost any solution of the scaled tent map (Eq. \ref{tent_scaled}) is uniformly
distributed in $[0, 2(1+s)]$ -- we can repeat our analysis
in section \ref{sec:tent} but represent our initial condition as
$x_0= (1+s) \sum_{j=0}^\infty x_{0,j} / 2^j$.  Its shadowing solution,
which satisfies the original tent map (Eq. \ref{tentmap}), can be obtained through the
conjugacy (Eq. \ref{conjugate_scaled}).  Thus, the distribution
of the shadowing solution can be obtained by mapping a uniform
distribution in $[0,2(1+s)]$ through the conjugacy map.
This leads to a uniformly distribution in $[0,2]$, which is
precisely the distribution of a physical solution of the tent map.

This simple example is useful in illustrating the concept of shadowing
and the utility of the conjugacy map.  But it is rather special
in that the shadowing solution is almost always physical.  Our next
example introduces a tilted version of the tent map.  Although
the tilting perturbation to the tent map seems as simple as the
scaling perturbation, the shadowing solution, as we will see,
is almost always a quasi-physical solution.

\subsection{An example of quasi-physical shadowing solution}
\label{sec:sub:quasi-physical-shadow}
The tilted tent map is defined as
\begin{equation} \label{tent_tilted}
    \tilde\varphi_s(x) := \begin{cases}
    \frac2{1+s} x \quad & x < 1+s \\
    \frac2{1-s}(2-x) \quad & x \ge 1+s
    \end{cases}
\end{equation}
When $s=0$, this map is identical to the tent map (Eq. \ref{tentmap}).
Let $\tilde{x}_{i+1}=\tilde\varphi_s(\tilde{x}_i), i=0,1,\ldots$ be a solution of 
$\tilde{\varphi}_s$ that is also a perturbed solution of $\varphi$.
As we explained in the previous subsection,
the corresponding 
shadowing solution, which solves $\varphi$, can be found through a conjugacy map $\tilde{h}_s$ that connects the solutions of $\varphi$ and $\tilde{\varphi}_s$.
If $\varphi\circ \tilde{h}_s \equiv \tilde{h}_s\circ\tilde\varphi_s$,
then $x_i=\tilde{h}_s(\tilde{x}_i), i=0,1,\ldots$ is the shadowing solution 
that satisfies $x_{i+1} = \varphi(x_i)$, corresponding to the perturbed solution 
$\tilde{x}_i,i=0,1,\cdots$.

The conjugacy map, although more complex than the one
in the last subsection, has the following closed form:
\begin{equation} \label{conjugate_tilted}
    \tilde{h}_s(x) = \sum_{j=0}^{\infty} \frac{\xi_{s,j}(x)}{2^j}\;,
\end{equation}
where
\begin{equation} \label{conjugate_tilted_helper}
    \xi_{s,j}(x) := \begin{cases}
    0 & j=0, x<1+s \\
    1 & j=0, x\ge 1+s \\
    \xi_{s,j-1}(x) & j>0, \tilde\varphi_s^j(x) < 1+s \\
    1 - \xi_{s,j-1}(x) & j>0, \tilde\varphi_s^j(x) \ge 1+s.
    \end{cases}
\end{equation}
In the above expression, $\tilde{\varphi}^j_s$ refers to the $j$-time 
composition of $\tilde{\varphi}_s.$ That is, if $\tilde{x}_{j+1} = \tilde{\varphi}_s(\tilde{x}_j), j = 0,1,\cdots,$ is a solution, $\tilde{\varphi}_s^i(\tilde{x}_j) = \tilde{x}_{j+i},\quad i,j = 0,1,\cdots$. 

To see why $\tilde{h}_s$ constructed in Eq. \ref{conjugate_tilted} satisfies the definition of a conjugacy map --
$\tilde{h}_s(\tilde\varphi_s(x)) = \varphi(\tilde{h}_s(x))$ for all $x$ --
we need to analyze two cases: $x<1+s$ and $x\ge 1+s$.
\begin{figure}
  \centering
    \includegraphics[width=0.48\textwidth]{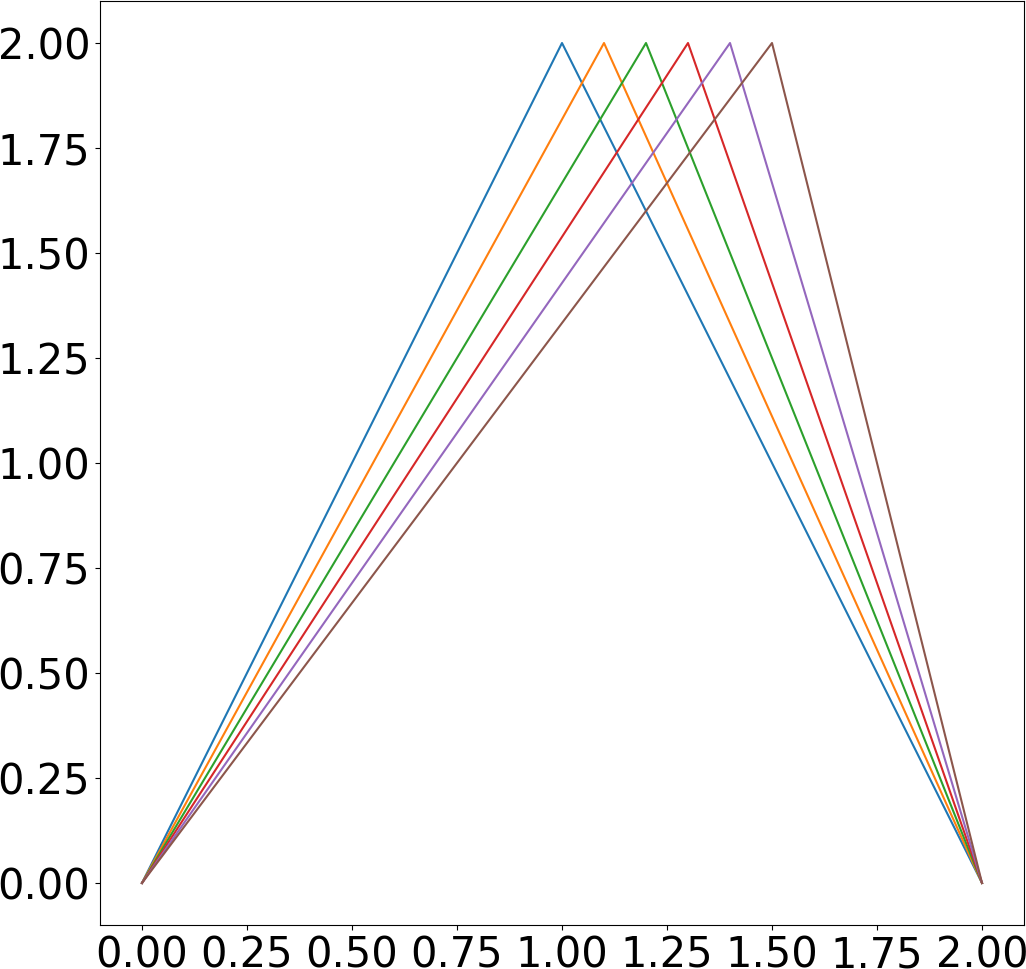}
    \hspace{0.02\textwidth}
    \includegraphics[width=0.48\textwidth]{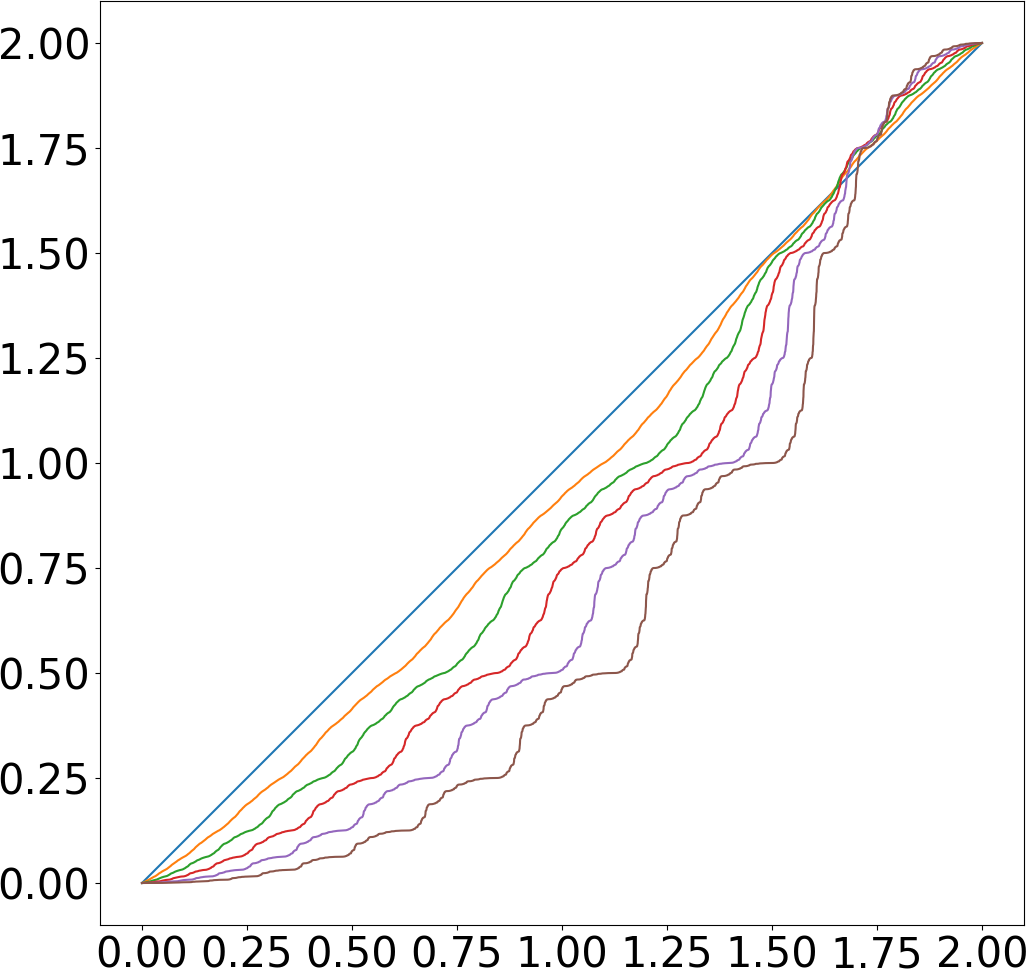}
    \\ \vspace{0.02\textwidth}
     \caption{L: the tilted tent map Eq. \ref{tent_tilted} for $s=0, 0.1, 0.2, 0.3, 0.4,$ and 0.5. R: the conjugacy $\tilde{h}_s$ between the tent map (Eq. \ref{tentmap})
    and the tilted tent map (Eq. \ref{tent_tilted}), evaluated using
    Eq. \ref{conjugate_tilted}-\ref{conjugate_tilted_helper}, for the same set of $s$ as the
    left plot.}
     \label{fig:tilted_tent_conjugate}
\end{figure}
When $x<1+s$, $\xi_{s,0}(x)=0$; thus
$\tilde{h}_s(x) = \sum_{j=1}^{\infty} \frac{\xi_{s,j}(x)}{2^j}$.
Inside this infinite series, $\xi_{s,1}(x)=\xi_{s,0}(x)=0$ if $\tilde\varphi_s(x)<1+s$,
or $\xi_{s,1}(x)=1-\xi_{s,0}(x)=1$ if $\tilde\varphi_s(x)\ge 1+s$.
Thus, $\xi_{s,1}(x) = \xi_{s,0}(\tilde{\varphi}_s(x))$ according to the definition of $\xi_{s,0}$.
Using this as the base case, one can inductively verify that
$\xi_{s,j+1}(x) = \xi_{s,j}(\tilde\varphi_s(x))$
for all $j>0$, using just the definitions of $\xi_{s,j}$ and $\xi_{s,j+1}$.
Therefore,
$\tilde{h}_s(x) = \frac12 \sum_{j=0}^{\infty} \frac{\xi_{s,j}(\tilde\varphi_s(x))}{2^j}
=\frac12 \tilde{h}_s(\tilde\varphi_s(x))$.
On the other hand, because
$\tilde{h}_s(x) = \sum_{j=1}^{\infty} \frac{\xi_{s,j}(x)}{2^j}\le1$,
$\varphi(\tilde{h}_s(x)) = 2\tilde{h}_s(x)$ according
to the definition of $\varphi$.  Thus,
$\varphi(\tilde{h}_s(x)) = \tilde{h}_s(\tilde\varphi_s(x))$ holds when $x<1+s$.

When $x\ge1+s$, $\xi_{s,0}(x)=1$; thus
$\tilde{h}_s(x) = 1+ \sum_{j=1}^{\infty} \frac{\xi_{s,j}(x)}{2^j}
                = 2- \sum_{j=1}^{\infty} \frac{1-\xi_{s,j}(x)}{2^j}$.
Inside this infinite series, $1-\xi_{s,1}(x)=1-\xi_{s,0}(x)=0$ if $\tilde\varphi_s(x)<1+s$,
or $1-\xi_{s,1}(x)=\xi_{s,0}(x)=1$ if $\tilde\varphi_s(x)\ge 1+s$.
Thus, $1-\xi_{s,1}(x) = \xi_{s,0}(\tilde{\varphi}_s(x))$ according to the definition of $\xi_{s,0}$.
Using this as the base case, one can inductively verify that
$1-\xi_{s,j+1}(x) = \xi_{s,j}(\tilde\varphi_s(x))$
for all $j>0$, using just the definitions of $\xi_{s,j}$ and $\xi_{s,j+1}$.
Therefore,
$\tilde{h}_s(x) = 2 - \frac12 \sum_{j=0}^{\infty} \frac{\xi_{s,j}(\tilde\varphi_s(x))}{2^j}
=2 - \frac12 \tilde{h}_s(\tilde\varphi_s(x))$.
On the other hand, because
$\tilde{h}_s(x) = 1+\sum_{j=1}^{\infty} \frac{\xi_{s,j}(x)}{2^j}\ge 1$,
$\varphi(\tilde{h}_s(x)) = 4 - 2 \tilde{h}_s(x)$ according
to the definition of $\varphi$.  Thus,
$\varphi(\tilde{h}_s(x)) = \tilde{h}_s(\tilde\varphi_s(x))$ also holds when $x\ge 1+s$.

\begin{figure}
   \includegraphics[width=0.48\textwidth]{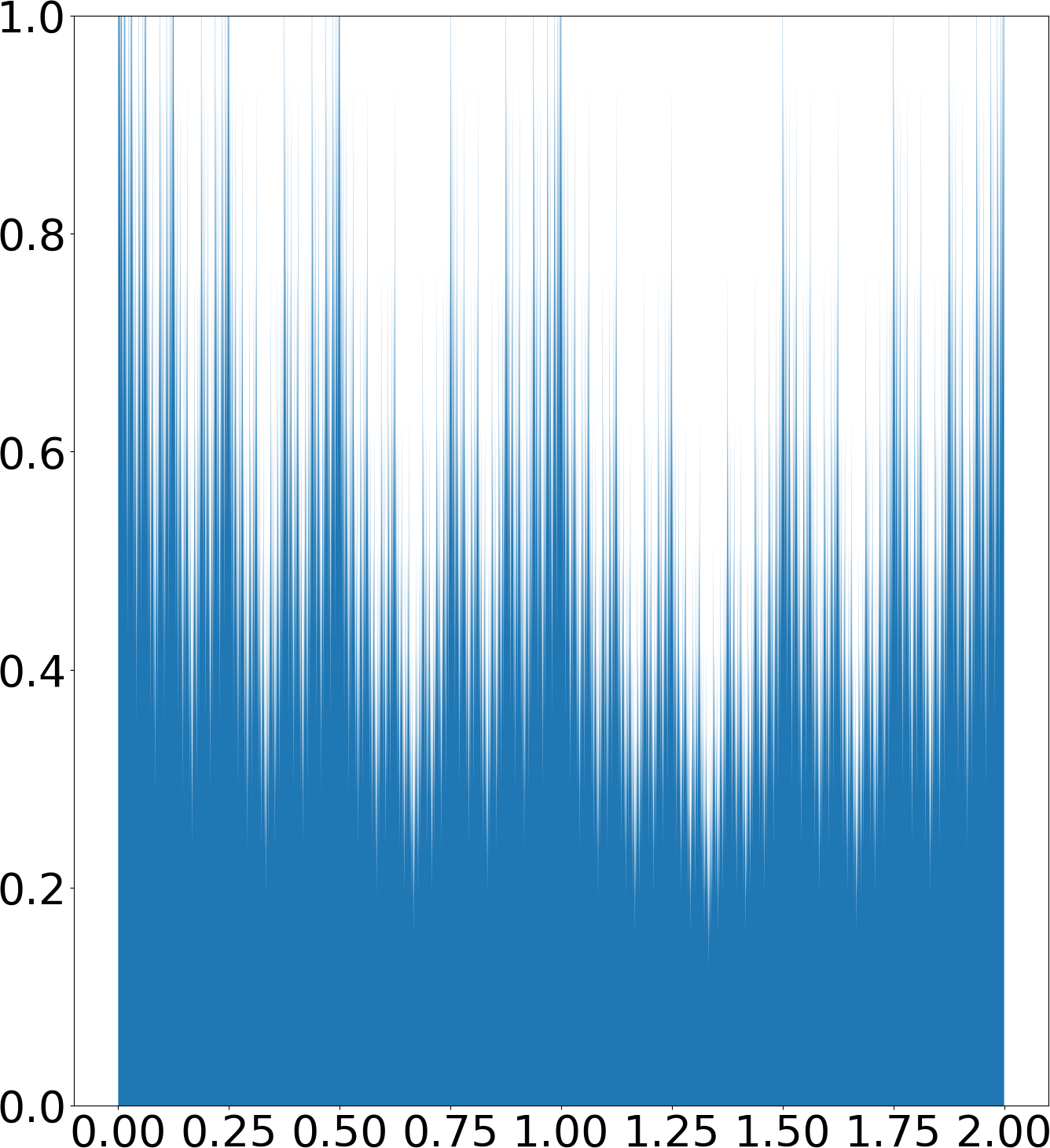}
    \hspace{0.02\textwidth}
    \includegraphics[width=0.48\textwidth]{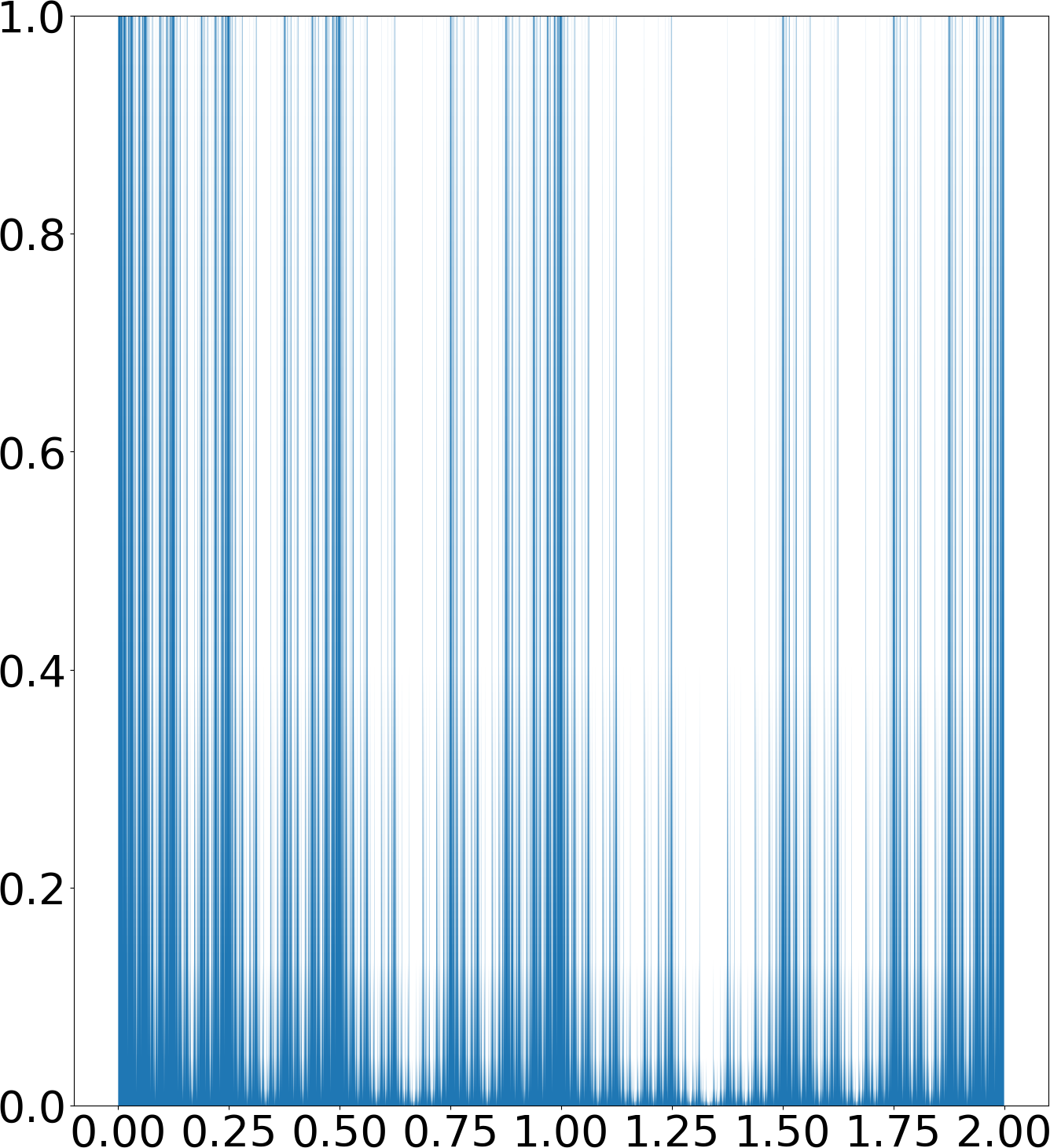}
   \caption{
     L: the density of a trajectory satisfying Eq. \ref{tentmap}
    that shadows a random trajectory satisfying Eq. \ref{tent_tilted} for $s=0.1$.
    R: the density of a trajectory satisfying Eq. \ref{tentmap}
    that shadows a random trajectory satisfying Eq. \ref{tent_tilted} for $s=0.5$.
    }
    \label{fig:tilted_tent_density}
\end{figure}

Figure \ref{fig:tilted_tent_conjugate} shows how fractal the conjugacy map is.
This fractal conjugacy map transforms a uniform distribution in $[0,2]$ into
a fractal distribution, shown in Figure \ref{fig:tilted_tent_density},
similar to the ones plotted in section \ref{sec:quasiphysical}.
In fact, we will show that the fractal distribution obtained by mapping a uniform
distribution through $\tilde{h}_s$ is exactly the family of distributions
shown in Figure \ref{fig:tent_quasiphysical}.  We will also show that
a physical solution of the tilted tent map (Eq. \ref{tent_tilted}) is uniformly
distributed in $[0,2]$.  Thus, for almost any physical solution of the
tilted tent map, a shadowing solution of the original tent map, obtained
through the conjugate map $\tilde{h}_s$, has a fractal distribution.
Such a shadowing solution is therefore a quasi-physical solution of
the tent map.

We first show that a physical solution of the tilted tent map 
(\ref{tent_tilted}) is uniformly distributed in $[0,2]$, for
any $0\le s<1$. For this, we only need to show that the uniform
distribution is stationary under the tilted tent map.
The function $\tilde{\varphi}_s$ stretches an infinitesimal
region around each $x$ by the absolute value of the derivative at $x$; 
hence, a uniform density on an infinitesimal region around $\tilde{\varphi}_s(x)$ 
is reduced by the same factor. In other words, let $\tilde{\rho}(x)$ 
be the map of a uniform density on $[0,2]$, through $\tilde{\varphi}_s$. Then, 
$\tilde{\rho}(\tilde{\varphi}_s (x)) = 0.5/|\tilde \varphi_s'(x)| + 0.5/|\tilde \varphi_s'(x')|,$ where
$x$ and $x'$ lie on two sides of $1+s$ and
$\tilde{\varphi}_s(x)=\tilde{\varphi}_s(x')$; the two terms express the conservation of probability 
from the two pre-images of $\tilde{\varphi}_s$. Substituting the piecewise constant 
derivative of $\tilde{\varphi}_s$ on the two intervals, $[0,1+s)$ and $[1+s,2]$,
we have $\tilde{\rho}(\tilde{\varphi}_s (x)) = 0.25(1+s) + 0.25(1-s) = 0.5,$ at 
all $x.$ Hence, a uniform density of 0.5 is unaltered by mapping through 
$\tilde\varphi_s(x)$.

We now show that the conjugacy $\tilde{h}_s$ maps a uniform
distribution to a fractal.  Here we use the closed forms Eqs. \ref{conjugate_tilted}-\ref{conjugate_tilted_helper}.
If $x$ is a random number drawn uniformly in $[0,2]$,
it has $\frac{1+s}2$ probability of being less than $1+s$.
Thus, $\xi_{s,0}(x)=0$ with probability $\frac{1+s}2$.
This means, according to Eq. \ref{conjugate_tilted},
$\tilde{h}_s(x)<1$ with probability $\frac{1+s}2$:
$\tilde{h}_s(x)$ is more likely to lie in the left half
of the domain $[0,2]$.
Furthermore, for $j\ge 0$, $\xi_{s,j+1}(x)=\xi_{s,j}(x)$
with probability $\frac{1+s}2$ since each $\tilde{\varphi}^j_s(x)$ is sampled
from the uniform distribution on $[0,2]$.  This probability has direct
implication on $\tilde{h}_s(x)$ since, again by Eq. (\ref{conjugate_tilted}),
$\xi_{s,j}$ is the $j$th bit in the binary representation of $\tilde{h}_s$ --
each bit repeats the previous one with probability
$\frac{1+s}2$.  For a uniformly random
$x$, $\tilde{h}_s(x)$ is exactly the kind of initial condition
we used to construct the quasi-physical solution in section \ref{sec:quasiphysical}, with $p=\frac{1+s}2$.
Since this analysis holds for any $x$ sampled uniformly on $[0,2]$, we can conclude that
the shadowing solution is almost surely nonphysical.  Let $x_{s,i},i=0,1,\ldots$ be a solution
to the tilted tent map $\tilde{\varphi}_s$ with $x_{s,0}$ chosen 
randomly in $[0,1+s]$.  Then, with a hundred percent probability,
it is a physical solution that uniformly visits the domain $[0,2]$.
Its shadowing solution $\tilde{h}_s(x_{s,i}),i=0,1,\ldots$, however,
is a quasi-physical solution, also with a hundred percent probability.
It explores $[0,2]$ at a nonuniform frequency with a fractal probability distribution.

\subsection{Are general shadowing solutions physical?}
\label{sec:tentPerts}
In the previous subsection, we examined a tilting perturbation to the tent map.
We derived the fractal conjugacy map, and showed that a typical shadowing trajectory
is exactly the quasi-physical solution we analyzed in section \ref{sec:quasiphysical}.
This example contrasts with section \ref{sec:sub:shadow-example}, in which the
shadowing solutions of a different perturbation, the scaling perturbation,
are physical solutions.  Which situation is more typical?  When we make other
types of perturbations, do we expect to observe physical or nonphysical
shadowing solutions?

\begin{figure}
    \centering
    \includegraphics[width=0.48\textwidth]{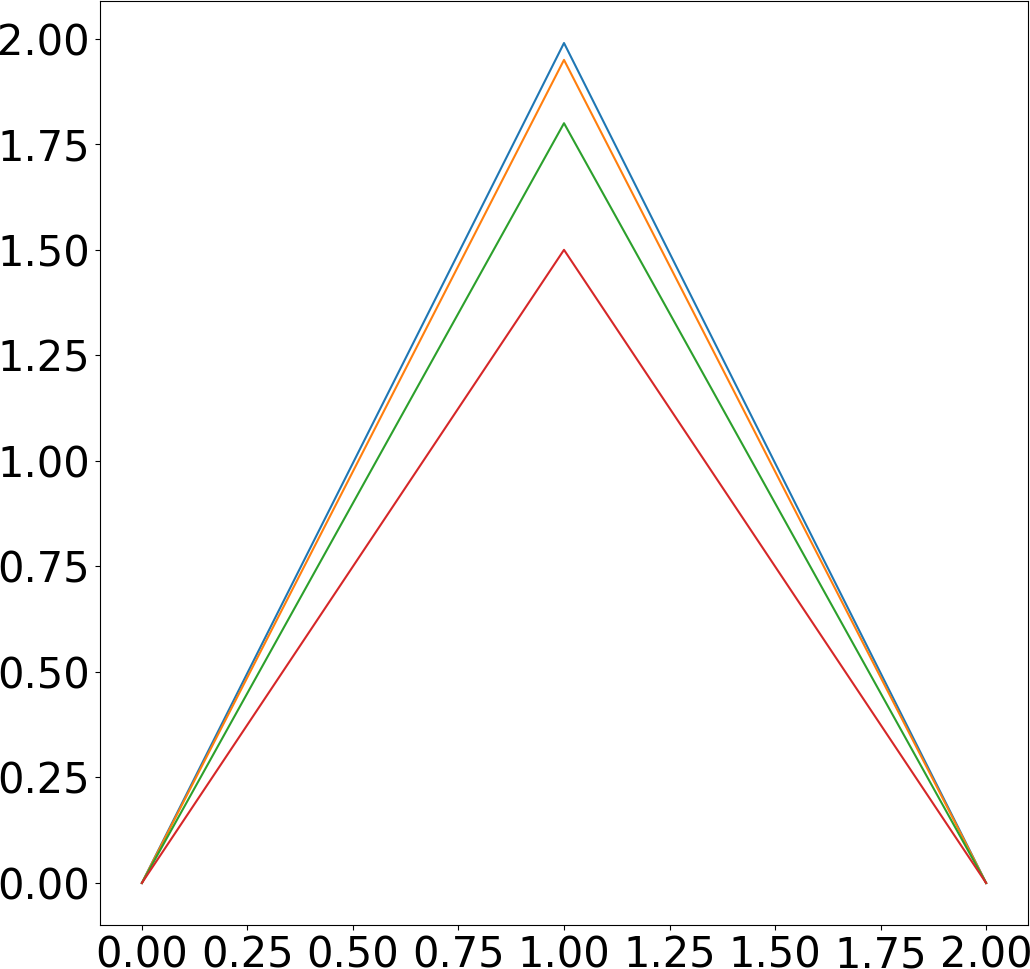}
    \caption{The ``squashed'' tent map (Eq. \ref{eqn:tent_squashed}) for $s=0.01, 0.05, 0,2,$ and 0.5.}
    \label{fig:tent_squashed}
\end{figure}

This section attempts to answer this question by investigating several other
perturbations, the first of which scales the height, but not the width,
of the tent map.  The resulting ``squashed'' tent map, illustrated
in Figure \ref{fig:tent_squashed}, has the closed form 
\begin{equation} \label{eqn:tent_squashed}
\hat{\varphi}^{\rm sq}_s(x) := \begin{cases}
    (2-s)x \quad & x < 1 \\
    (2-s)(2 - x) \quad & 2 >  x \ge 1.
    \end{cases}
\end{equation}

\begin{figure}
    \centering
    \includegraphics[width=0.48\textwidth]{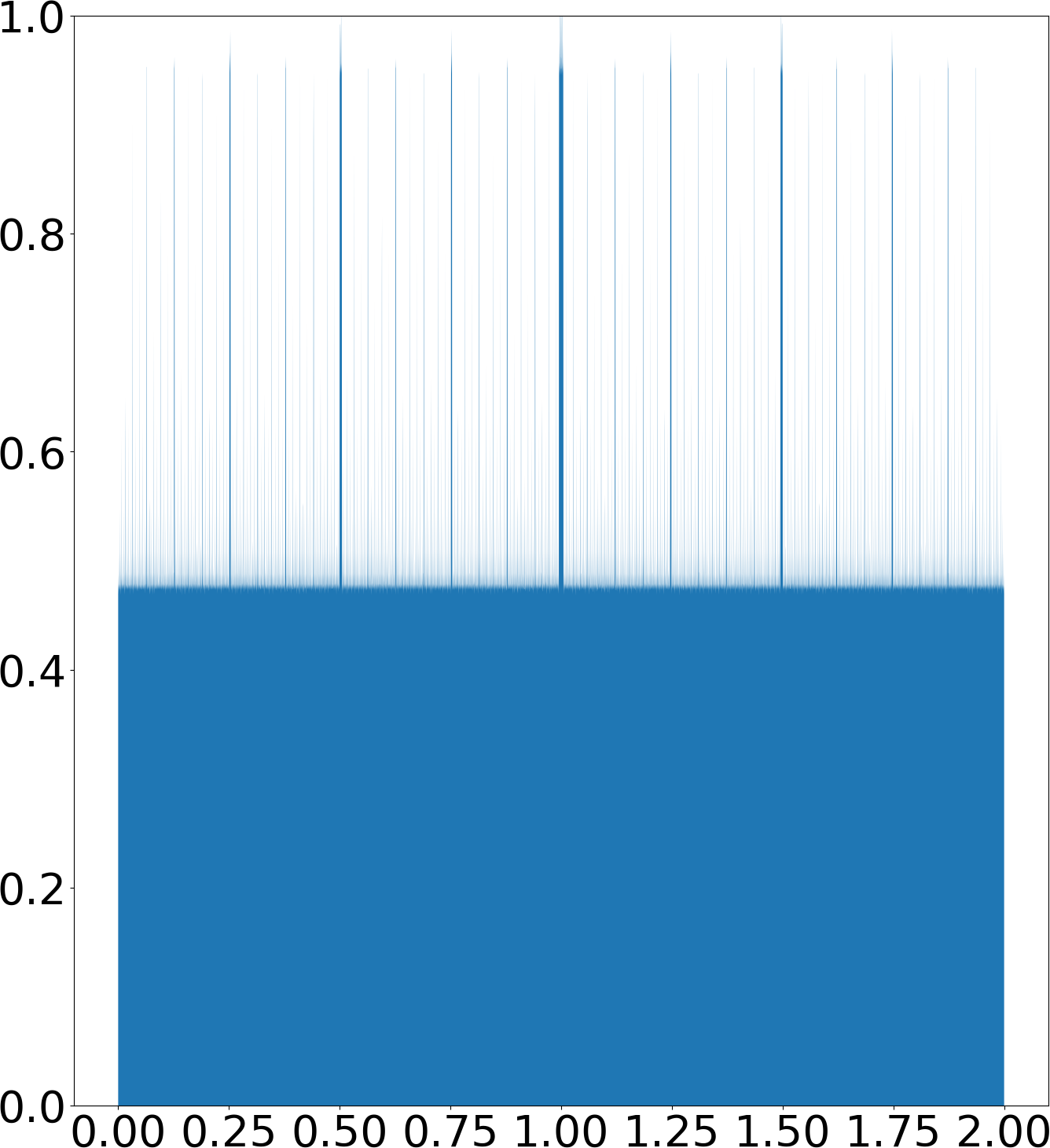}
    \hspace{0.02\textwidth}
    \includegraphics[width=0.48\textwidth]{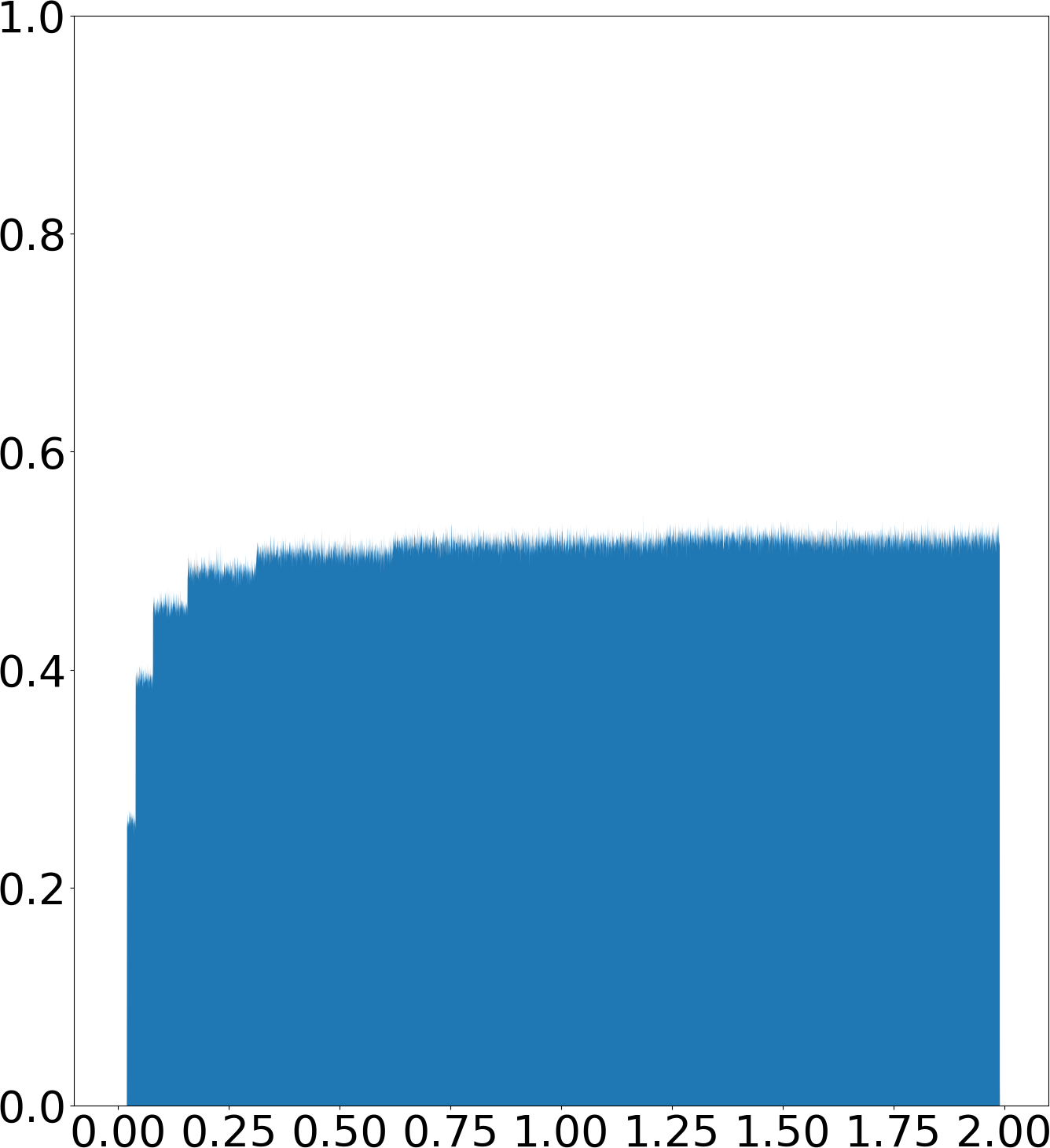}
    \\ \vspace{0.02\textwidth}
    \includegraphics[width=0.48\textwidth]{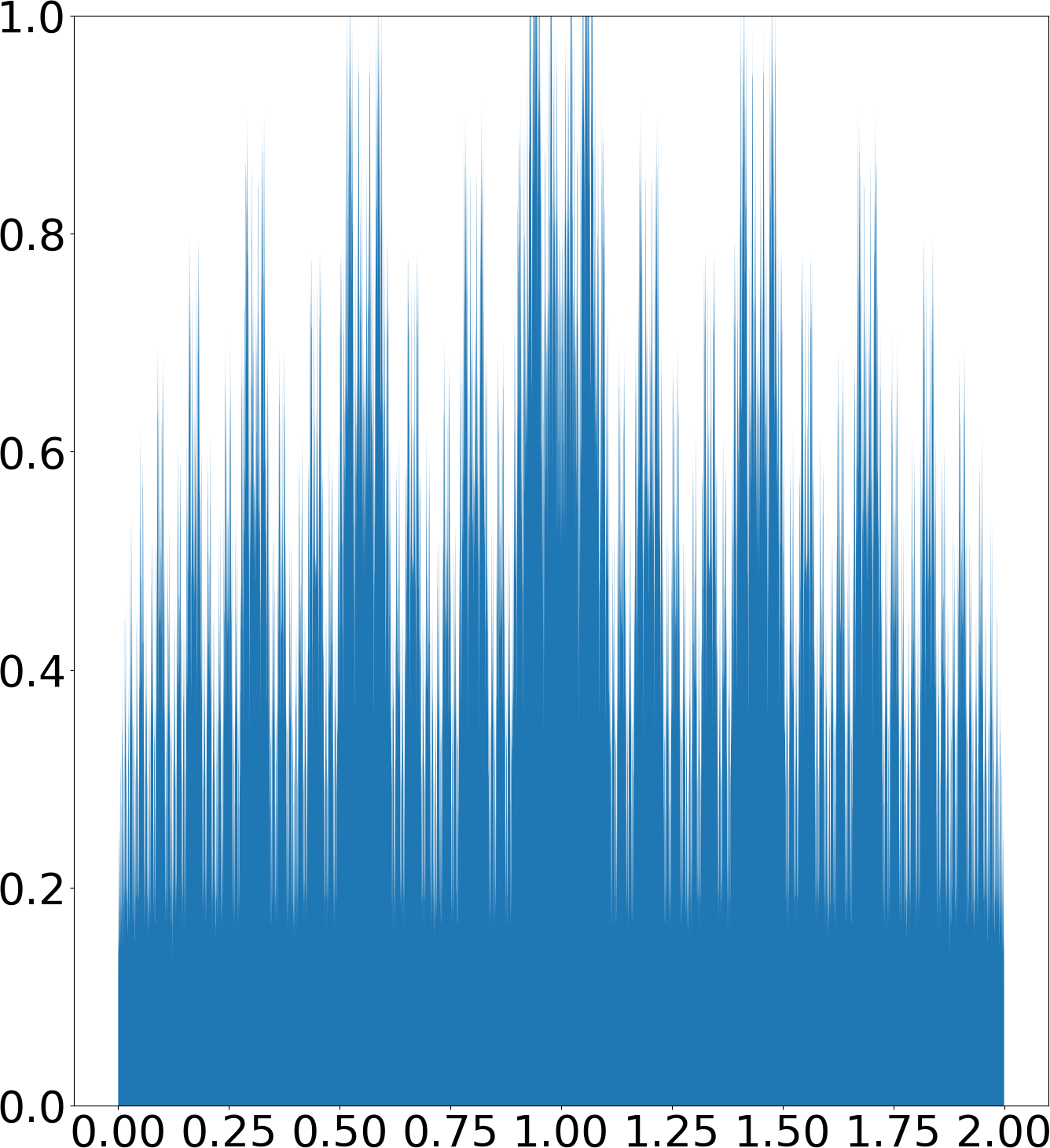}
    \hspace{0.02\textwidth}
    \includegraphics[width=0.48\textwidth]{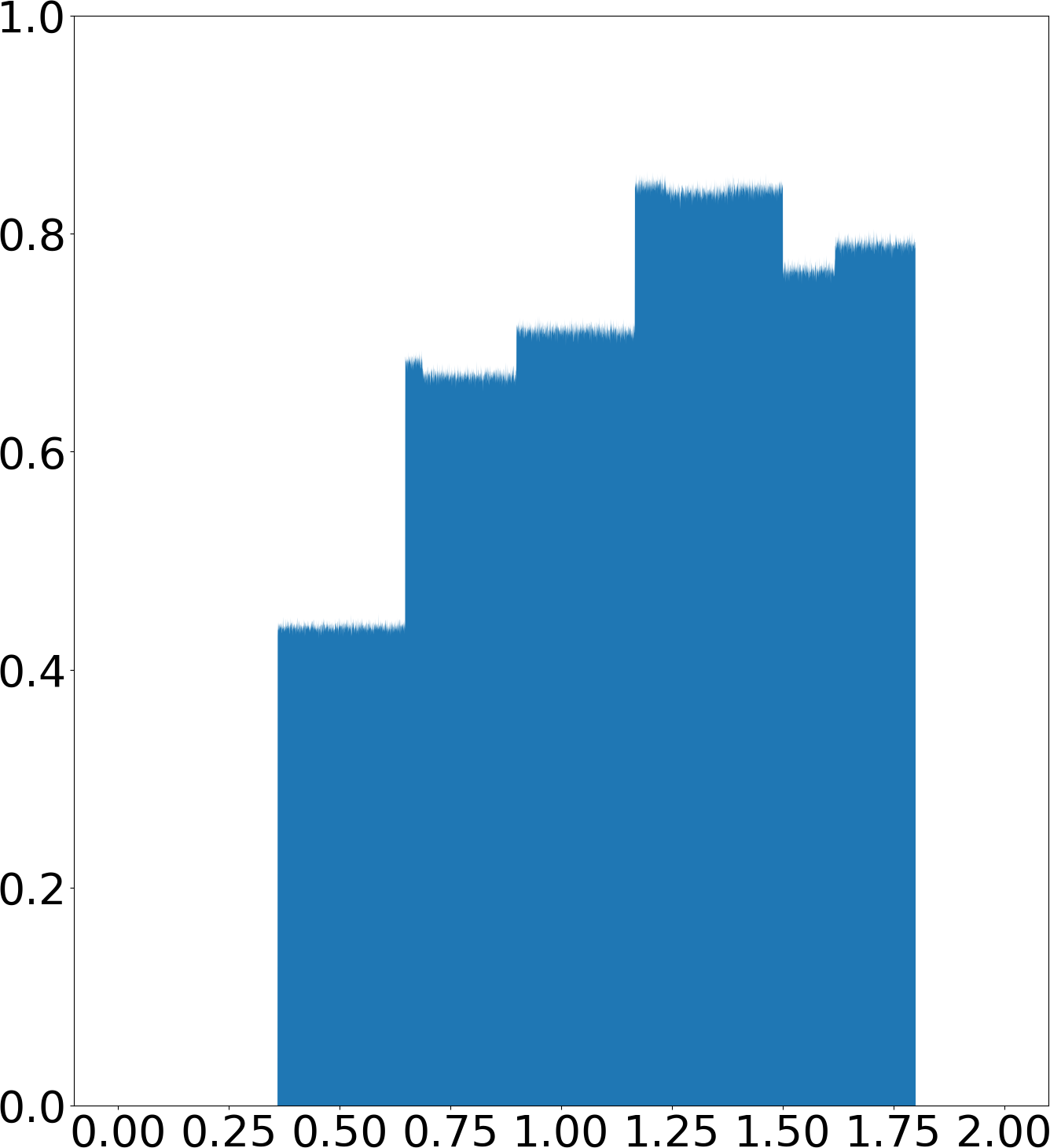}
    \caption{The left column shows the empirical probability distribution of the shadowing solution for the squashed tent map at $s=0.01$ (top) and at $s=0.2$ (bottom). The right column shows the empirical probability distribution of a physical solution at the same two values of $s = 0.01$ (top) and $s = 0.2$ (bottom).}
    \label{fig:tent_squashed_shadow}
\end{figure}
In Figure \ref{fig:tent_squashed_shadow}, we show side-by-side the probability distributions generated by observing a long solution (of length 10 billion) 
of $\hat{\varphi}^{\rm sq}_s$ (left) and its corresponding shadowing solution (right)
at two different values of $s.$ The left column showing the probability distribution of the shadowing solution is reminiscent of the hairy probability distributions of 
the quasi-solutions of the tent map (section \ref{sec:quasiphysical}). The reader is referred to the Supplementary Material section \ref{sec:computingShadowing} for the computational method used in this paper, for the shadowing solutions. Note that, unlike the tilted tent maps, the family of squashed tent maps do not have the uniform distribution as their SRB distribution, but have a different regular probability distribution as indicated on the right column of Figure \ref{fig:tent_squashed_shadow}. On the other hand, the shadowing solution is distributed neither like a physical solution of the original tent map nor of the squashed tent map, as suggested by its fractal-like probability distribution on the left column of Figure \ref{fig:tent_squashed_shadow}.

\begin{figure}
    \centering
    \includegraphics[width=0.48\textwidth]{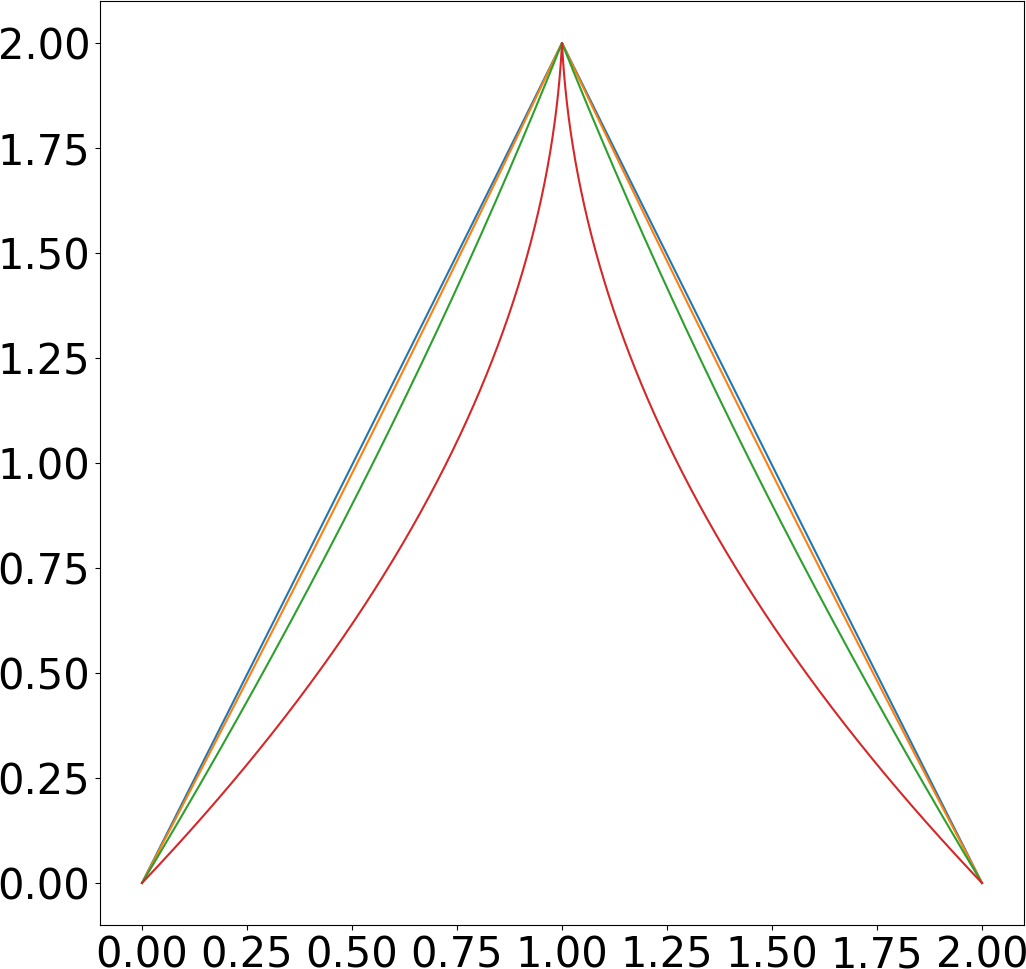}
    \caption{The pinched tent map (Eq. \ref{eqn:tent_pinched}) for $s = 0.01$ (blue), $s=0.05$ (orange), $s=0.2,$ (green) and $s=0.5$(red).}
    \label{fig:tent_pinched}
\end{figure}

\begin{figure}
    \centering
    \includegraphics[width=0.48\textwidth]{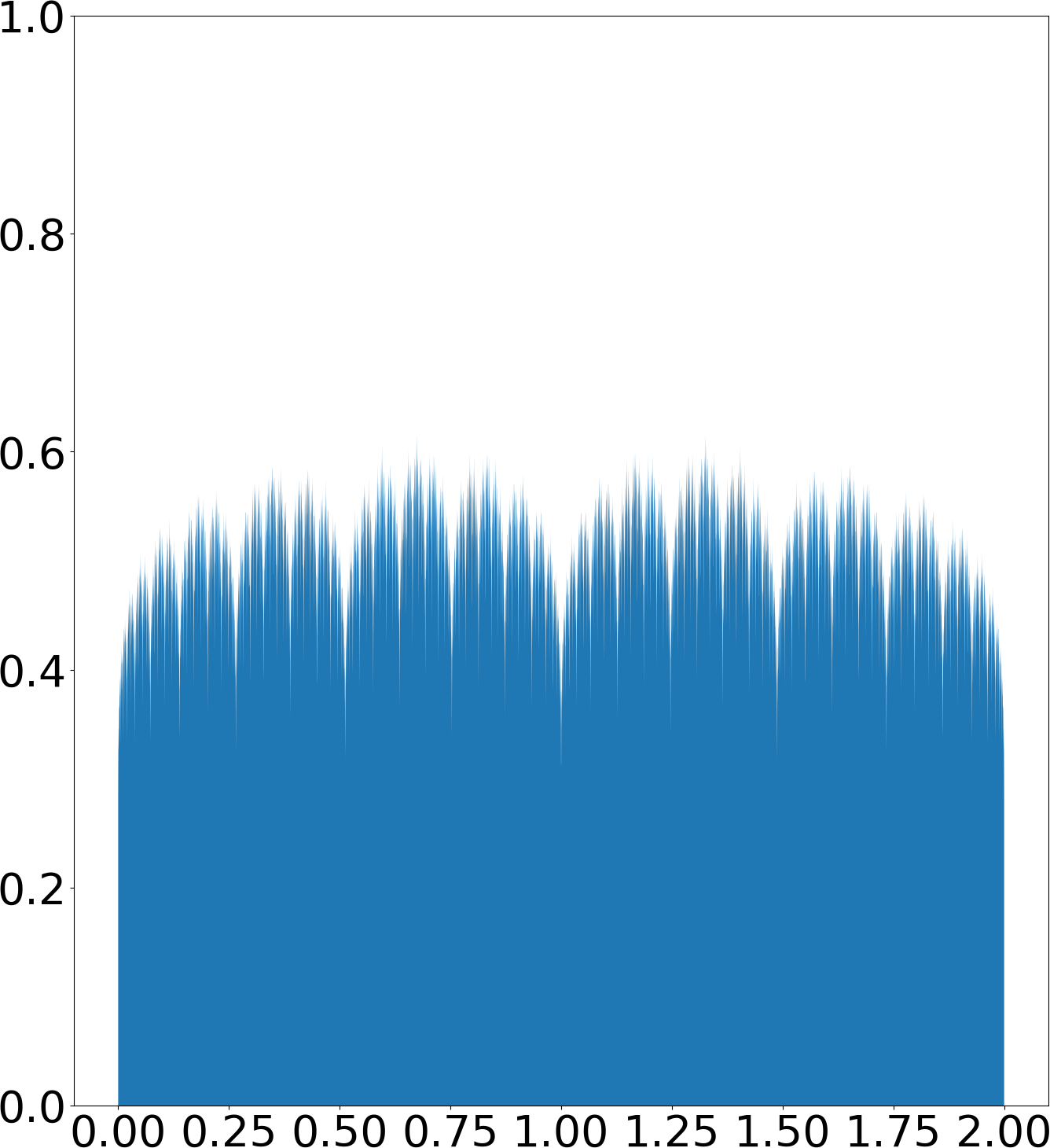}
    \hspace{0.02\textwidth}
    \includegraphics[width=0.48\textwidth]{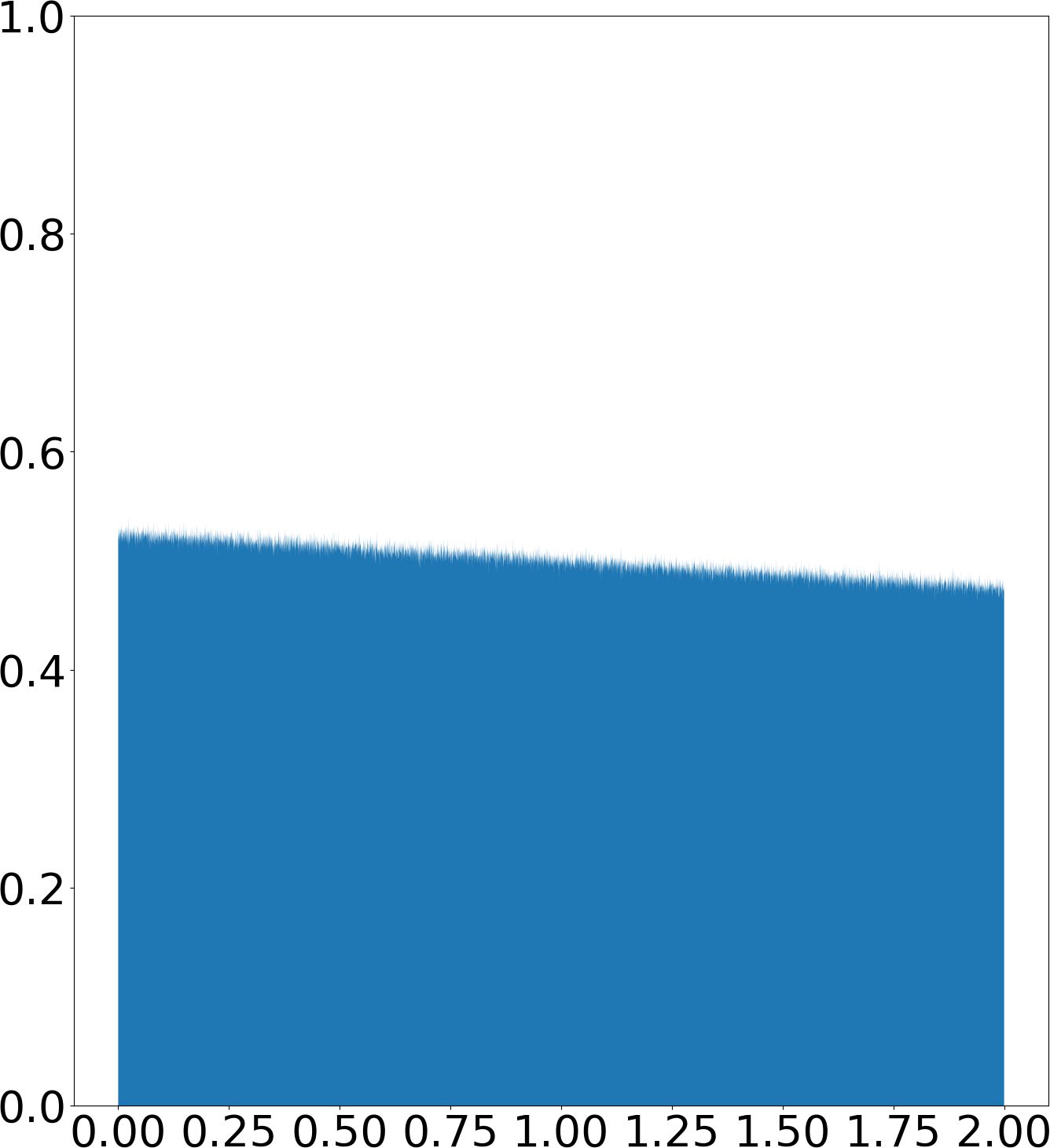}
    \\ \vspace{0.02\textwidth}
    \includegraphics[width=0.48\textwidth]{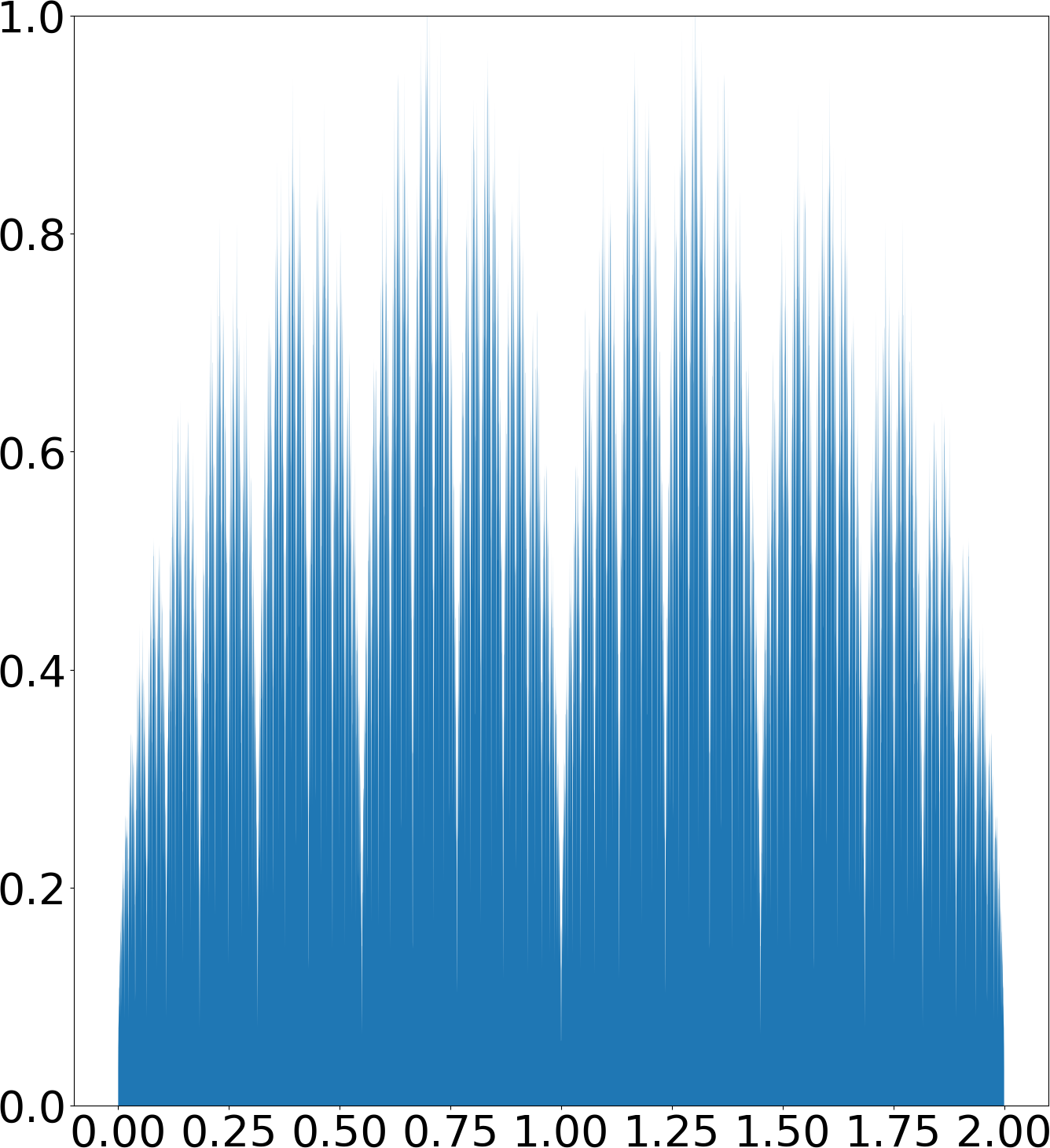}
    \hspace{0.02\textwidth}
    \includegraphics[width=0.48\textwidth]{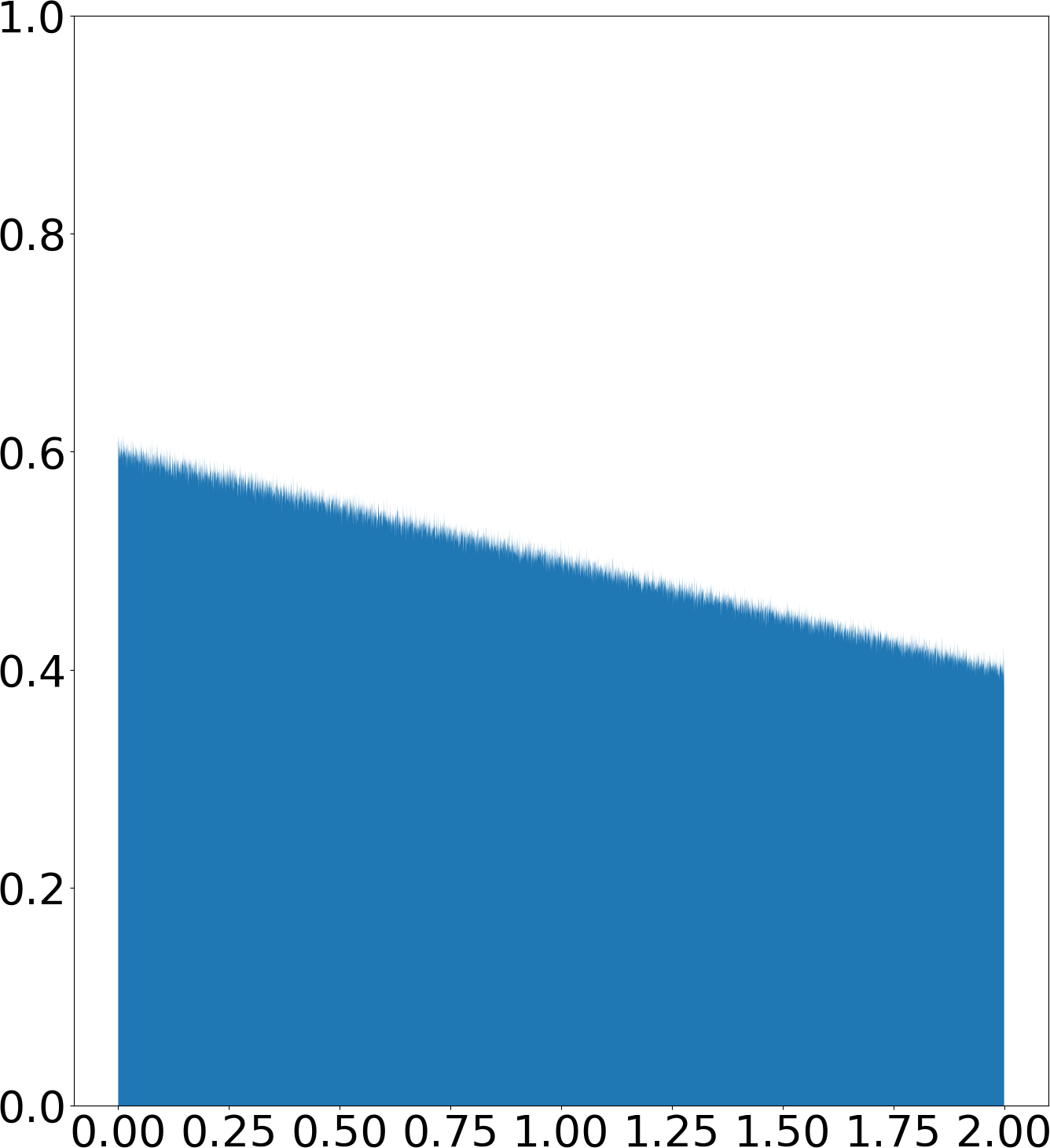}
    \caption{The left column shows the empirical distribution of the shadowing solution for the pinched tent map at $s=0.05$ (top) and at $s=0.2$ (bottom). The right column shows the probability distribution of a physical solution at the same two values of $s = 0.05$ (top) and $s = 0.2$ (bottom).}
    \label{fig:tent_pinched_shadow}
\end{figure}

\begin{figure}
    \centering
    \includegraphics[width=0.48\textwidth]{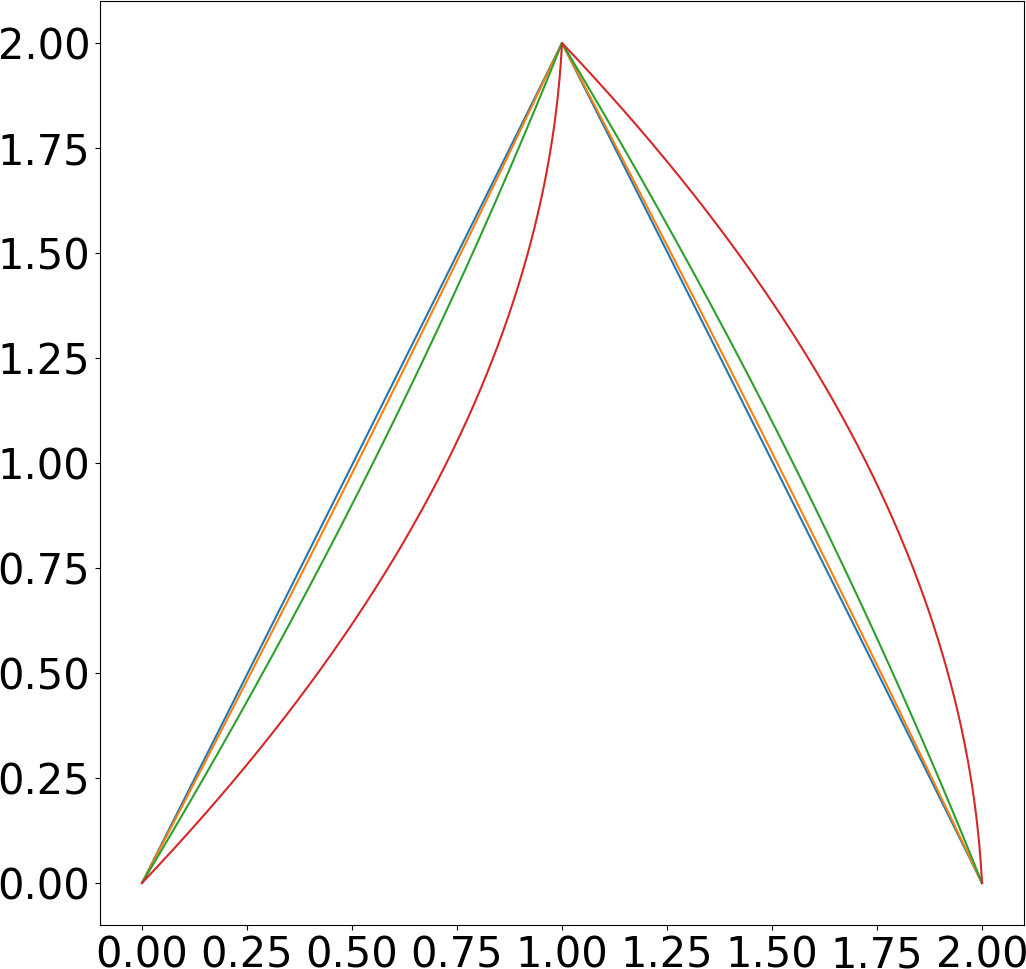}
    \caption{The wave tent map (Eq. \ref{eqn:tent_wave}) for $s = 0.01, 0.05, 0.2,$ and 0.5.}
    \label{fig:tent_wave}
\end{figure}

\begin{figure}
    \centering
    \includegraphics[width=0.48\textwidth]{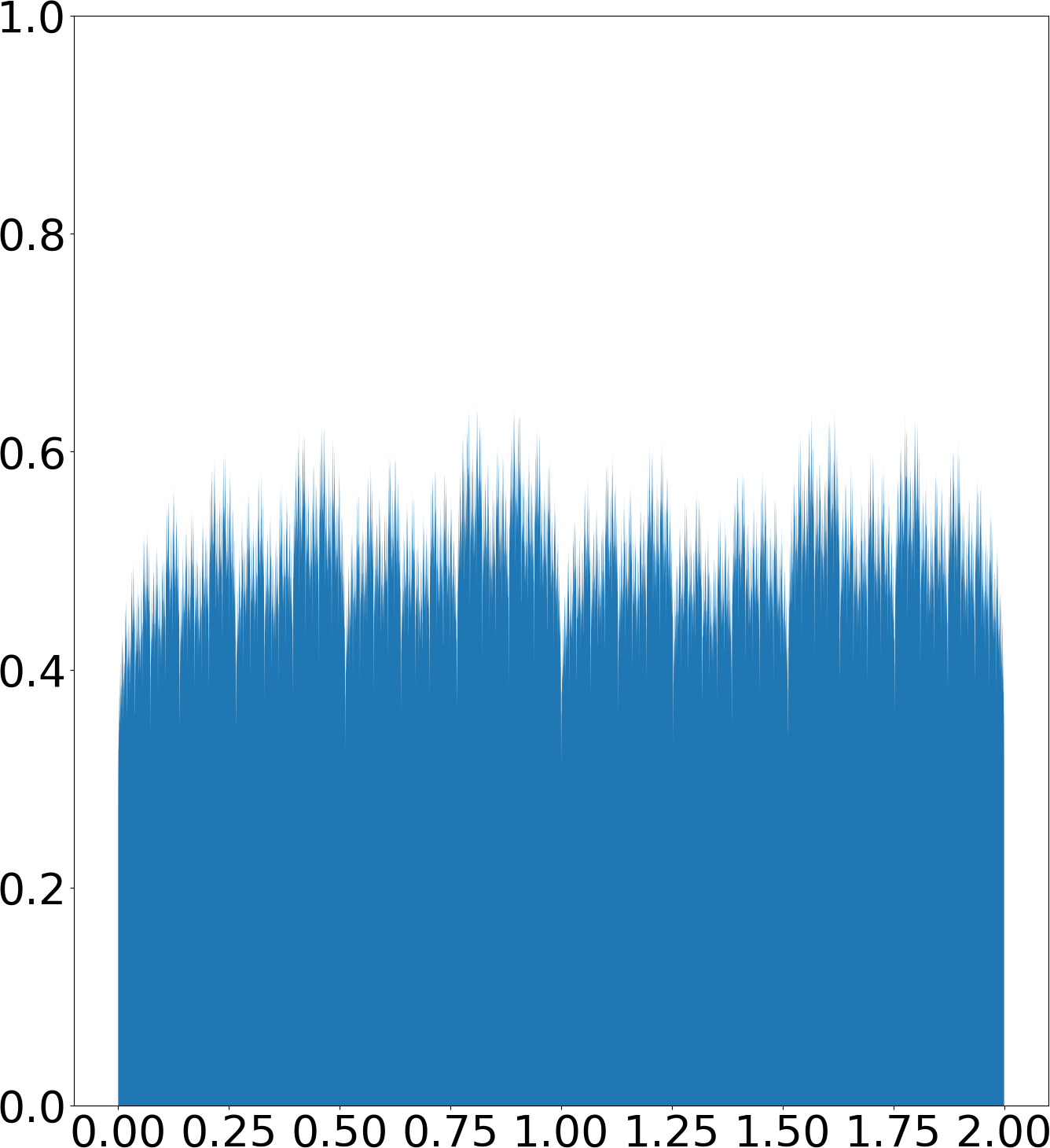}
    \hspace{0.02\textwidth}
    \includegraphics[width=0.48\textwidth]{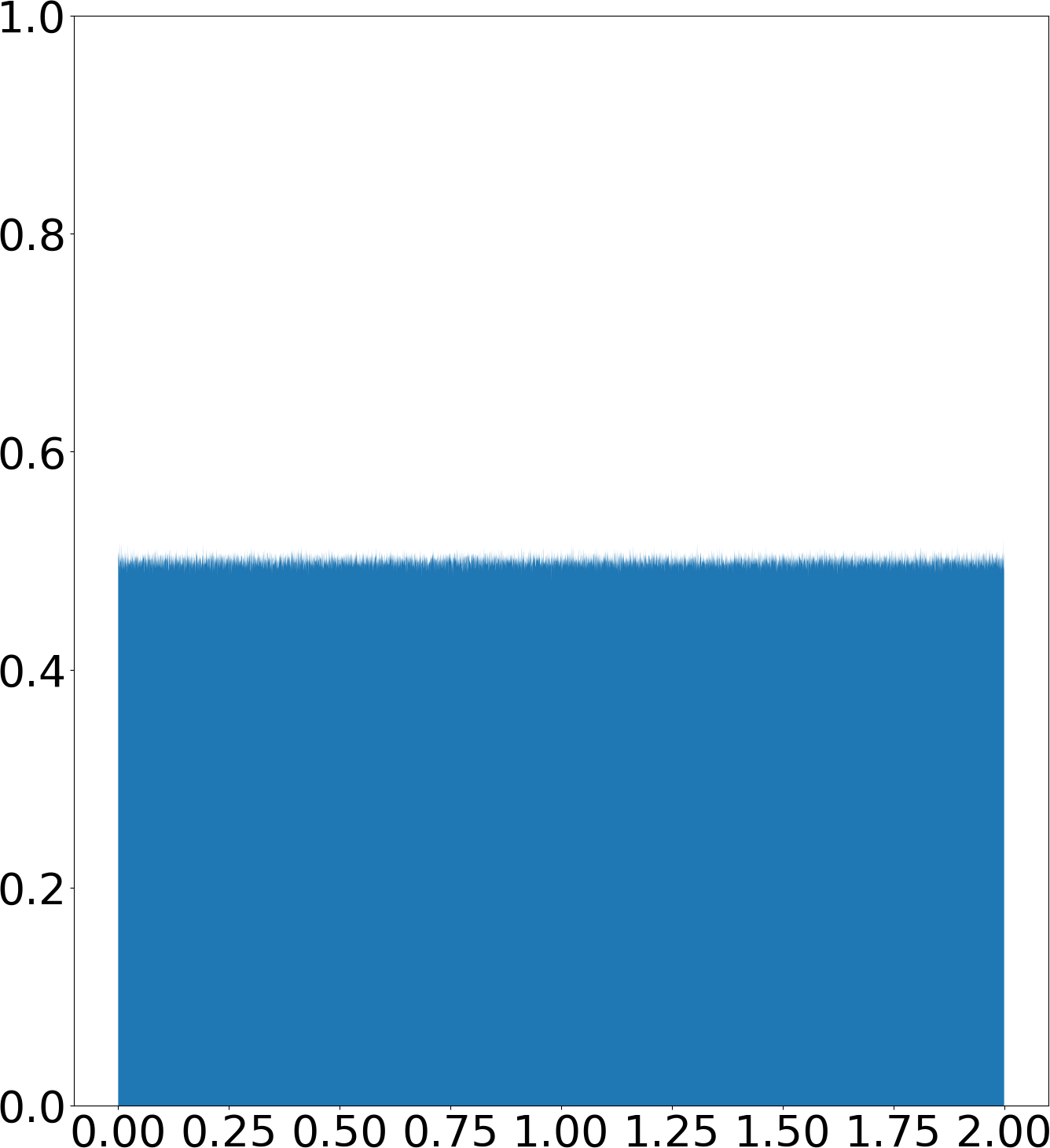}
    \\ \vspace{0.02\textwidth}
    \includegraphics[width=0.48\textwidth]{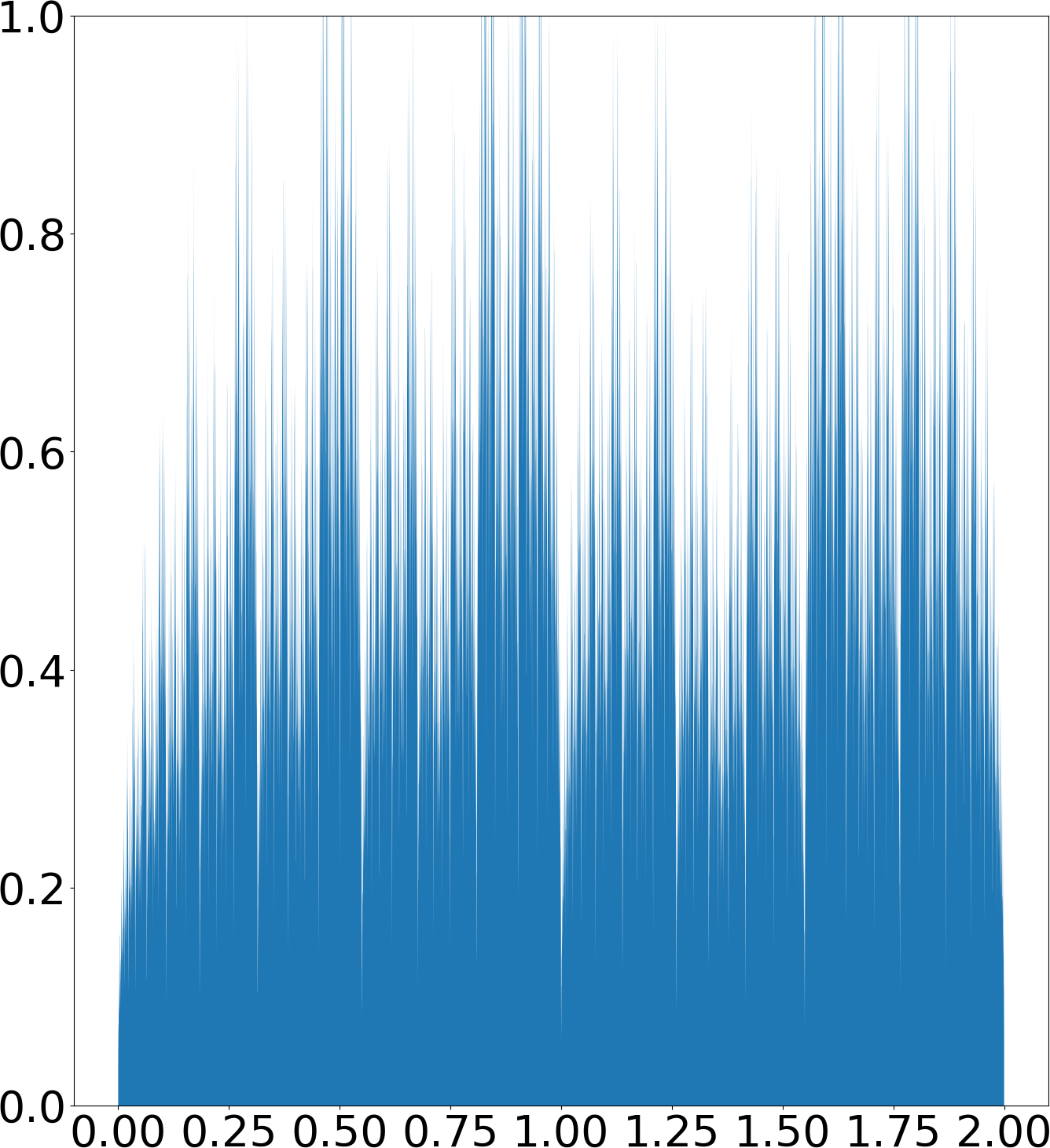}
    \hspace{0.02\textwidth}
    \includegraphics[width=0.48\textwidth]{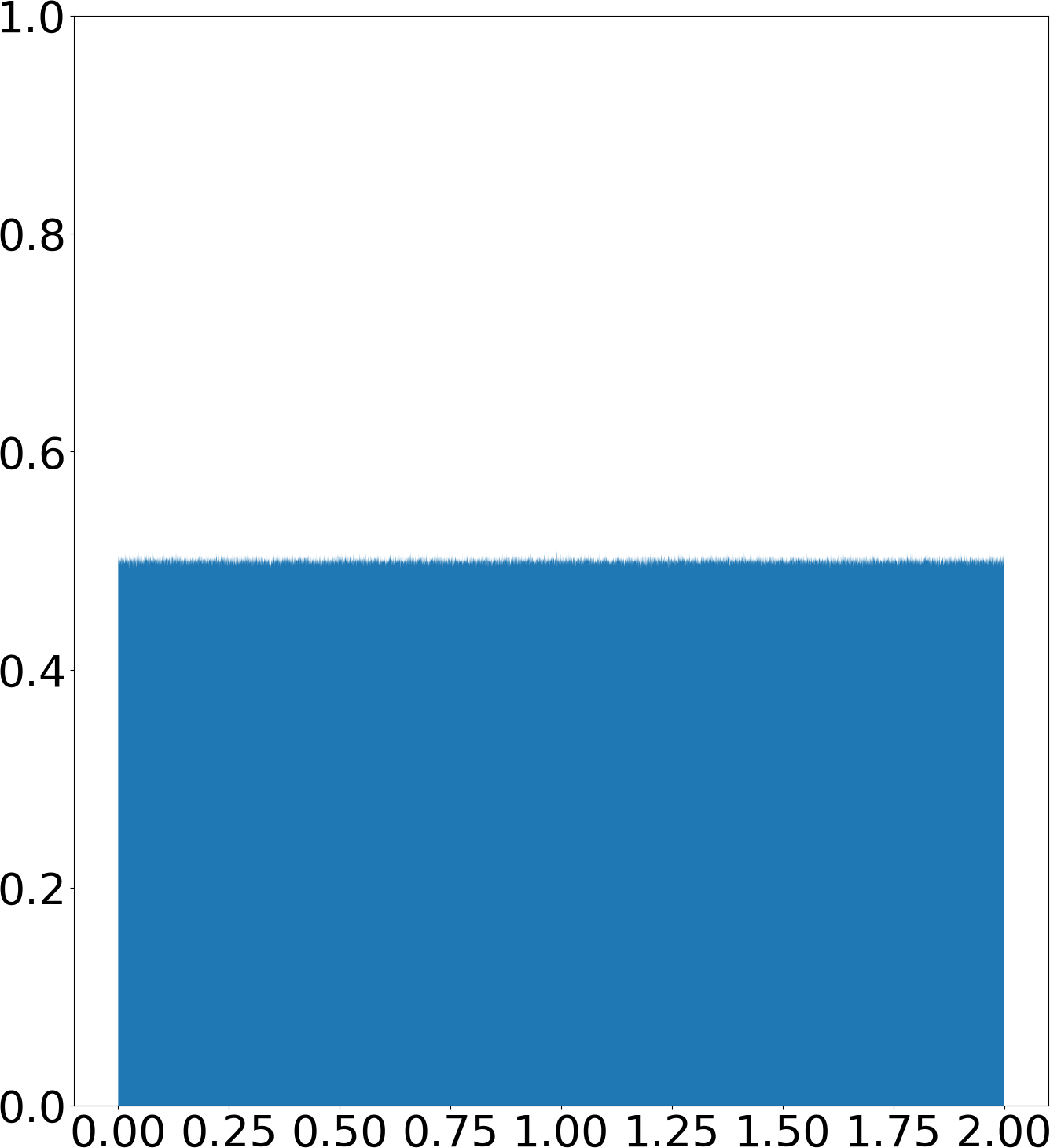}
    \caption{The left column shows the empirical distribution of the shadowing solution for the wave tent map (Eq. \ref{eqn:tent_wave}) at $s=0.05$ (top) and at $s=0.2$ (bottom). The right column shows the probability distribution of a physical solution at the same two values of $s = 0.05$ (top) and $s = 0.2$ (bottom).}
    \label{fig:tent_wave_shadow}
\end{figure}

Next we consider a second perturbation of the tent map, the \emph{pinched} tent map, which has the following closed form
\begin{align}
    \hat{\varphi}^p_s(x) = \begin{cases}
        \dfrac{4x}{1 + s + \sqrt{(1+s)^2 - 4s x} }, & x < 1 \\
        \dfrac{4(2-x)}{1 + s + \sqrt{(1+s)^2 - 4s(2- x)} }, & 2 \leq x \leq 1.
    \end{cases}
    \label{eqn:tent_pinched}
\end{align}
Again, at $s=0$, the original tent map is recovered, and at other values of $s,$ the tent map is ``pinched'' by perturbations that are symmetric around $x = 1$, and zero at the end points. Figure \ref{fig:tent_pinched} shows the pinched tent map at different $s$ values.  
In Figure \ref{fig:tent_pinched_shadow}, we show the corresponding plots of the 
physical and shadowing distributions for the pinched tent map. From the right 
column, which shows the probability density of a long physical solution, we 
can see that the pinching perturbation changes the uniform density of the original tent map to a linearly varying density. The more pronounced the perturbation -- the higher the value of $s$ -- the steeper the slope. The shadowing distribution, in this case as well, appears fractal. Once again, with probability one -- or, for any randomly chosen initial condition of the solution of the pinched tent map -- a physical solution is shadowed by a quasi-physical solution.
We consider yet another perturbation of the tent map with the closed form
\begin{align}
\hat{\varphi}^w_s(x) = \begin{cases}
        \dfrac{4x}{1 + s + \sqrt{(1+s)^2 - 4s x} }, & x < 1 \\
        \dfrac{4(2-x)}{1 - s + \sqrt{(1-s)^2 - 4s(2- x)} }, & 2 \leq x \leq 1.
    \end{cases}
    \label{eqn:tent_wave}
\end{align}
The above map, called the wave tent map, is illustrated in Figure \ref{fig:tent_wave}.
To construct the pinched tent map (Eq. \ref{eqn:tent_pinched}), we perturbed the original tent map in such a way that the perturbations on the two halves of the interval $[0,2]$ are mirror images of each other about the $x=1$ line. In the wave tent map, the perturbations on the two halves are a reflection of each other about the 
$x$-axis.
The wave perturbation does not noticeably alter the SRB density of the original tent map. As we show in Figure \ref{fig:tent_wave_shadow} (right), a solution of the 
wave tent map, starting from almost any point on $[0,2]$ -- a physical solution -- has a density that looks almost identical to a uniform density of 0.5. On the left column of Figure \ref{fig:tent_wave_shadow}, we observe that the distributions of the shadowing solutions corresponding to the physical solutions on the right once again appear fractal.  
Hence, we can once again conclude that almost every solution, which is physical, 
has a corresponding shadowing solution that is nonphysical. In this section,
we have shown that the same conclusion holds for several different perturbations 
of the tent map. The physical solutions were distributed nonuniformly 
and differently from one another. However, we observed a commonality in the different perturbations: the shadowing solutions corresponding to almost every physical solution had a fractal-like probability distribution that did not resemble the distributions of the physical solutions in any case. Thus, we have shown significant evidence, through analytical constructions of perturbed tent maps in this section and 
section \ref{sec:quasiphysical} that shadowing solutions can almost surely be nonphysical.

\subsection{What about the Lorenz equation?}
In order to analyze the behavior of the shadowing solutions of the Lorenz'63 system 
of equations, we first make a closed form approximation of the Lorenz map (see section \ref{sec:lorenzMapApprox} of the supplementary material). Except at rare parameter values, the Lorenz system does not have 
solutions that shadow perturbed orbits for all time, but rather only for a finite time 
\cite{lorenzPOTP}. Hence, the Lorenz map also does not have infinitely long shadowing solutions. This implies that a numerical solution of the Lorenz map approximates 
a true solution, as accurately as desired, only for a finite time.

A closed form approximation is constructed by regression performed using long solutions 
of the Lorenz map (Figure \ref{fig:lorenz_map}). 
\begin{figure}
    \centering
    \includegraphics[width=0.32\textwidth]{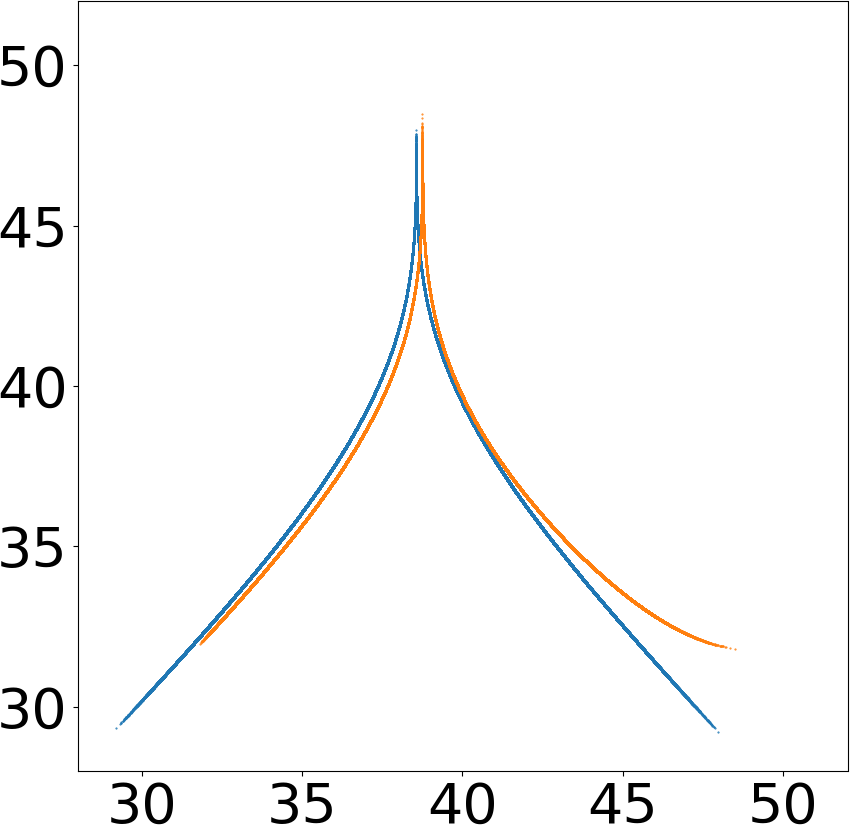}
    \hspace{0.005\textwidth}
    \includegraphics[width=0.32\textwidth]{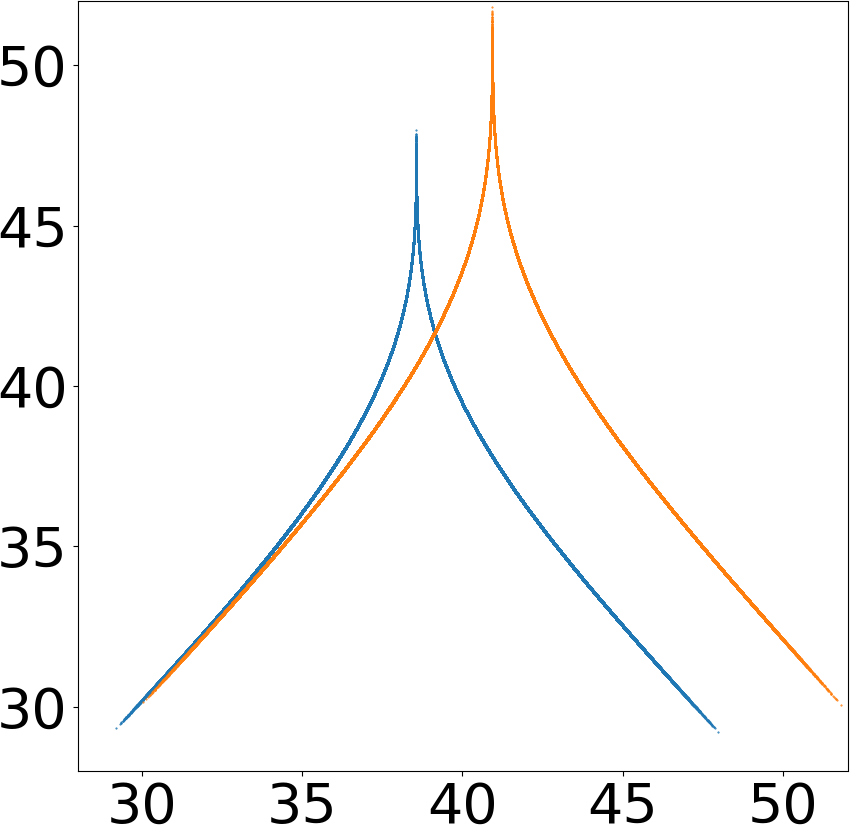}
    \hspace{0.005\textwidth}
    \includegraphics[width=0.32\textwidth]{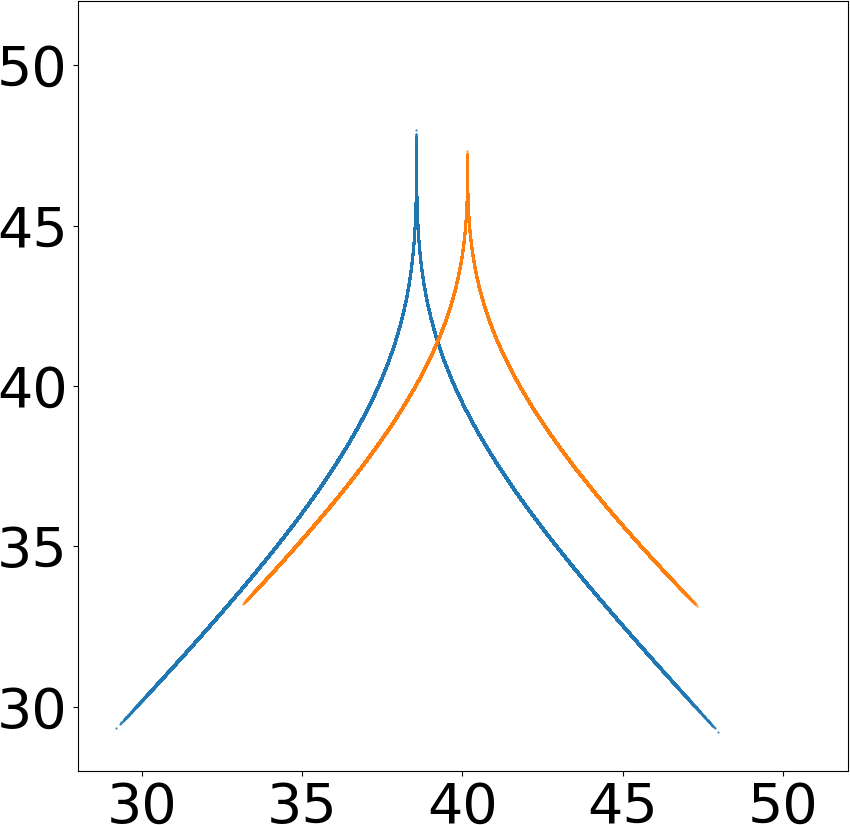}
    \caption{
The left, center, and right plots show the effect of the
    parameters $\sigma, \rho$, and $\beta$ on the Lorenz map, respectively.
    In the left plot, the blue and orange lines represent
    $\sigma=10$ and $12$, respectively, while $\rho  = 28$ and $\beta = 8/3.$
    In the center plot, the blue and orange lines represent
    $\rho=28$ and $30$, respectively, while $\sigma = 10$ and $\beta  = 8/3.$ 
    In the right plot, the blue and orange lines represent
    $\sigma=8/3$ and $10/3$ respectively, while $\sigma = 10$ and 
		$\rho = 28.$}
    \label{fig:lorenz_params}
\end{figure}
The Lorenz map approximation obtained this way is illustrated in Figure \ref{fig:lorenz_params}, for variations of the parameters around their standard values of 
$\rho = 28, \sigma = 10,$ and $\beta = 8/3.$ Using the approximate Lorenz map, we calculate next the shadowing solutions. 
\begin{figure}
    \centering
    \includegraphics[width=0.48\textwidth]{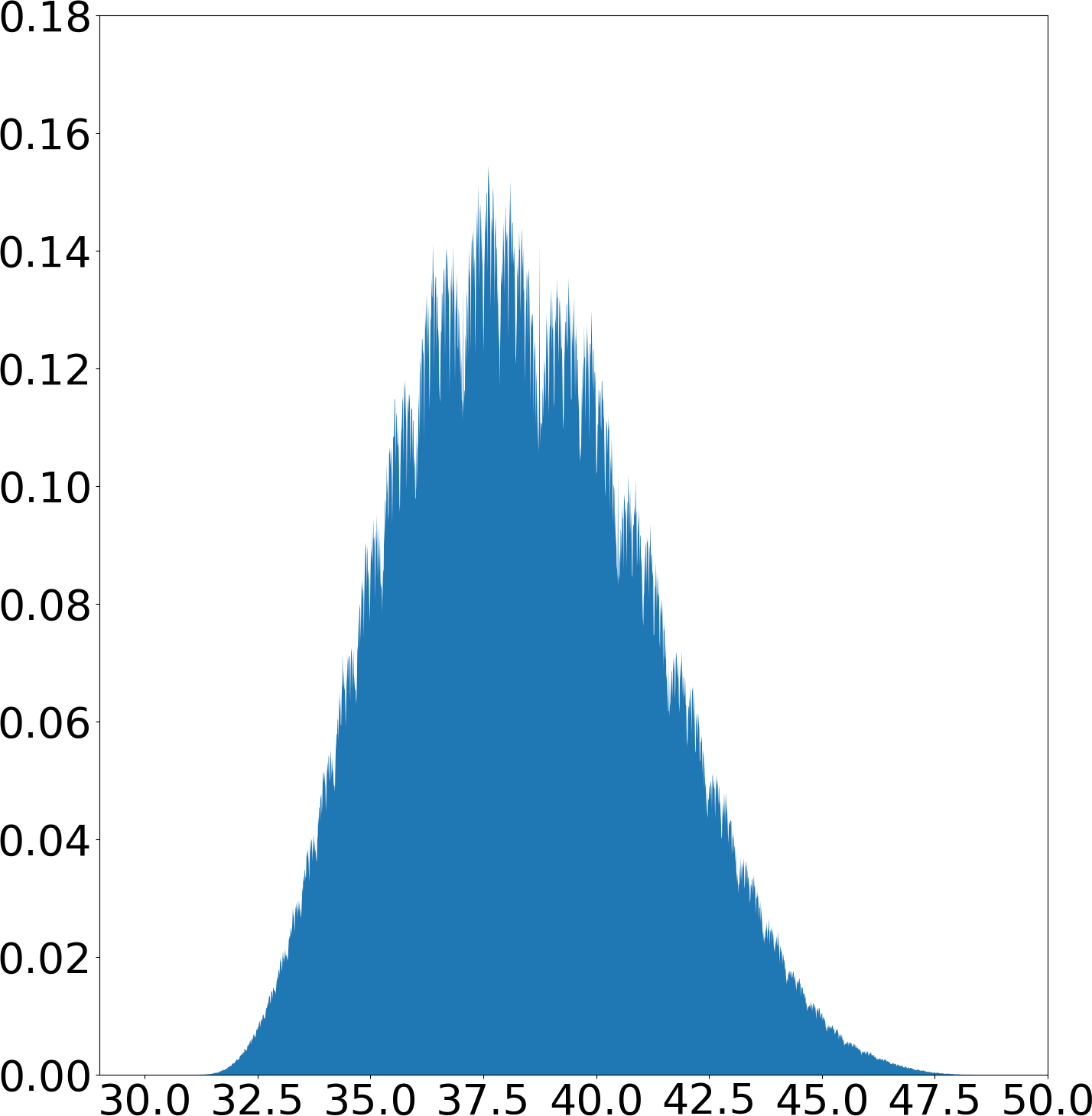}
    \hspace{0.005\textwidth}
    \includegraphics[width=0.48\textwidth]{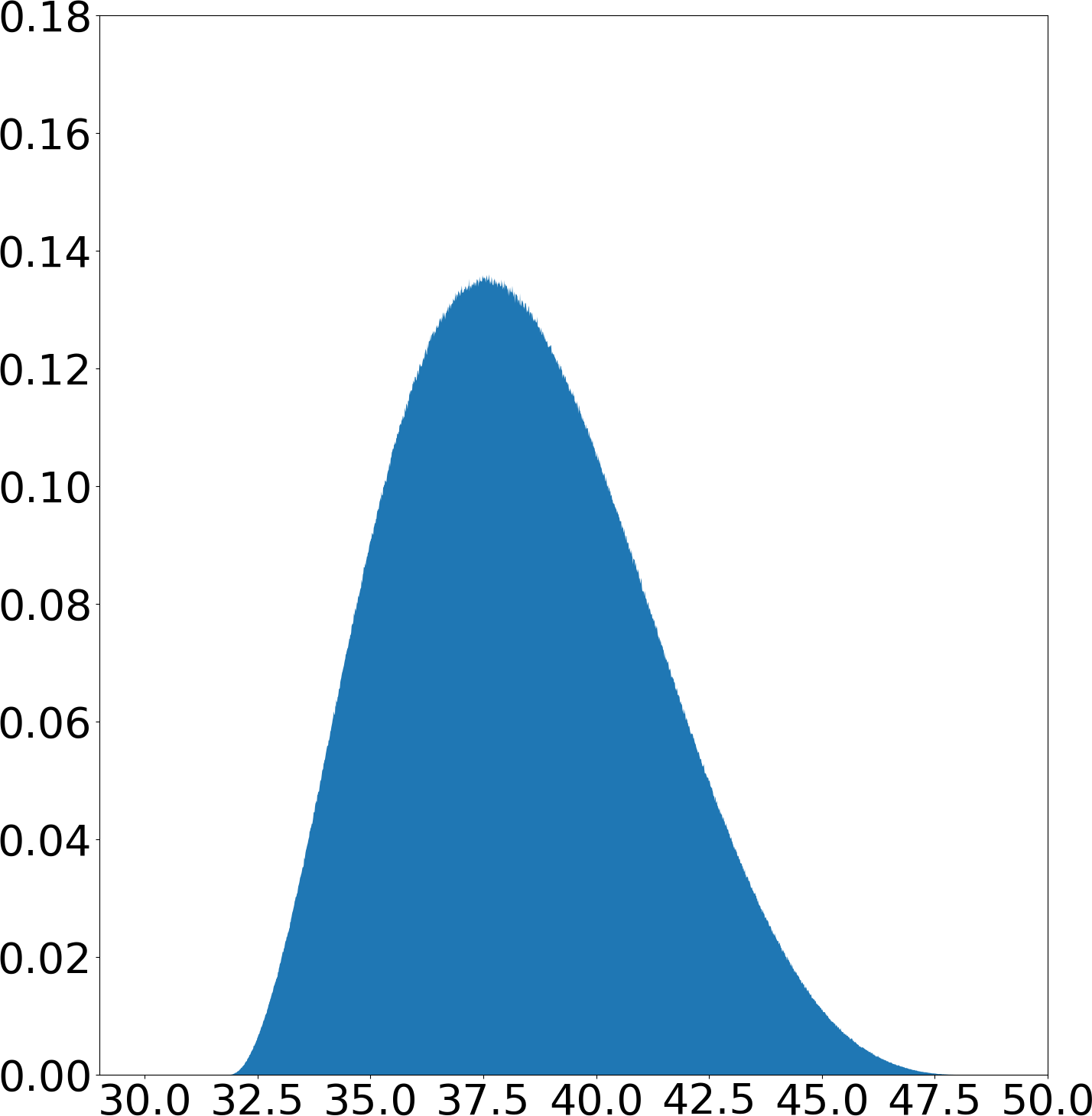}
    \\
    \includegraphics[width=0.48\textwidth]{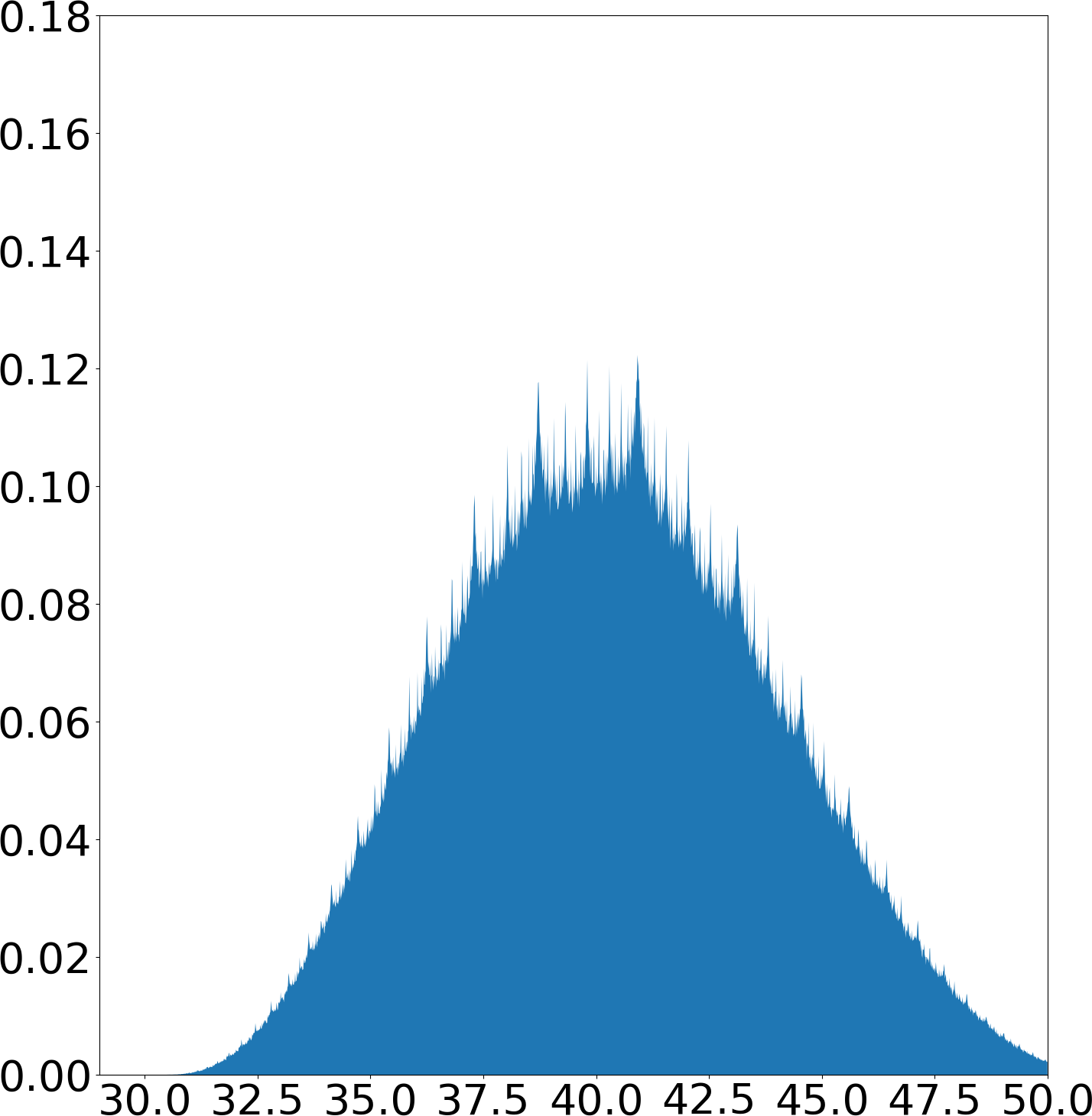}
    \hspace{0.005\textwidth}
    \includegraphics[width=0.48\textwidth]{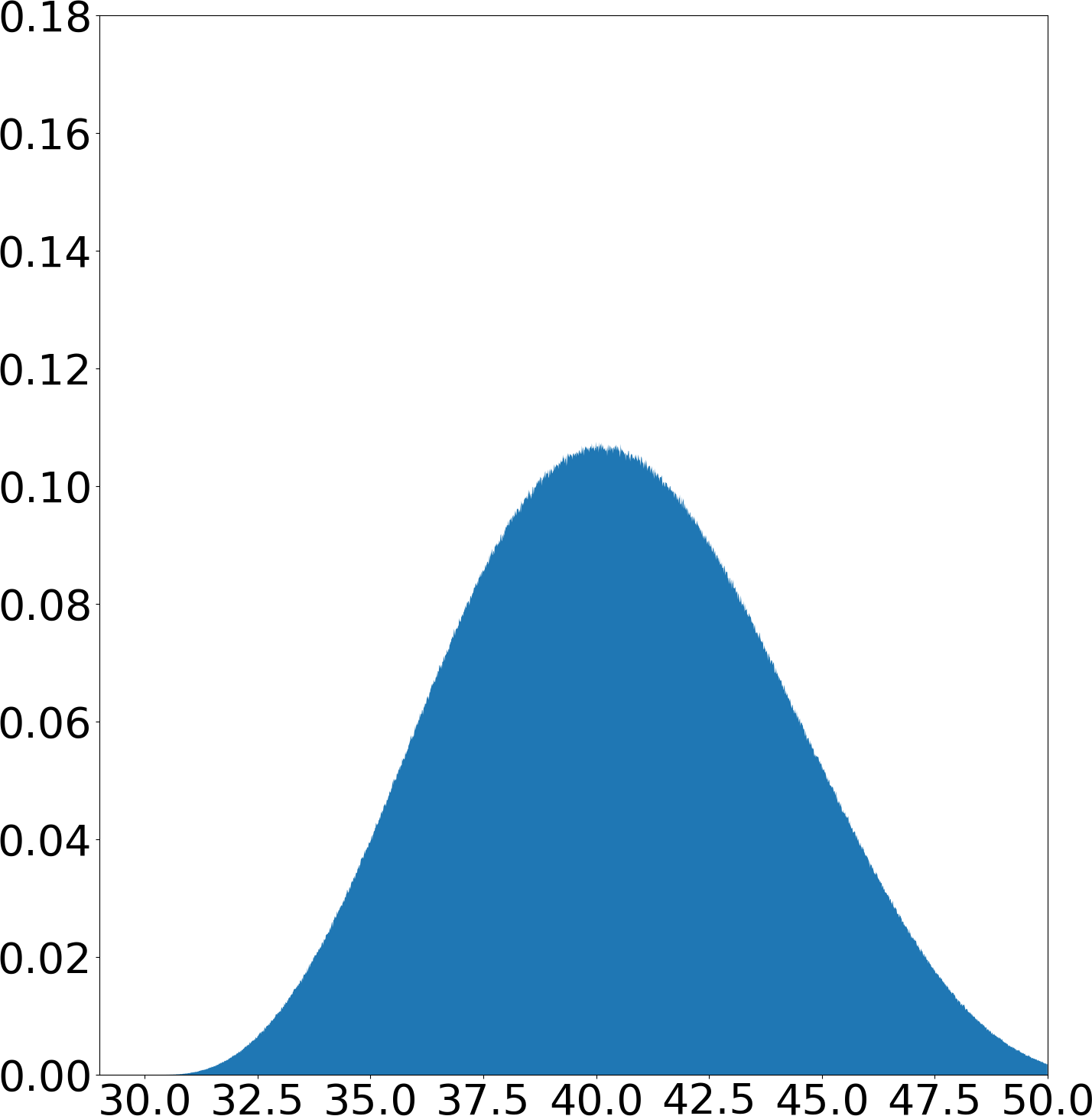}
    \\
    \includegraphics[width=0.48\textwidth]{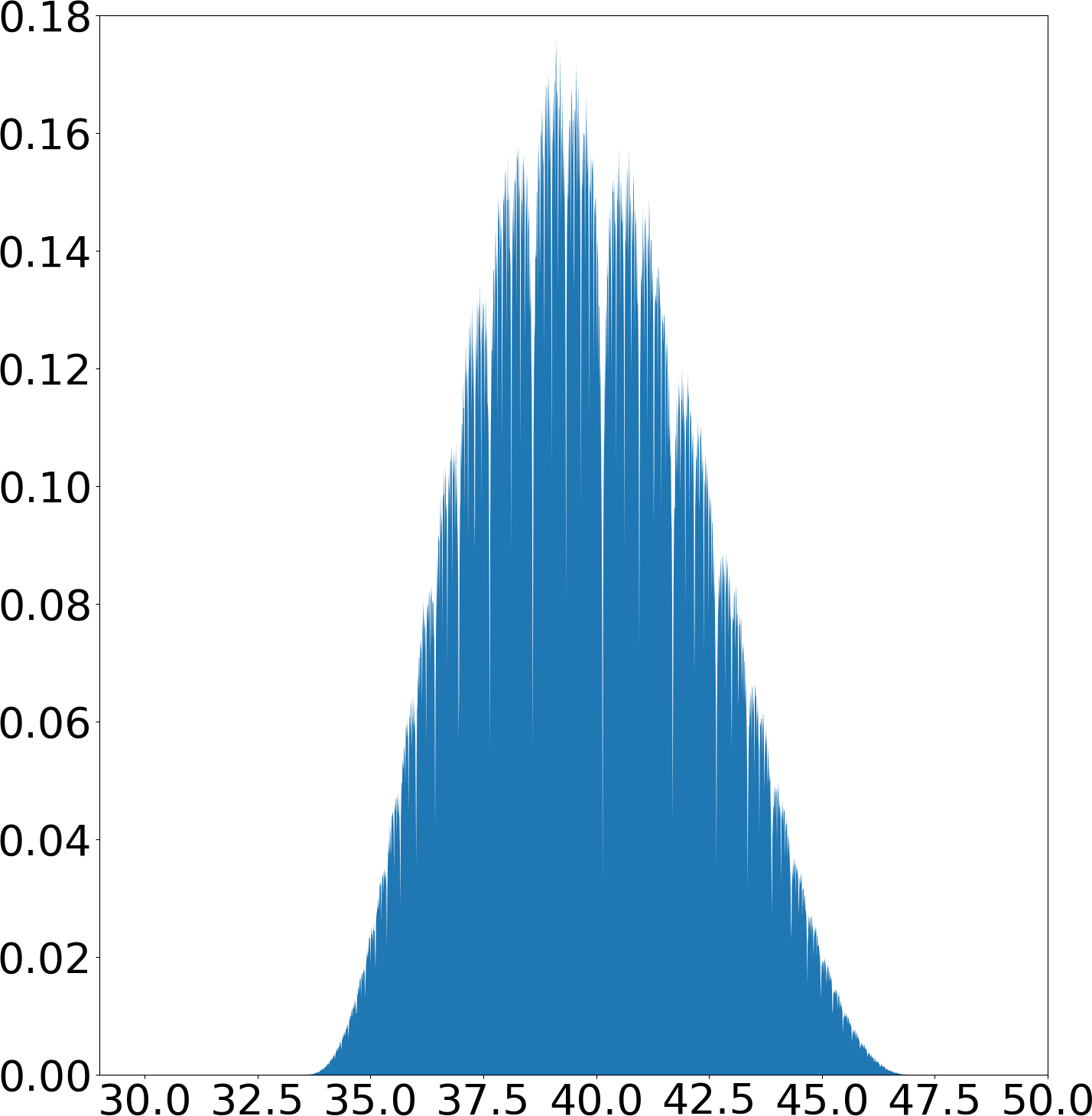}
    \hspace{0.005\textwidth}
    \includegraphics[width=0.48\textwidth]{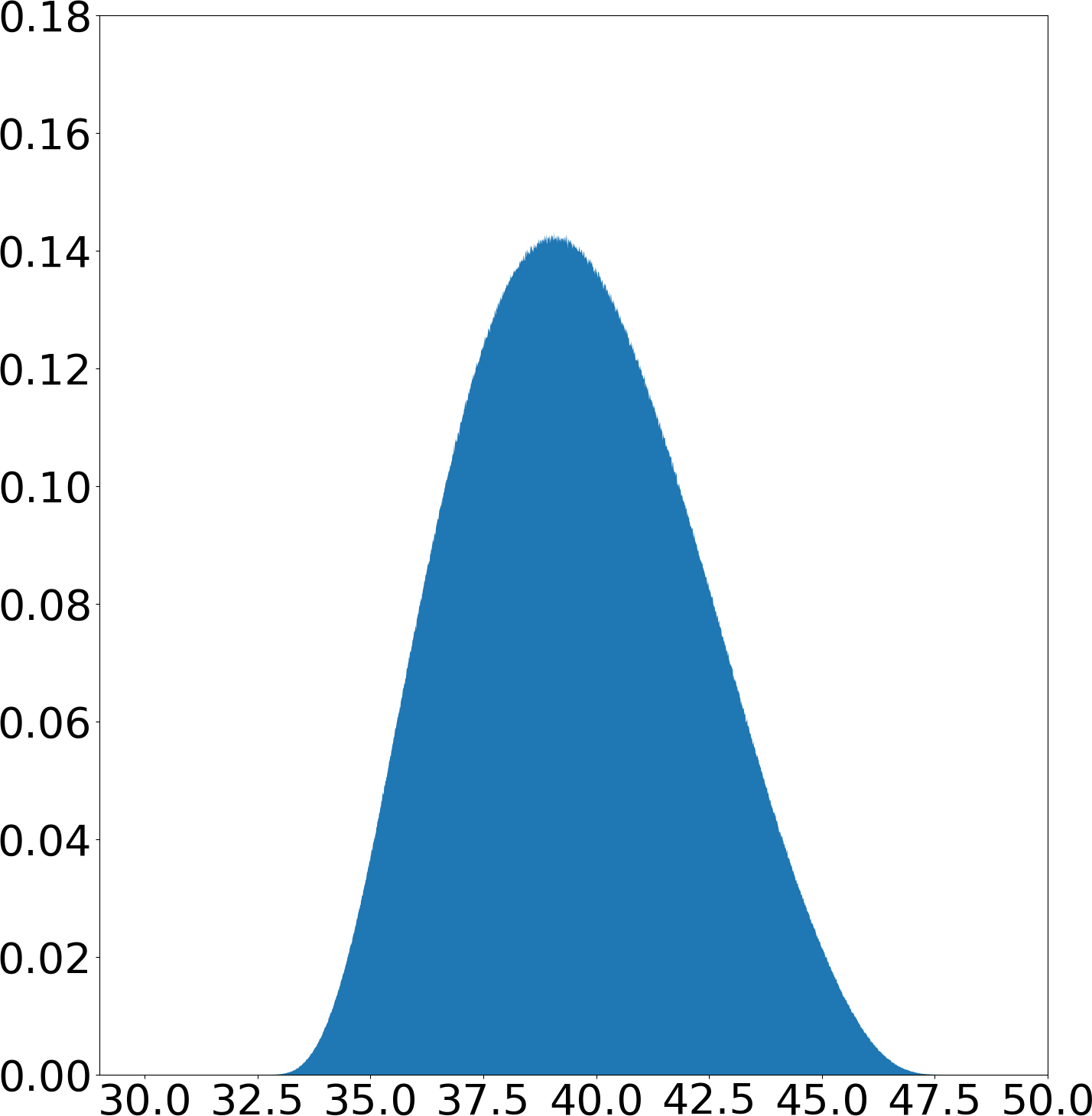}
    \caption{The left column shows the empirical distribution of the shadowing solution, and the right column shows the physical distribution of the Lorenz map at the following sets of parameters: top row: $\sigma = 15, \rho = 28, \beta = 8/3$, $\rho = 29$, middle row: $\sigma = 10, \rho = 30, \beta = 10/3$, and bottom row: $\sigma = 10, \rho = 28, \beta = 10/3$.
    }
    \label{fig:lorenz_params_scaled}
\end{figure}
The empirical distribution computed from long shadowing solutions, of length 10 billion, are shown in the left column of Figure \ref{fig:lorenz_params_scaled} at three different sets of parameter values, which are also different from the standard values of $\rho = 28, \sigma = 10$, and $\beta = 8/3$. 
In each row, one of the parameters is perturbed from the standard values. On the right column, we plot the distribution of physical solutions, the SRB distribution, at the set of parameters corresponding to each row. We again notice that, while the physical distributions appear smooth, the shadowing solutions clearly exhibit roughness. Thus, we draw the same conclusion in the Lorenz map as in the various perturbations of the tent map treated in section \ref{sec:tentPerts}: shadowing solutions do not have the same distribution as a typically observed solution of the governing equation. In other words, shadowing solutions are nonphysical.

\section{Discussion and outlook}
\label{sec:discussion}
\begin{figure}
    \centering
    \includegraphics[width=0.3\textwidth,height=1.35in]{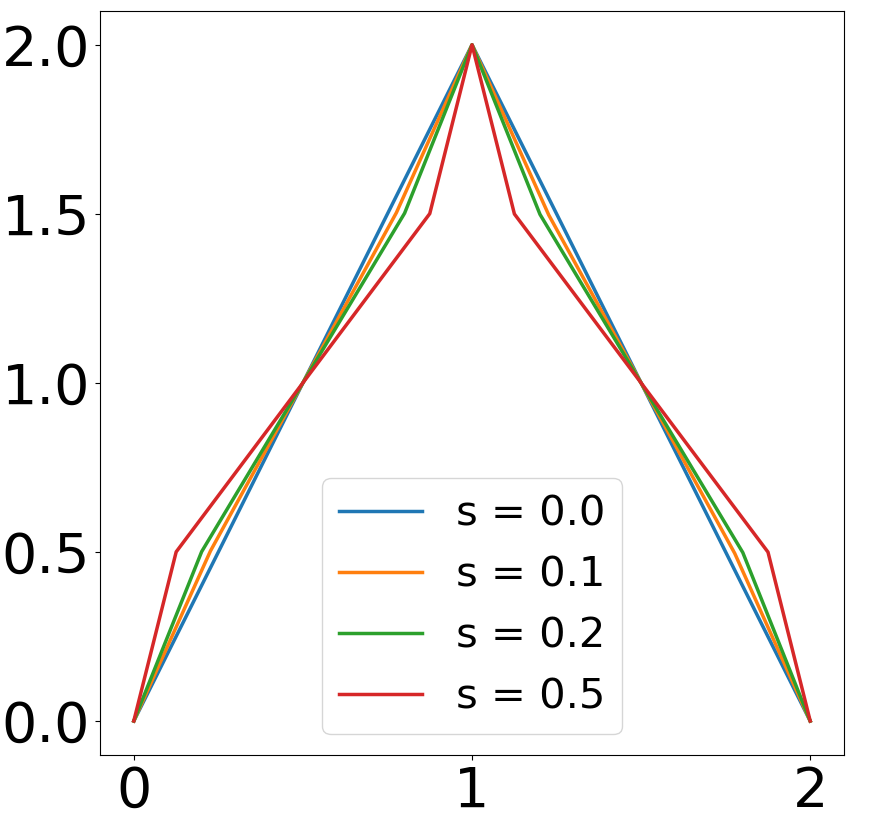}
     \includegraphics[width=0.6\textwidth,height=1.35in]{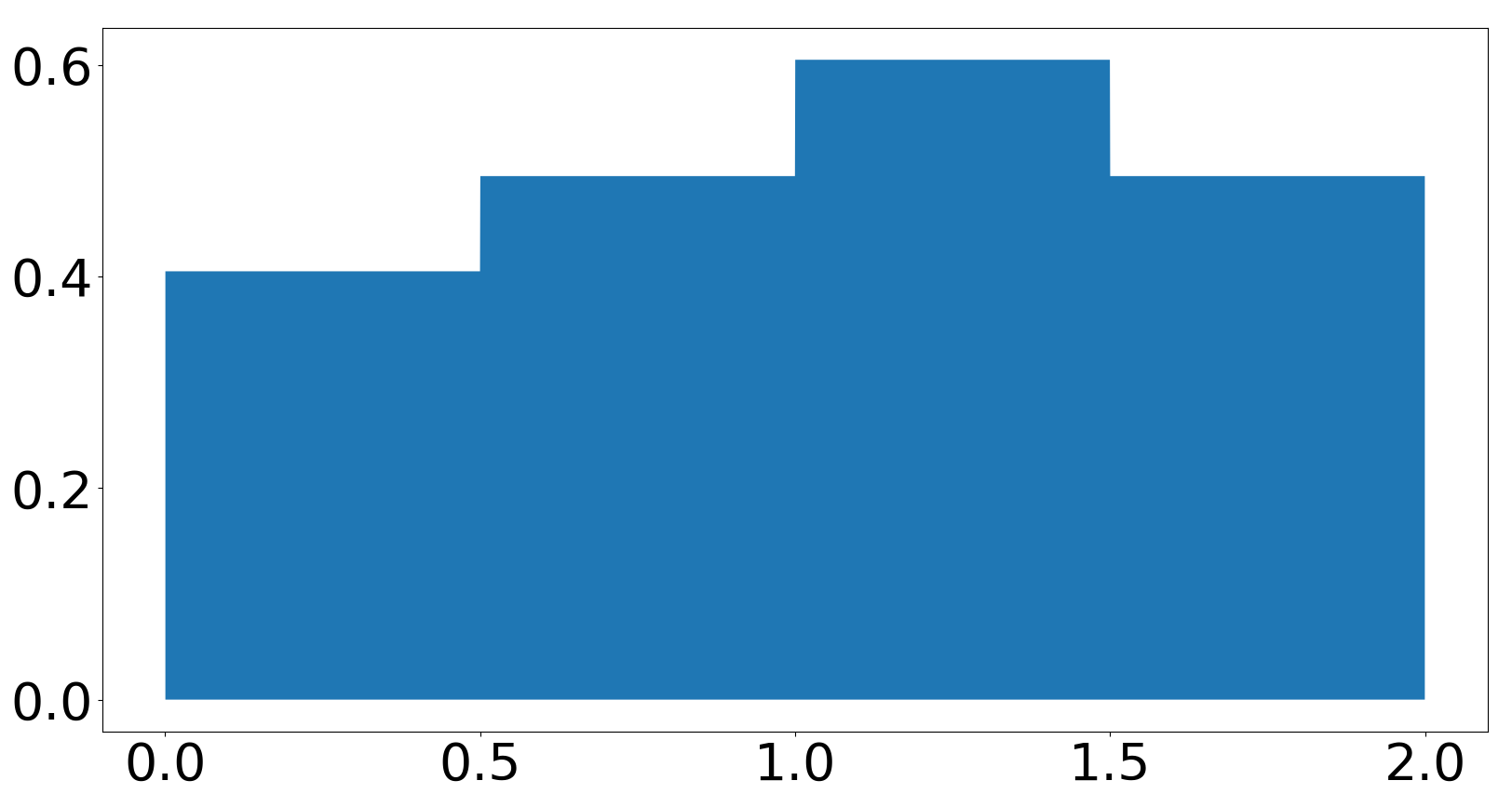}
    \includegraphics[width=0.3\textwidth,height=1.35in]{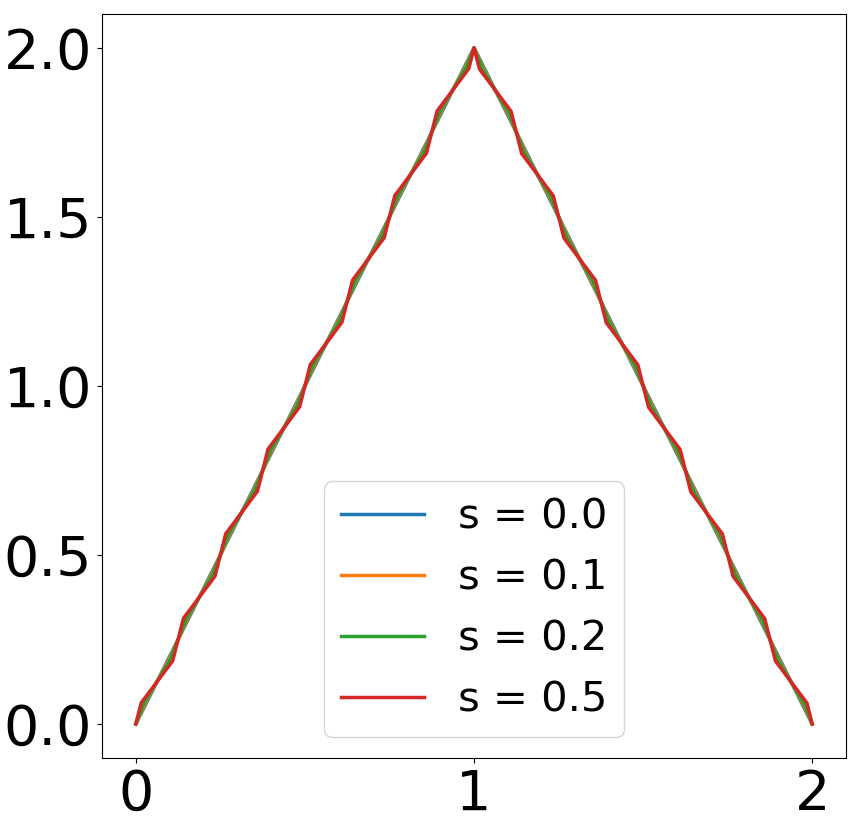}
     \includegraphics[width=0.6\textwidth,height=1.35in]{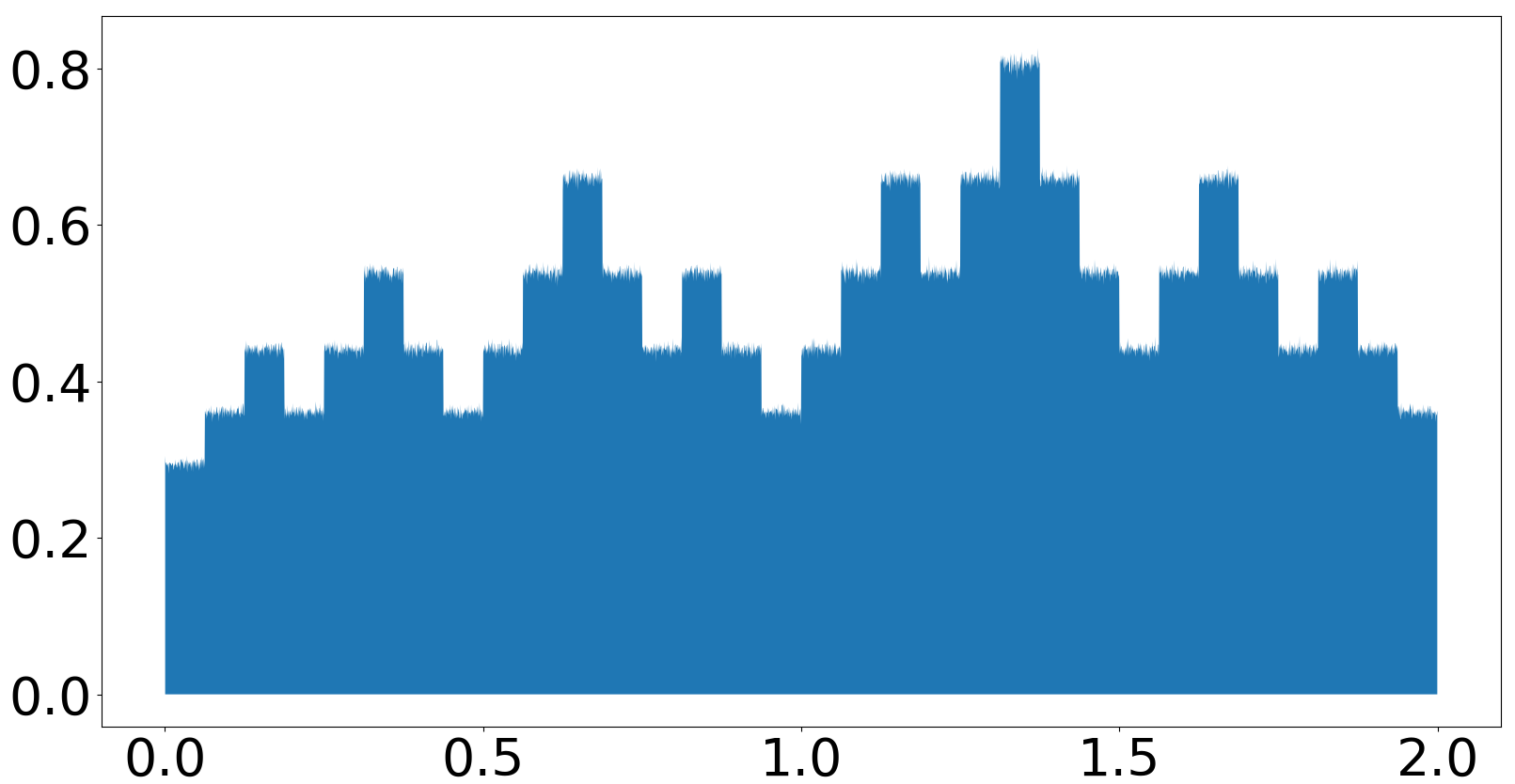}
    \includegraphics[width=0.3\textwidth,height=1.35in]{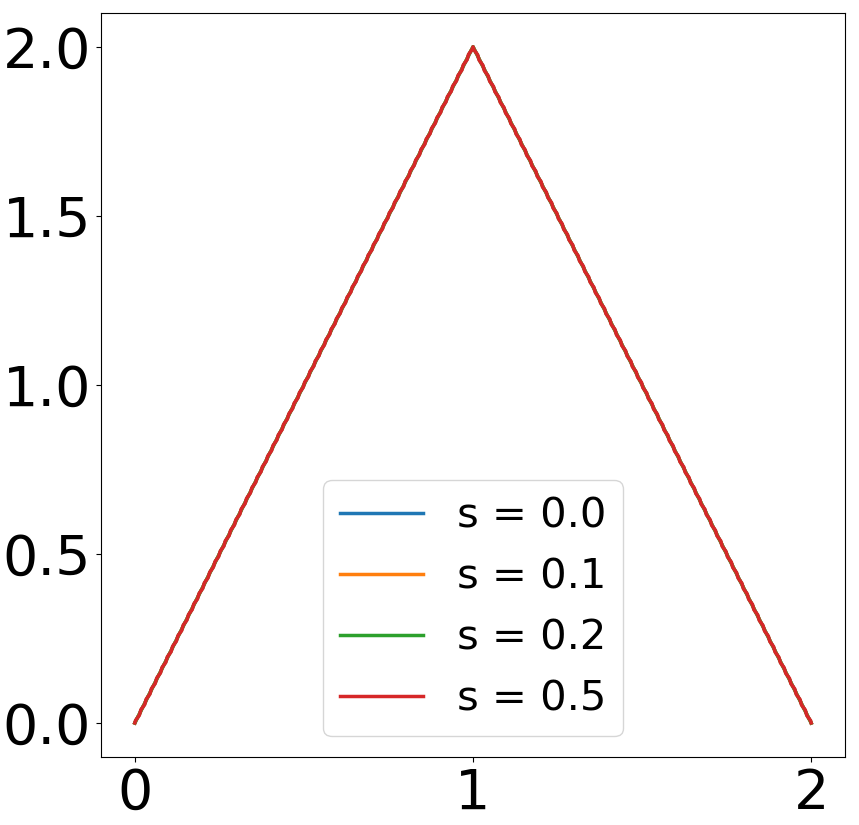}
     \includegraphics[width=0.6\textwidth,height=1.35in]{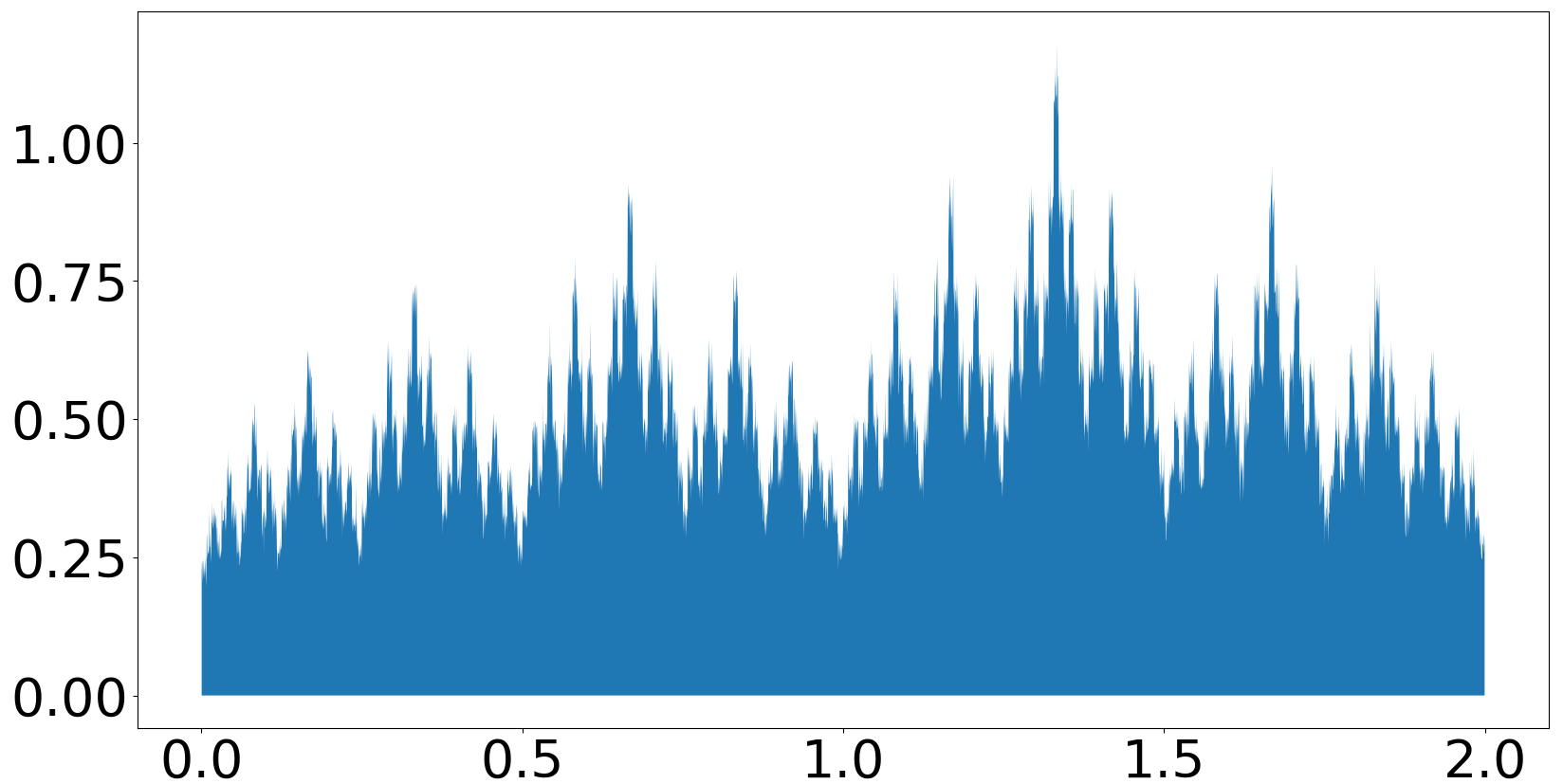}
    \caption{
    The effect of increasing $n$ on the physical probability distribution associated to the plucked tent map. The left column shows the plucked tent map at 
    different values of $s$ and $n = 0$ (top), $n=3$ (middle), $n=6$ (bottom). The original tent map is at $s=0$ on each plot.
    The right column shows the stationary, physical probability distribution of the plucked tent map at $s=0.1$ and $n = 0$ (top), $n=3$ (middle), $n = 6$ (bottom).}
    \label{fig:pluckedTentMap}
\end{figure}

Through rigorous counterexamples, we show that shadowing can lead
to nonphysical solutions. This has a troubling implication for numerical simulations of chaotic
governing equations.  Through the examples in this paper, it is clear that a numerical solution, satisfying
a governing equation perturbed due to numerics, is not 
expected to be shadowed by a
physical solution of the real governing physics.  What is, if any, the
relation of a numerical solution, such as DNS of turbulent flows, to
the true physics? For an ${\cal O}(10^{-18})$ 
perturbation (due to numerical error), 
we may get completely different statistics from a true, physical solution. How then can we trust numerical solutions when they are not guaranteed to share the long-term statistical behavior of the governing equation? 
We conclude with an example that illustrates a counterintuitive feature of some chaotic systems: a small perturbation to the 
governing equation can significantly change the statistics/long-term 
behavior of its physical solutions. We construct yet another 
perturbed tent map -- the \emph{plucked} 
tent map -- in which an oscillatory perturbation is introduced (see Supplementary Material section \ref{sec:pluckedTentMap} for the map equation). In Figure \ref{fig:pluckedTentMap}, we show that the magnitude of the oscillatory perturbation is controlled by parameters $s$ and $n$, and its frequency is controlled by the parameter $n$; at $s=0$, we recover the original tent map at all $n$. The physical probability distributions -- computed empirically over a trajectory of length 10 billion -- at $s=0.1$, are shown on the right column of Figure \ref{fig:pluckedTentMap}. On the top row, $n=0$, and there is already a marked asymmetry developed in the physical distribution compared to the uniform distribution, which is the physical distribution seen at $s=0, n=0$. The figure shows that by increasing $n,$ despite the fact that the magnitude of the perturbation becomes smaller with $n$, we see a dramatic change in the appearance of the stationary probability distribution: we observe an apparent fractal distribution reminiscent of that of the quasi-physical solutions in 
section \ref{sec:tentPerts}. By construction of the map (Supplementary Material section \ref{sec:pluckedTentMap}), the nonuniformities of the probability distribution at $n=0$ are transferred to smaller and smaller scales, 
as $n$ is increased. It is worth emphasizing that these remarkable changes in the physical distribution (shown on the bottom-right of Figure \ref{fig:pluckedTentMap}) are effected with a tiny perturbation -- the perturbed map at $n=6$ and the original tent map appear indiscernible on the bottom-left of Figure \ref{fig:pluckedTentMap}. 

In view of the plucked tent map, we can further question the
validity of numerical solutions. Can numerical solutions play 
the role of the plucking perturbation to the 
governing equation? That is, can numerical solutions represent 
slight perturbations to the governing equations for which the physical 
distribution is drastically different from the physical distribution 
of the original governing equation? Sauer \cite{sauer} has shown 
examples in which numerical error due to computation in double-precision floating point arithmetic causes significant change in the stationary probability distribution on the attractor; other works \cite{cns} further analyze the unreliability due to double-precision arithmetic of statistical measures such as power spectra and correlation functions. The plucked tent map adds to this list of examples by demonstrating a mechanism to produce an extreme non-smooth response (Supplementary Material section \ref{sec:pluckedTentMap}). Uniformly hyperbolic 
dynamical systems exhibit \emph{linear response} 
(\cite{ruelle}\cite{baladi}) by which small parameter perturbations 
lead to small changes in statistics, which can be expanded as Taylor series around the reference parameter value. But, uniform hyperbolicity is a mathematical idealization, and although some physical systems have been observed to behave as if they were uniformly hyperbolic \cite{gallavotti}, a violation of uniform hyperbolicity is more likely 
\cite{wormell}\cite{luca}. 

Both the nonphysicality of shadowing solutions, and the existence of extremely non-smooth statistical response, undermine the validity of using shadowing for sensitivity analysis of statistics \cite{qiqi-lss}\cite{angxiu-lss}\cite{lasagna}. When the goal is to compute derivatives of ensemble averages, where the ensemble
is distributed according to the physical measure, i.e., the SRB measure, shadowing-based methods can give wrong results. This is because shadowing-based methods compute the sensitivities of ensemble statistics along shadowing solutions, but these may not be physical solutions that reproduce ensemble statistics. The error in shadowing sensitivities has been observed
before and appropriately attributed as ``ergodicity breaking error'' \cite{patrick} \cite{angxiu-error}. 

Violations of smooth response, like in the plucked tent map, have an immediate implication for shadowing sensitivities as well. In particular, if perturbed solutions are shadowed by physical solutions for all time, the change in the statistics of the perturbed solutions must be small. In other words, the physicality of shadowing predicts that there cannot be a large change in statistics due to small parameter perturbations; the reality is that, as illustrated by the plucked tent map, the effect of small parameter changes on the statistics can be drastic.

In this paper, we have constructed several counterexamples that 
dispel the notion that shadowing solutions are physical solutions, i.e., that their 
statistical distribution is the same as that typically observed for almost every solution 
of a governing equation. The existence of long-time shadowing solutions \cite{grebogi} has historically been used to address the issue of whether numerical simulations, which are 
perturbed solutions, represent the true physics implied by the governing equation. In light of the evidence in this paper, we must reopen this issue. Even when numerical simulations are 
shadowed by a true solution in that the difference between them is small for a long time, this shadowing solution may not represent the long-term or ensemble behavior of the physical system. The nonphysicality of shadowing solutions also indicates that shadowing-based methods can lead to incorrect values of sensitivities of statistics to parameter changes. 

\vspace{1in}

\textbf{Funding:} This work was supported by Air Force Office of Scientific Research Grant No. FA8650-19-C-2207.
\textbf{Conflict of Interest:} The authors declare that they have no conflict of interest.

\bibliographystyle{amm}
\bibliography{refs.bib}
\section{Supplementary material}
\label{sec:supp}
The supplementary material for this paper including the code and data to generate the figures can be found on Github \cite{supp}. The code can be found under the \verb+code+ subdirectory inside which section-wise code is separated into further subdirectories. The data used to plot the figures can be found under \verb+data+. The files referred to in this section can be found in the appropriate subdirectory under \verb+code+. 
\subsection{Approximation of the Lorenz Map}
\label{sec:lorenzMapApprox}
The motivation for approximating the Lorenz map is that a 
closed form expression for the map is necessary for our numerical shadowing procedure (section \ref{sec:computingShadowing}). In a small region around the cusp, we approximate the map using an exponential function. The tails on both sides are fitted with a sum of a cubic polynomial and a rational function. Thus, the approximate Lorenz map has the 
following closed form expression:
\begin{align}
    \varphi(x) = \begin{cases}
        z_{\rm max} - f_R(x - z_{\rm sep})   &  x > z_{\rm sep} \\
        z_{\rm max} -  f_L(z_{\rm sep} - x)   &  x \leq z_{\rm sep}
    \end{cases}
\end{align}
where 
\begin{align}
    f_R(y) = \Big( (1000 y)^p + 
        \sum_{n=0}^3 p_{R,n} y^n + \dfrac{a_{R,0} + a_{R,1} y}{\sum_{n=0}^3 b_{R,n} y^n}\Big),
\end{align}
and,
\begin{align}
    f_L(y) = \Big((1000 y)^p 
        + \sum_{n=0}^3 p_{L,n} y^n + \dfrac{a_{L,0} + a_{L,1}y}{\sum_{n=0}^3 b_{L,n} y^n}\Big).
\end{align}
\begin{figure}
		\includegraphics[width=0.45\textwidth]{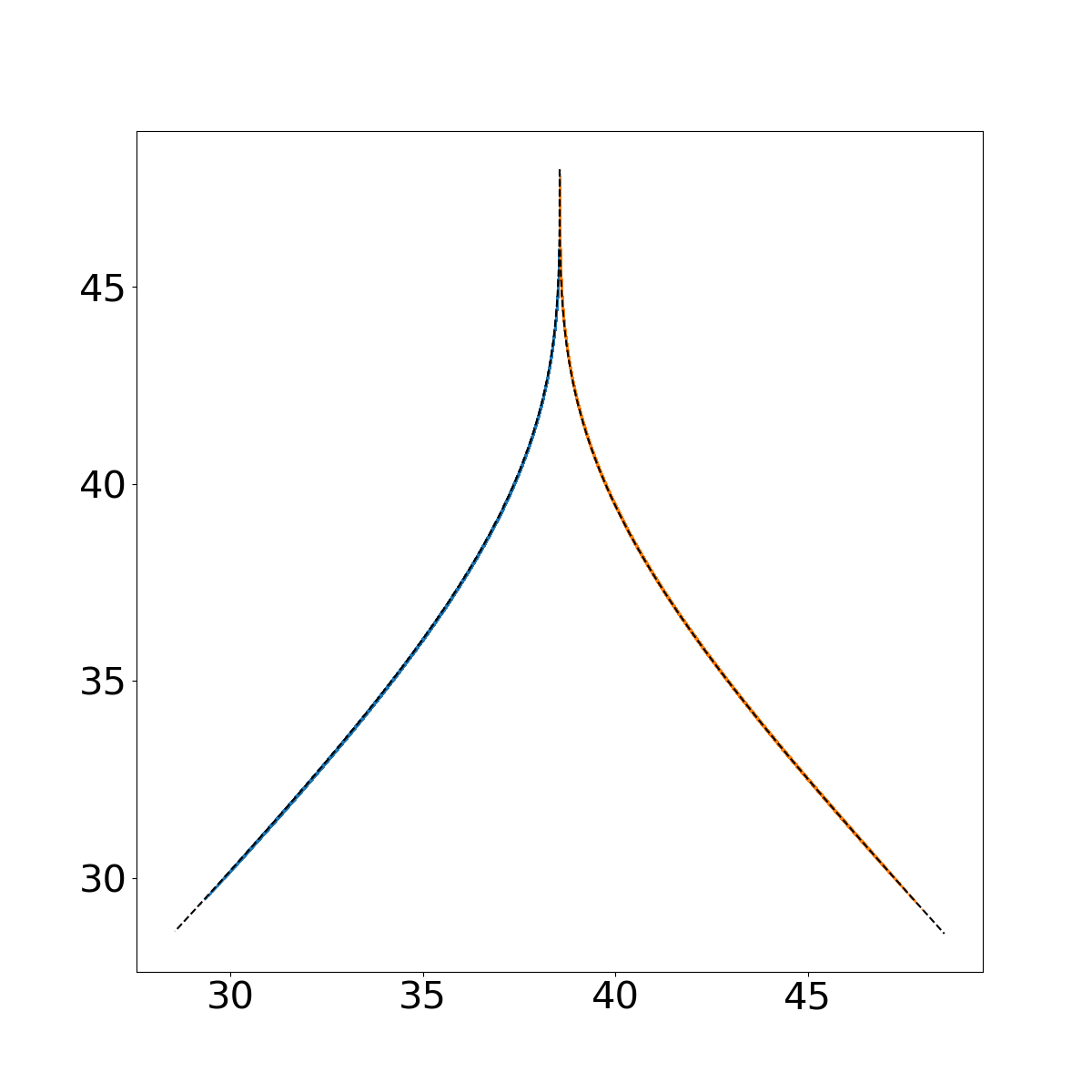}
		\includegraphics[width=0.45\textwidth]{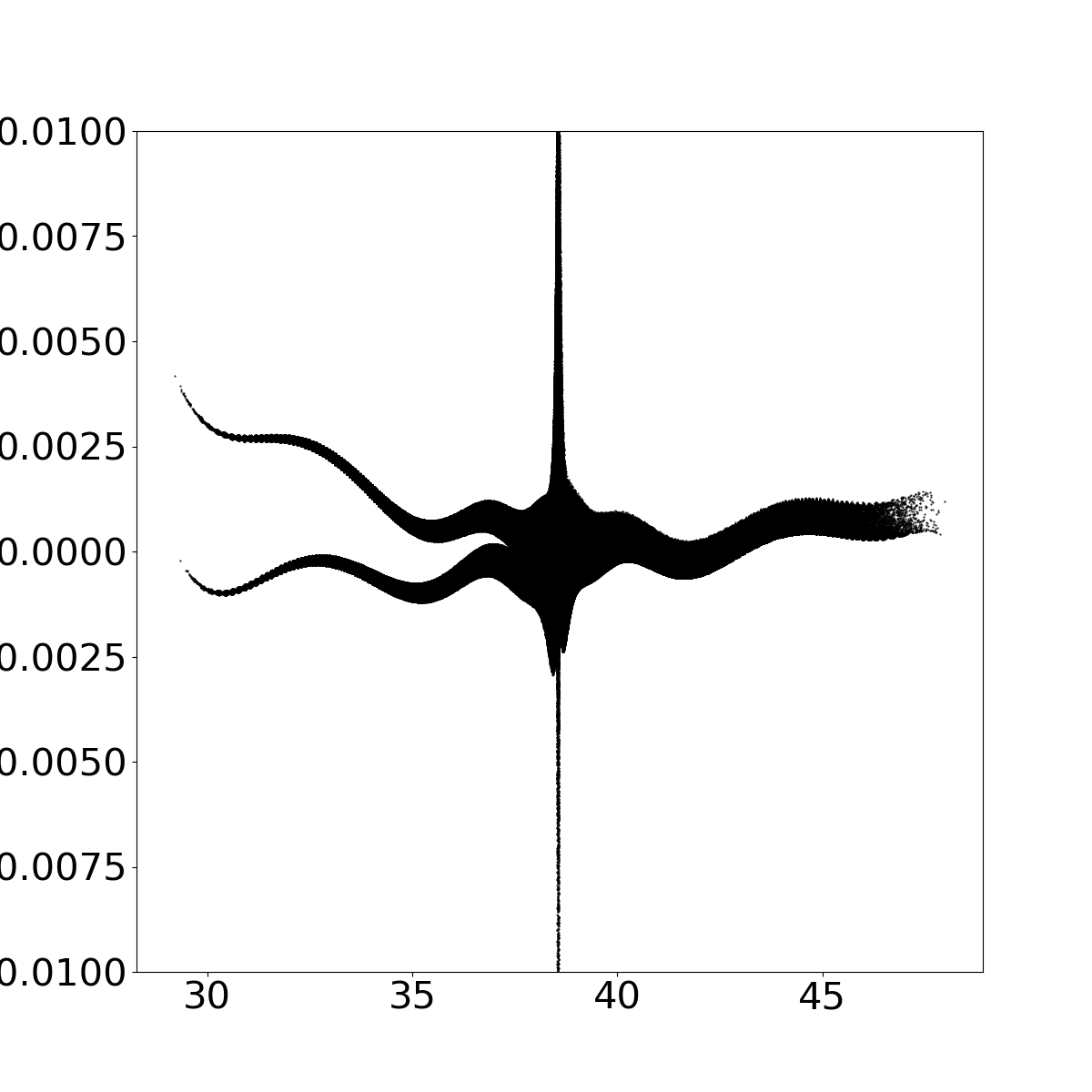}
		\caption{L: the blue and orange lines are the Lorenz map at the standard parameters, and the dotted black lines indicate the approximate Lorenz map (Table \ref{tab:lorenzFit}). R: the regression error is shown as a function of $z$} 
	\label{fig:lorenzMapComp}
\end{figure}
The location of the cusp is denoted $z_{\rm sep}$ (in which ``sep'' stands 
for \emph{separation}), and the maximum and minimum values encountered 
in the Lorenz map iterates are denoted $z_{\rm max}$ and $z_{\rm min}$, respectively. The exponent of the cusp is denoted $p$. The coefficients of the cubic polynomial modelling the left (right) tail are  denoted $p_{L,n}, n = 0,1,2,3$ ($p_{R,n}, n = 0,1,2,3$). The coefficients of the numerator and denominator of the rational function modelling the left (right) tail are denoted $a_{L,0}, a_{L,1}$ ($a_{R,0}, a_{R,1}$) and $b_{L,n}, n=0,1,2,3$ ($b_{R,n}, n=0,1,2,3$), respectively. The values of these coefficients, which are obtained by
regression, are shown for the map at standard parameters, in the table below. As shown in Figure \ref{fig:lorenzMapComp}, the fit obtained matches the Lorenz map closely.
\begin{table}[H]
    \centering
    \begin{tabular}{|c|c|}
    \hline 
         $z_{\rm sep}$ = 38.55302437476555 &
         $z_{\rm min}$ =  29.213182255013322 \\ 
         $z_{\rm max}$ =  47.978140718671284 & 
         $p$ = 0.28796740575434676   \\
         $p_{L,3}$ = -0.00024683786275242047 & $p_{L,2} = $ 
         0.016174566354858824 \\
         $p_{L,1} = $ 0.40179772568004946 & 
         $p_{L, 0}  = $ -0.24612651488351725 \\
         $p_{R,3} = $ -0.00020712463321688308 & 
         $p_{R,2} = $ 0.017130843276711716 \\
         $p_{R, 1} = $ 0.3930080703420676 & 
         $p_{R, 0} = $ -0.23471384266765036 \\
         $a_{L,1}$ = -0.05405742075580959 & $a_{L,0}$ = -0.05405742075580959 \\
         $a_{R,1}$ = -0.05351127783496397 & $a_{R,0}$ = 0.22489891059122896 \\
         $b_{L,3}$ = 0.5609397451213353 & $b_{L,2}$ = -0.3491184293228338 \\  $b_{L,1}$ = 2.419972619058592 & $b_{L,0}$ = 1.0 \\
         $b_{R,3}$ = 0.6456076059873844 & $b_{R,2}$ = -0.34840383986411055 \\ $b_{R,1}$ = 2.6035438510917692 &  $b_{R,0}$ = 1.0 \\
         \hline
    \end{tabular}
    \caption{The fitting parameters of the Lorenz map at $\sigma = 10, \beta = 8/3$, and $\rho = 28.$}
    \label{tab:lorenzFit}
\end{table}

\subsection{Computing shadowing solutions}
\label{sec:computingShadowing}
		In general, to numerically compute the shadowing solutions, one could use existing methods such as 
		the least squares shadowing method \cite{qiqi-lss}. However, since the  maps we consider are all one-dimensional chaotic systems that have in common non-invertibility with two inverse \emph{branches}, we devise a simpler approach. The map $\varphi_s$ in this section refers to any of the perturbations of the tent map (section \ref{sec:tentPerts}) or the Lorenz map. Suppose we are given a perturbed solution $x_n, n =0,1,\cdots,N$ that we must compute a shadow of. 
The shadowing solution, $y_n, n=0,1,2,\cdots,N$ 
must a) satisfy the governing equation: $y_{n+1} = \varphi_s(y_n)$
and b) lie close to the given perturbed solution at all times up to 
$N$, i.e., $|x_n - y_n| < \epsilon$, for some $\epsilon > 0$, for all $n \leq N$. Suppose we set $y_0 = x_0$, the difference $y_1 - x_1$ will amplify on further iterations under $\varphi_s$ for any perturbed solution. On the other hand, backward iteration of the map is contracting, and thus we set $y_N = x_N$. Then, we proceed backward in time to 
construct a solution of $\varphi_s$, noting that any difference that emerges at a given time $n$ will be made smaller starting at $n-1$. Each point has two pre-images under $\varphi_s$, and thus there are two possible choices for $y_{N-1}$, each lying in one of two fixed sub-intervals separated by the cusp of $\varphi_s$. Due to the contraction of errors backward in time, as long as we choose a pre-image in the same subinterval as the perturbed solution, we are guaranteed to approximate a shadowing solution. Marching backward by choosing at each step, the pre-image $y_n$ in the same subinterval as $x_n$, $y_n$ approximates the shadowing solution better as $n$ decreases. Thus, the procedure to find a shadowing solution simply reduces to solving for a backward trajectory (specifically one among the possible $2^N$), starting at a given final condition,  $x_N$.
	
		Hence, it is clear that all we need is the inverse of 
		$\varphi_s$, which is propagated backward by choosing the same \emph{branch} of the inverse as $x_n$ at time $n$. This logic is implemented for each map of section \ref{sec:tentPerts} in the function \verb+shadow+ that can be found in the files named for each map (for example, the shadowing solution of the pinched tent map can be found by executing the \verb+shadow+ function of \verb+tent_shadow/tent_shadow_pinched.py+). These functions use the analytical inverses of the maps, which are easy to derive for the tent map perturbations of section \ref{sec:tentPerts}. For the Lorenz map, we use Newton's method to solve for the inverse, and this is implemented in the file \verb+lorenz_map/shadow.py+. Note that we need a closed form expression of the map, for the Newton's method, and this is indeed the reason why we approximate the map, as described in section \ref{sec:lorenzMapApprox}.

\subsection{The \emph{plucked} tent map}
\label{sec:pluckedTentMap}
We provide a recursive definition of the plucked tent map, which 
is illustrated at different values of $n$ and $s$ in Figure \ref{fig:pluckedTentMap}. First we define a function $f_s(x)$, which 
creates a bend in the tent map that increases with $s$, around 
$((1-s)/2,1)$:
\begin{align}
		f_s(x) = 		{\rm min}\Big(
				\dfrac{2x}{1-s}, 2 - \dfrac{2(1-x)}{1+s}\Big),
				\;\;\; x < 1. 
\end{align}
Then, we introduce oscillations using repetitions of
the above map within the unit interval,
\begin{align}
		o_s(x) = \begin{cases}
				f_s(2x)/2, & \; x < 0.5 \\
				2 - f_s(2-2x)/2, & \; x \ge 0.5.
		\end{cases}
\end{align}
We can modulate
the frequency of repetitions -- proportional to $n$ -- 
of the oscillations through
\begin{align}
		\label{eqn:frequency}
		\lambda_{s,n}(x) = \dfrac{o_s(2^n x - \floor*{2^n x})}{2^n} 
		+ 2 \frac{\floor*{2^n x}}{2^n},
\end{align}
where $\floor*{x}$ is the greatest integer less than or equal to $x$.
Finally, the plucked tent map is defined as the above function 
in the unit interval, and as its reflection about $x=1$,
in the interval $[1,2)$.
\begin{align}
		\label{eqn:plucked}
		\varphi_{s,n}(x) = {\rm min}\left(
		\lambda_{s,n}(x), \lambda_{s,n}(2-x)\right), \; 0< x < 2. 
\end{align}
Recall that our motivation is to construction a slight perturbation of the tent map whose stationary probability distribution is not just 
non-uniform (e.g. like the other tent map perturbations of section \ref{sec:tentPerts}), but in which the nonuniformity can be controlled, and made skewed as desired. Thus, we choose to construct a base perturbation of the tent map, ${\rm min}(o_s(x), o_s(2-x))$ -- 
which is also obtained by setting $n=0$ in $\varphi_{s,n}$ --  
in which an asymmetry is produced in the probability distribution (top-right of Figure \ref{fig:pluckedTentMap}).
Upon repeating, with appropriate scaling, the oscillation $o_s$, 
the frequency of which is controlled by $n$, we obtain $\varphi_{s,n}$, 
as indicated by Eq. \ref{eqn:frequency}-\ref{eqn:plucked}. Thus, the asymmetric probability distribution, through this process of scaled repetition, acquires an apparent fractal-like structure seen on increasing $n$, as shown on the bottom-right of Figure \ref{fig:pluckedTentMap}.

One way to see the asymmetry in the probability distribution 
about $x=1$, at $n=0$, is to construct a Markov chain with nodes 00, 01, 10, and 11, that indicate the subintervals $(0,0.5)$, $[0.5,1)$, 
$[1,1.5)$ and $[1.5,2)$, respectively. From Eq. \ref{eqn:plucked}, we can see that the application of $\varphi_{s,0}$ yields the following transition matrix:
\[
\begin{bmatrix}
		(1-s)/2\quad & (1+s)/2 \quad & 0 \quad & 0 \\
		0 \quad & 0 \quad & (1+s)/2 \quad & (1-s)/2 \\
		0 \quad & 0  \quad & (1+s)/2 \quad & (1-s)/2 \\
		(1-s)/2 \quad & (1+s)/2 \quad & 0 \quad & 0
\end{bmatrix}
.\] 
The stationary probability distribution of this Markov chain, which 
is the left eigenvector of the transition matrix corresponding to eigenvalue 1, is $[(1-s)^2/2, \; (1-s^2)/2, \; (1+s)^2/2,\; (1-s^2)/2]^T$.
One can verify this is consistent with the top-right of Figure \ref{fig:pluckedTentMap} for $\varphi_{0.1,0}$. Clearly, the time spent by an infinitely long physical solution in the second quarter interval, $[0.5,1]$ is about 1.2 times the time spent in the first quarter, $[0,0.5]$; the time spent in the third quarter $[1,1.5]$ is about 1.5 times that spent in the first quarter. This nonuniformity is more pronounced with increasing $s$. For instance, at $s=0.5$, the left subinterval, [0,1),is three times less likely to be visited by a long trajectory compared to $[1,2)$. 

But, as mentioned in the main text, even a small perturbation -- very small $s$ -- is sufficient to trigger a remarkable variation in the statistics. This is because this particular construction transfers 
the nonuniformities in the probability distribution at larger scales to smaller scales, with increasing $n.$ The nonuniformity at the largest 
scale of half intervals is retained at higher values of $n.$ That is,
the probability of visiting the interval $[0,1]$ is the same: $(1-s)$, at all $n$. However, with increasing $n,$ the nonuniformity emerges within smaller subintervals, due to the construction of the map that relies on repeating the behavior at larger scales at smaller scales (Eq. \ref{eqn:frequency}). This repetitive construction is responsible for the apparent fractal structure of the probability distribution (Figure \ref{fig:pluckedTentMap}, bottom-right). The scripts that generate the plucked tent map and its 
physical probability distributions can be found under \verb+tent_sens_stat+. 
\end{document}